\documentclass[aps,twocolumn,a4paper,10pt]{revtex4}
\usepackage{mathrsfs}
\usepackage{graphicx}
\usepackage{latexsym}
\usepackage{amsmath}
\usepackage{amssymb}
\usepackage{textcomp}
\usepackage{amsbsy}
\usepackage{graphics}
\usepackage{braket}
\usepackage{graphicx}
\usepackage{graphicx}
\usepackage{amsmath,amssymb}

\newcommand\beq{\begin{equation}}
\newcommand\eeq{\end{equation}}
\newcommand\bea{\begin{eqnarray}}
\newcommand\eea{\end{eqnarray}}
\usepackage{graphicx}
\usepackage{graphicx}
\usepackage{xcolor}
\begin{document}
\title{\Large Canonical Transformations and Loop Formulation\\ of SU(N) Lattice Gauge Theories}
\author{Manu Mathur${}^*$   and T. P. Sreeraj}  
\email{manu@bose.res.in, sreerajtp@bose.res.in}
\affiliation{S. N. Bose National Centre for Basic Sciences,\\ Salt Lake, JD Block, Sector 3, Kolkata 700098, India} 
\begin{abstract}
  We construct canonical 
 transformations to reformulate  SU(N) Kogut-Susskind lattice gauge theory in terms of a set of fundamental   loop \& string  flux operators along with  their canonically conjugate loop \& string electric fields. 
 The canonical relations between the initial SU(N) link operators and the final SU(N)  loop \& string operators, consistent with SU(N) gauge transformations,  are explicitly constructed  over the entire lattice. We show that as a consequence of SU(N) Gauss laws all SU(N) string  degrees of freedom become cyclic and decouple from the physical Hilbert space ${\cal H}^p$.    
 %There are no local gauge or redundant loop degrees of freedom. 
 The Kogut-Susskind Hamiltonian rewritten in terms of the fundamental physical loop operators has global SU(N) invariance. There are no gauge fields.
  We further show that the $(1/g^2)$ magnetic field terms on plaquettes create and annihilate  the  fundamental plaquette loop fluxes while the $(g^2)$ electric field terms describe all their interactions. In the weak coupling ($g^2 \rightarrow 0$) continuum limit the SU(N) loop dynamics  is described by SU(N) spin Hamiltonian with nearest neighbour interactions.  
 %These  interaction)s reduce to  nearest neighbour plaquette loop interactions  in the weak coupling, $g^2\rightarrow 0$, continuum limit.      
  %The SU(N) loop Hamiltonian is shown to have a simple weak coupling ($g^2\rightarrow 0)$ limit. 
  %The physical loop  Hilbert space ${\cal H}^p$ is characterized by $(N^2-1)(d-1)$ gauge invariant SU(N) quantum number per lattice site.nerato ture
 In the simplest SU(2) case, where  the canonical transformations map the SU(2) loop Hilbert space 
  into the Hilbert spaces of hydrogen atoms, we analyze the special role of the hydrogen atom dynamical symmetry group $SO(4,2)$ in the  loop dynamics and
 the  spectrum. A simple tensor network ansatz 
 %for SU(2) lattice gauge theory low energy states  
 in the SU(2) gauge invariant 
 hydrogen atom loop  basis is discussed.
\end{abstract} 
%\abstract{\noindent We construct canonical 
%transformations to rewrite  the
%SU(N) lattice gauge theory Kogut-Susskind Hamiltonian in terms of a set of fundamental SU(N) plaquette loop  flux operators and their conjugate loop electric fields. We show that after canonical transformations, as a consequence of Gauss laws, the  local SU(N) gauge degrees of freedom become cyclic and drop out.  The canonical relations between the initial SU(N) Kogut Susskind link operators and the final  SU(N) loop operators  over the entire lattice are worked out 
%%over the entire lattice 
%in a self consistent manner. 
%%There are no local gauge or redundant loop degrees of freedom. 
%The Kogut-Susskind Hamiltonian  in terms of the loop operators has global SU(N) invariance. 
% The $(1/g^2)$ magnetic field term simply creates and annihilates  the  fundamental plaquette loops while the $(g^2)$ electric field term describes their interactions. These loop interactions
% simplify significantly in the weak coupling, $g^2\rightarrow 0$, limit.      
% %The SU(N) loop Hamiltonian is shown to have a simple weak coupling ($g^2\rightarrow 0)$ limit. 
% %The physical loop  Hilbert space ${\cal H}^p$ is characterized by $(N^2-1)(d-1)$ gauge invariant SU(N) quantum number per lattice site. 
% In the special SU(2) case, where  the canonical transformations connect the loop Hilbert space 
% to the Hilbert spaces of hydrogen atoms, we discuss the special role of the dynamical symmetry group $SO(4,2)$ of hydrogen atom in the SU(2) loop dynamics.   

\maketitle
%\flushbottom

%\newcommand{\nsum}[1][1.4]{% only for \displaystyle
%    \mathop{%
%        \raisebox
%            {-#1\depthofsumsign+1\depthofsumsign}
%            {\scalebox
%                {#1}
%                {$\displaystyle\sum$}%
%            }
%    }
%}

\section{Introduction}
\label{sec:intro}
Loops carrying non-abelian fluxes as the fundamental dynamical variables provide an alternative and interesting approach to study Yang Mills theories directly in terms of gauge invariant variables. Their importance in understanding long distance non-perturbative physics of non-abelian gauge theories  has been amply emphasized by  Wilson \cite{wilson}, Mandelstam \cite{mans2}, Yang \cite{yang}, Nambu \cite{nambu} and Polyakov \cite{poly}. In fact,  
%the loop studies are  interesting in their own right as they  also go beyond  providing a gauge invariant description of gauge theories. 
after the work of Ashtekar on  loop quantum gravity, loops carrying SU(2) fluxes have also become relevant in the  quantization of  gravity \cite{ashtekar} where they describe basic quantum excitations  of geometry.  The formulation of  
 gauge field  theories on lattice by Wilson \cite{wilson} and  Kogut-Susskind \cite{kogut} 
 is  also a step  towards  the  
loop  formulation  of gauge theories as one  directly  works  with  the  gauge covariant link operators 
or holonomies (instead of the gauge  field) which are joined together successively to get  Wilson loops.  However,  in spite of extensive work in the past, a systematic  transition from the standard  
SU(N) Kogut-Susskind lattice Hamiltonian formulation (involving link operators with spurious gauge degrees of freedom) to a  SU(N) loop formulation (involving  loop operators without local  gauge and redundant  loop degrees of 
freedom) is  still missing in the literature. 
%This paper is concerned with a systematic reformulation of pure SU(N) Kogut-Susskind Hamiltonian lattice gauge theories 
%\cite{kogut} in terms of a set of SU(N) fundamental loop operators and their conjugate loop electric fields in $(2+1)$ and $(3+1)$ dimensions.
This 
is the motivation for the present work. We obtain 
a set of fundamental, mutually independent SU(N) loop flux and their conjugate loop electric field operators  by gluing together the standard SU(N) Kogut-Susskind link operators along certain loops  (see Figure \ref{fpls})  through a series of iterative canonical transformations over the entire lattice. The canonical transformations also simultaneously produce a set of SU(N) string flux  and their conjugate string electric field operators. 
%The canonical transformations ensure that the total degrees of freedom of the theory remain unchanged at every stage.    
We show that as a consequence of SU(N) Gauss laws
at every lattice site, all string degrees of freedom become cyclic or unphysical  and completely decouple. 
As  canonical transformations keep the total degrees of freedom intact at every step, we are left only with the relevant, physical and  mutually independent SU(N) loop degrees of freedom without any local gauge or loop redundancy.  Hence, these canonical transformations also enable us to completely evade the serious problem of Mandelstam constraints (see below and section \ref{shf}) confronted by loop approaches to non-abelian lattice  gauge theories.

%We further show that the Kogut-Susskind Hamiltonian, rewritten in terms of the loop operators, has SU(N) global invariance. There are no gauge fields.   The ($1/g^2$)  magnetic field terms simply create-destroy SU(N) fundamental loop fluxes. All SU(N) loop-loop interactions are through their electric fields and are contained in the ($g^2$) electric field terms.   We also show that the SU(N) loop Hamiltonian  has a simple weak coupling $g^2 \rightarrow 0$ continuum limit. 
 
 In the past few decades there have been a number of approaches  proposed to reformulate SU(N) Yang Mills 
theories %\cite{wilson,yang,mans,mans2,nambu,poly,kogut,goldjack,gravity,gravity2,pehkj,gravity3,sharat,sharat2,ramesh,ramesh2,nair,nair5,nair6,pullin,rest5,rest9,brugmann,kolawa,gambini,gambini2,loll,loll1,watson,watson2,migdal,ruhl,bishop,sharat1,sharat12,manunpb,manuplb,ani,robson,manujp,ms1,rmi2}
 \cite{wilson,yang,mans2,nambu,poly,kogut,goldjack,gravity,sharat,ramesh,nair,pullin,rest5,loll,migdal,bishop,manuplb,ani,robson,kolawa,pietri,manujp,ms1,rmi2}
 directly in terms of loops or gauge invariant variables. 
%One of the main and long pursued  problems in the   
%Hamiltonian formulation of gauge theories is to find projections from the the  phase space of canonical 
%link variables $\left(\vec E(x; \hat i), \vec {\sf A}(x; \hat i)\right)$, constrained 
%by the non-abelian local Gauss laws,   
%to a smaller physical space of unconstrained gauge invariant variables which are related to traces of 
%Wilson loop operators.  
%In the past few decades many approaches have been developed towards this end \cite{}. 
All these approaches attempt to 
solve the non-abelian Gauss laws by first reformulating the theory in terms of operators which transform covariantly under 
gauge transformations and then exploiting this gauge covariance to define gauge invariant operators  %\cite{goldjack,gravity,gravity2,gravity3,nair,nair2,nair3,nair4,nair5,nair6} 
and  gauge invariant states.  In one of the earliest 
approaches \cite{goldjack}, a polar decomposition of the covariant electric 
fields was used to solve the SU(2) Gauss law constraints. However, the resulting 
magnetic field term in the SU(2) gauge theory Hamiltonian is technically involved and difficult to work with. Also such a polar decomposition for SU(3) or higher SU(N)  gauge group is not clear.
In  approaches motivated by gravity \cite{sharat,gravity,ramesh}, a gauge invariant metric or dreilbein tensor is constructed out of the covariant SU(2) electric or magnetic field. The problem with  such approaches is the exact equivalence between the initial and final (gauge invariant) coordinates is  not simple \cite{gravity}. Further, the gauge group SU(2) plays a very special role 
and  generalization of  these ideas to SU(N) gauge theories is not straightforward. In Nair-Karabali \cite{nair} approach the SU(N) vector potentials
enable us to define gauge covariant 
matrices leading to gauge invariant coordinates 
which are then quantized  to 
analyze the theory directly in the physical Hilbert space ${\cal H}^p$. 

An old and obvious choice for the gauge covariant 
operators \cite{pullin,rest5,loll,migdal,bishop,manuplb,ani} in any dimension is the set of all possible holonomies around closed loops (see section \ref{shf}). These loop operators  transform covariantly under gauge transformations, commute amongst themselves and their traces (Wilson loop operators) are gauge invariant. 
%They are product of Kogut-Susskind link operators \cite{kogut} along closed oriented 
%loops and transform covariantly under gauge transformations. 
In SU(N) lattice gauge theories, one easily obtains a gauge invariant (Wilson) loop basis  in ${\cal H}^p$ by applying all possible   SU(N) Wilson loop 
operators on the gauge invariant strong coupling vacuum \cite{pullin,rest5,loll,migdal,bishop,manuplb,ani}.  However, this simple construction again over describes 
lattice  gauge theories. Now the  over-description is because  all possible  Wilson loop operators are not mutually independent but satisfy notorious Mandelstam constraints  \cite{pullin, migdal,bishop,loll,manuplb,ani} discussed briefly in section \ref{shf}. These constraints  are extremely difficult to solve due to  their large number and   non-local nature (see section (\ref{shf})).
In fact, as also mentioned in \cite{pullin},  the loop approach advantages of solving the non-abelian Gauss law constraints become far less appealing  due to the presence of these non-local Mandelstam constraints. 
 In  general, a common  and widespread  belief is that  loop formulations  of gauge theories, though aesthetically appealing, are seldom practically rewarding. As an example relevant for this work, 
 in the simplest SU(2) lattice gauge theory case the Mandelstam constraints can be exactly solved 
in arbitrary dimension using the (dual) description where  electric fields or equivalently the angular momentum operators are diagonal \cite{ani, robson,pietri,manuplb,kolawa,manujp}. The resulting  gauge invariant (loop) basis, also known as the spin network basis,  is orthonormal as well as complete. Thus there are no redundant loop states or SU(2) Mandelstam constraints. The loop basis  is characterized by a set of angular momentum or equivalently electric flux quantum numbers. The action of 
the important $1/g^2$ magnetic field term on this gauge invariant loop or spin network basis (labelled by angular momentum quantum numbers) is highly geometrical  
%.  The corresponding Schr\"odinger equation in loop space \cite{manu}  involves 
$3n$-$j$ Wigner coefficients (see section \ref{shf}) ($n=6,~10$ for space dimension $d=2,~3$ respectively \cite{manuplb}). However, the corresponding loop Schr\"odinger equation involving these Wigner coefficients over the entire lattice is extremely  complicated to solve. Further, there are numerous (angular momentum) triangular constraints at each lattice site and  local abelian constraints on each link  \cite{manuplb,ani,robson,kolawa,pietri,manujp}. All these issues  make  this dual approach less viable for any practical calculation  even for the simplest SU(2) case. These dual loop approaches, when generalized to SU(3) or higher SU(N) lattice gauge theories, further suffer from the  problem of multiplicities involved with  SU(N) representations \cite{rmi2} for $N \ge 3$. 

As mentioned earlier, the loop formulation of SU(N) lattice gauge theory discussed in this work completely evades the problem of redundancy of loops or equivalently the problems of Mandelstam constraints by defining a complete set of fundamental SU(N) loop operators. All SU(N) loop flux operators start and end at the origin of the lattice.  There are no local or non-local constraints and there are no gauge fields.
The SU(N) loop dynamics is described by a generalized SU(N) spin Hamiltonian. 
The 1-1 canonical relations between the   initial Kogut-Susskind SU(N) link operators and the final   SU(N) loop \&  string operators are explicitly worked out in a self consistent manner.  
%There are no local constraints and the loop dynamics in the weak coupling limit is decribed by SU(N) spin Hamiltonian.  
%The canonical transformations  transform 
The important $(1/g^2)$ plaquette magnetic field terms,
describing SU(N) flux interactions (discussed above in terms of $3n$-$j$ Wigner coefficients) transform  or simplify  into  SU(1,1) raising and lowering 
operators of the fundamental plaquette loop  fluxes (see ((\ref{sf}) and (\ref{mft})). This is the simplest and most elementary form of a plaquette magnetic field term on lattice. Therefore, they have the simplest possible action in the loop space which is  extensively discussed in section {\ref{ldsp} and section \ref{sldyn}. All local and non-local  interactions amongst the fundamental loops  are described by $(g^2)$ electric field terms. We further show that in the weak coupling ($g^2\rightarrow 0$) continuum limit, the SU(N) loop Hamiltonian reduces to SU(N) spin model with nearest neighbour interactions.  The global SU(N) invariance of spin model is the residual SU(N) gauge transformations at the origin. 

Throughout this paper we  find it convenient to explain the  ideas using  the simplest SU(2) lattice gauge theory as an example. 
Remarkably, in this simple SU(2) case, the canonical transformations also  establish 
an  exact and completely unexpected equivalence between  bare essential (physical) loop degrees of freedom of SU(2) lattice gauge theory and  hydrogen atoms. This novel correspondence was the focus of our preceding work \cite{ms1} where we emphasized a possible   wider scope of loop approaches.      
%In section \ref{sso42}, 
We further discuss this equivalence in this work in the context of hydrogen atom dynamical symmetry group SO(4,2) and its special role in the SU(2) loop dynamics and the spectrum. 
% of SU(2) lattice gauge theories.
% connect the loop Hilbert space ${\cal H}^p$ to the Hilbert spaces of hydrogen atoms \cite{ms1}.
%We show that the SU(2) loop dynamics is completely 
%governed by the dynamical symmetry group SO(4,2) of hydrogen atoms.  We briefly discuss the recent tensor network approaches within our loop formulation 
%%of lattice gauge theories 
%to explore the loop Hilbert space ${\cal H}^p$. 

The plan of the paper is as follows. We start with a very brief introduction to Kogut Susskind Hamiltonian formulation of SU(N) lattice gauge theory in section \ref{shf}. This section is included to  set up the notations, conventions required to maintain consistency and completeness through out the presentation. We also briefly discuss SU(2)   Mandelstam constraints and difficulties associated with them  to highlight the importance of the canonical transformations  in the loop approach to SU(N) lattice gauge theory. In  section \ref{ctlols}, we discuss these canonical transformations. We show how the strings, associated with gauge  degrees of freedom, become cyclic and drop out as a consequence of Gauss laws.  We then describe  the SU(2) loop Hilbert space in terms of  Hilbert spaces of hydrogen atoms. In section \ref{sso42}, we 
discuss  the hydrogen atom dynamical symmetry group SO(4,2) and 
show the origin of its 15 generators in the context of SU(2) lattice gauge theories. The   section \ref{sldyn}, is devoted to SU(N) loop dynamics written directly in terms of the fundamental  loop operators. At the end we briefly describe a variational and a tensor network ansatz within  the present loop formulation. 
All technical details of the SU(N) canonical transformations are worked out in detail at the end in the appendices A and B. 
To keep the discussion simple and transparent, we will mostly work in two space dimension on a finite square lattice with open boundary conditions. The lattice sites 
and links are denoted by $n \equiv (x,y)$ and $(n; \hat i)$ 
respectively with $x,y =0,1,2\cdots ,\sf N; ~ i =1,2$. There are ${\cal N} \left(=(\sf N+1)\times (\sf N+1)\right)$ sites, ${\cal L} \left(= 2{\sf N}({\sf N}+1)\right)$ 
links and ${\cal P} \left(= {\sf N}^2\right)$ plaquettes satisfying:
$${\cal L} \equiv {\cal P} + \left({\cal N}-1\right).$$ 
%A link in the 
%$i^{th}$ direction at lattice site $n (=0,1,2, \cdots ,\sf N)$ will be denoted by $(n,i)$. 
We will often use $p=1,2,\cdots ,{\cal P}$ as a 
plaquette index without specifying their locations. 
%No matter is included in this work.%We choose open boundary conditions.   
%In the first half, we discuss the kinematical issues involved in canonical transformations leading to SU(N) loop operators and their conjugate SU(N) electric fields.In the second half, we discuss the dynamical issues in SU(N) lattice gauge theories.We will use $SU(2)$ gauge theory in 2+1 dimensions as a prototypical example. 

\section{SU(N) Hamiltonian formulation on lattice}
\label{shf} 
\begin{figure}
%\begin{center}
%\includegraphics[width=0.4\textwidth\height=0.4\textwidth]{1.pdf} 
\includegraphics[scale=.7]{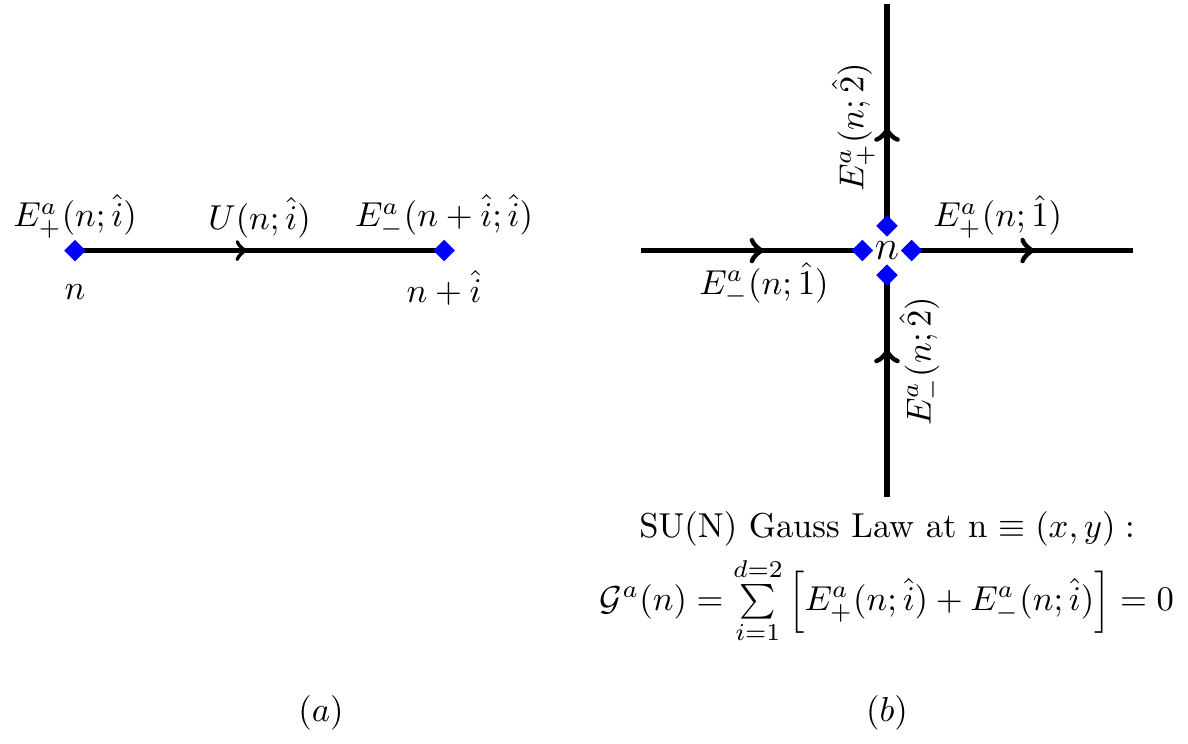}
%\end{center}
\caption{The  location of the left and  right electric fields
$E_+(n;\hat i)$ and $E_-(n+\hat i; \hat i)$ of a flux operator $U(n;i)$:  (a) on a link $(n;i)$, (b) around a lattice site $n=(x,y)$. The SU(N) Gauss law is also pictorially shown in (b).}
\label{flef} 
\end{figure}
The kinematical variables involved in Kogut and Susskind Hamiltonian formulation \cite{kogut} of lattice gauge theories are $SU(N)$ link operators $U(n; \hat i)$ and the corresponding conjugate link electric fields $E^a_+(n; \hat i)$ and $E^a_-(n+\hat i; \hat i)$. These electric fields rotate $U(n; \hat i)$ from left and right as shown in Figure \ref{flef}-a and satisfy the following canonical commutation relations
\begin{eqnarray}
\label{ccr11} 
\left[E^{a}_{+}(n; \hat i),U_{\alpha\beta}(n; \hat i)\right] & = &  - \left(\frac{\lambda^a}{2}~ U(n; \hat i)\right)_{\alpha\beta}  \\ 
\left[E^{a}_{-}(n+\hat i; \hat i),U_{\alpha\beta}(n; \hat i)\right] & = & ~~\left(U(n; \hat i)~\frac{\lambda^a}{2}\right)_{\alpha\beta} \nonumber  
\end{eqnarray} 
In (\ref{ccr11}), $\lambda^{a} ~(a=1,2, \cdots , (N^2-1))$ are the representation  matrices in 
the SU(N) fundamental representation satisfying $Tr \left(\lambda^a \lambda ^b\right) = ({1}/{2})\delta^{ab}$. The above SU(N) transformations 
imply that 
\begin{eqnarray} 
\left[E^{a}_{+}(n; \hat i),E^{b}_{+}(n; \hat i)\right]  &=&  if^{abc} E^{c}_{+}(n; \hat i), \nonumber  \\
\left[E^{a}_{-}(n; \hat i),E^{b}_{-}(n; \hat i)\right] &= & if^{abc} E^{c}_{-}(n; \hat i) 
\label{sunla}
\end{eqnarray} 
Above $f^{abc}$ are the SU(N) structure constants. We also define the strong coupling vacuum state  $|0\rangle$ by demanding $E_\mp^a(n;\hat i)|0\rangle =0$ on every link.
The link operators $U(n; \hat i)$ satisfy the following SU(N) conditions:  
\begin{eqnarray} 
U(n; \hat i)~U^{\dagger}(n; \hat i) = {\cal I}, ~~ U^{\dagger}(n; \hat i)~U(n; \hat i) = {\cal I}. 
\label{det1}
\end{eqnarray}
Above ${\cal I}$ is an $N \times N$ identity operator. Further, the determinant of the unitary matrix is also unity on every link:  $|U(n; \hat i)| ={\cal I}$. %$|U| \equiv {\rm det} U$.
Acting on strong coupling vacuum, the flux operator
$U(n;\hat i)$  creates and annihilates SU(N) fluxes on the link $(n;\hat i)$. The quantization relations (\ref{ccr11}) show that electric field operators  
$E^a_+(n; \hat i)$ and $ E^a_-(n; \hat i)$ are the generators of the 
left  and  right gauge transformations on  
$U(n; \hat i)$ and $U(n-\hat i; \hat i)$ respectively. 
This is also illustrated in  Figure \ref{flef}-b. The left and right electric fields of  link operator $U(n;\hat i)$ in (\ref{ccr11}) are related by 
% The left and right electric fields are not independent and are related by: 
\begin{eqnarray}
E_-^a(n+ \hat i; \hat i)= - R_{ab}(U^\dagger(n; \hat i))~ E_+^b(n; \hat i).
%E_{-}(n+\hat \hat i; \hat i) = - U^\dagger (n; \hat i)E_{+}(n; \hat i)U(n; \hat i). 
\label{elerrel}
\end{eqnarray} 
In (\ref{elerrel}) $R_{ab}(U) 
\equiv ({1}/{2}) Tr \left(\lambda^a U \lambda^b U^{\dagger}\right)$ is 
the rotation matrix satisfying $\tilde R(U) R(U) = R(U) \tilde R(U) = {\cal I}$  
where $\tilde R$ is the transpose of R.  
The relations (\ref{elerrel}) 
% consistent with the commutation relations 
%(\ref{ccr11}) and 
show that $E^a_-(n; \hat i)$ and $E^b_+(m,j)$  mutually  commute:
$[E^a_{-}(n; \hat i),E^b_{+}(m; \hat j)] =0$ 
and their magnitudes are equal
%at the left, right electric fields  satisfy the kinematical c
\begin{eqnarray} 
 \sum_{a=1}^{N^2-1} {E}^{a}_{+}(n; \hat i){E}^{a}_{+}(n; \hat i) 
& = &\sum_{a=1}^{N^2-1} E^{a}_{-}(n+ \hat i; \hat i)E^{a}_{-}(n+\hat i; \hat i) 
\nonumber  \\ 
& \equiv & E^{2}(n; \hat i), ~~~~~ \forall ~(n; \hat i) 
\label{consn} 
\end{eqnarray}
  The SU(N) gauge transformations are: 
% correspond to 
%rotating the body, space fixed frames of the rigid rotator \cite{kogut}: 
\begin{eqnarray}
\label{gt1n}
E_{\pm}(n; \hat i) & \rightarrow & \Lambda(n) ~E_{\pm}(n; \hat i) ~\Lambda^{\dagger}(n), \\ 
U(n; \hat i) & \rightarrow & \Lambda(n)~U(n; \hat i)~\Lambda^{\dagger}(n+\hat i). \nonumber 
\end{eqnarray}
In (\ref{gt1n}) we have defined $E_{\pm} \equiv \sum_{a}E_\pm^a ~\frac{\lambda^a}{2}$. 
The commutation relations (\ref{ccr11}) along with the gauge transformations (\ref{gt1n}) 
imply that 
%\begin{enumerate} 
%\item 
the generators of SU(N) gauge transformations at a lattice site n  are: 
\begin{eqnarray} 
{\cal G}^{a}(n) = \sum_{i=1}^{d}\Big(E_{-}^{a}(n; \hat i) + E_{+}^{a}(n; \hat i)\Big), ~~\forall ~~n,~a. 
\label{su2gln} 
\end{eqnarray}
The SU(N) Gauss law (\ref{su2gln}) is  illustrated in Figure \ref{flef}-b.
The Gauss law constraints, ${\cal G}^{a}(n) \approxeq 0,$ are imposed on the states to get physical states in  ${\cal H}^p$. %Hence, the canonical variables $\left(E_{\pm}(n,i), U(n,i)\right)$ are not free and are 
%constrained by (\ref{su2gln}). 
%\begin{figure}[t]
%\centering
%\includegraphics[width=0.3\textwidth,height=0.3\textwidth]{fig7.eps} 
%\caption{ The $18j$ ribbon diagram representing exact SU(2) loop dynamics in $d=2$. The interior (exterior) edge carries the initial (final) angular momenta and the six bridges carry the angular momenta which are invariant under the action of $\mathrm{Tr} U_{abcd}$. }
%% The six bridges 
%%along with the respective four angular momenta attached represent  
%%the six $6j$ symbols appearing  in (\ref{dyna3}) with $\bar{j}_i = j_i \pm \frac{1}{2}$.
As discussed in the introduction, the obvious and the simplest gauge invariant basis in  ${\cal H}^p$  is obtained by acting all possible Wilson loop operators $Tr {W}_\gamma$ on the strong coupling vacuum. Here, ${ W}_\gamma$ is the holonomy operator corresponding to a closed, oriented loop $\gamma$. 
However, not all Wilson loop operators are mutually independent and therefore the above basis is 
over-complete. This over-completeness 
can be appreciated  by considering the simplest SU(2) example
\cite{loll,manuplb}:  
 %of Mandelstam constraint is in SU(2) lattice gauge theory: 
%{\footnotesize 
\begin{align} 
\hspace {-0.16cm} Tr \left({ W}_{\gamma} { W}_{\bar \gamma}\right) |0\rangle \equiv Tr {W}_\gamma Tr { W}_{\bar \gamma} |0\rangle - Tr \left({ W}_\gamma { W}_{\bar \gamma}^{-1}\right)|0\rangle
\label{mc}
\end{align}
%}
involving any two arbitrary closed oriented loops denoted by $\gamma, \bar \gamma$ with a common 
starting and end point which can be anywhere on the lattice.
This trivial example shows that the three Wilson loop 
 states in (\ref{mc})  are not mutually independent.
 In the entire loop Hilbert space, involving all possible loops,  there are 
 numerous such relations  even on a 
 small lattice \cite{loll,manuplb}. Therefore, 
 the gauge theory rewritten in 
the Wilson loop basis contains many redundant and 
spurious loop degrees of freedom \cite{manuplb}.
These  mutual dependence  of  loop states are 
expressed by Mandelstam constraints like (\ref{mc})
in the case of SU(2) lattice gauge theory.   
These Mandelstam  constraints are difficult to solve 
in terms of independent loop 
coordinates \cite{loll} because of their large number
and their non-local nature. As mentioned earlier, the problem of over-completeness of Wilson loop states becomes more and more difficult as we go to higher dimension and larger SU(N) groups \cite{migdal,bishop,manuplb,rmi2}.  
% In the  case of SU(2) lattice gauge theories this problem can be elegantly resolved in the dual description \cite{sharat1,robson,manu,manu1}. In this dual approach the covariant operators are  chosen to be the  electric fields. One  diagonalizes a complete set of commuting electric field operators to get
% a local description of orthonormal and complete loop basis. These approaches though  solve the  Gauss law as well as the Mandelstam constraints, lead to highly complicated  magnetic field term in terms  Wigner coefficients. 
% As a result, the SU(2) 
% loop Schr\"odinger equation in the dual space involves 
%  18-j  and 30-j Wigner coefficients  in $d=2$ and $d=3$  respectively (see  (\ref{18j})).  
%  This makes the dual approach 
%  less viable for any practical computations. Further, there are numerous non-trivial angular momentum  triangular constraints to be satisfied by the wavefunctions in the dual space at every lattice site. 
  
 In the next section, using 
 canonical transformations, we construct a complete set 
 of  fundamental SU(N) loop operators which are mutually 
 independent. Thus the problems
 associated with SU(N) Mandelstam constraints, namely too 
 many loop degrees of freedom, are completely bypassed for any N. At the same time, unlike the dual approaches 
 mentioned above, the important SU(2) magnetic field terms reduce to a  sum of gauge invariant SU(1,1) creation-annihilation operators (see eqn. (\ref{mft}) and (\ref{s6jj})). These SU(1,1) operators  $(k_0, k_+,k_-)$ count, create and destroy the fundamental plaquette loops respectively as discussed in the next sections.   
 \begin{figure}
 \centering
\includegraphics[scale=.6]{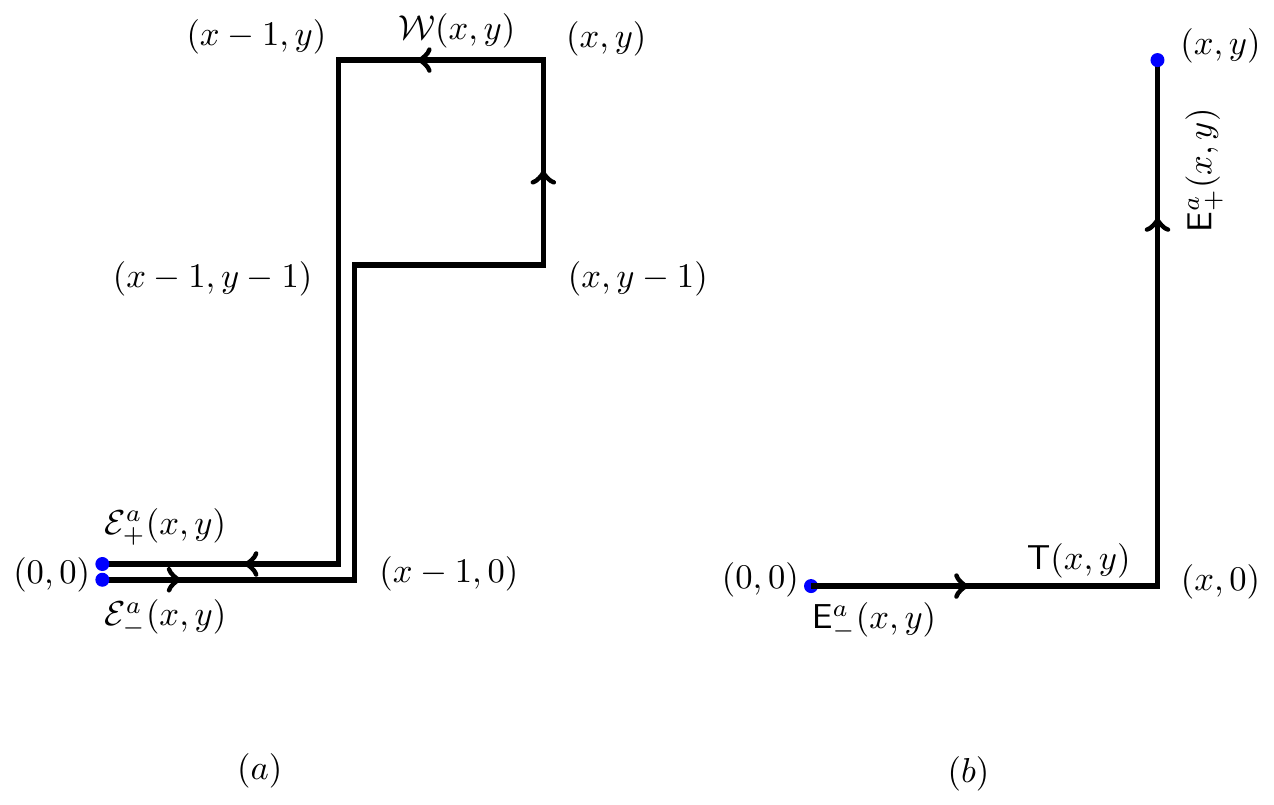}
 \caption{The plaquette loop operator ${\cal W}(x,y)$ \& the string flux  operator ${\sf T}(x,y)$  and their electric fields ${\cal E}^a_\mp(x,y)$ \& ${\sf E}^a_\mp(x,y)$ respectively. Note that the 
 electric fields ${\cal E}^a_\mp(x,y), ~{\sf E}^a_\mp(x,y)$ are located at the initial and final points of the loops \& strings respectively.}  
 \label{fpls}
 \end{figure}

\section{SU(N) Canonical Transformations: ~From links to loops \& strings}
% Loop Operators \& Loop States}
\label{ctlols}
%\section{From Links to Loops via Canonical Transformations

We start with a set of ${\cal L}$  standard SU(N) Kogut-Susskind flux and their left, right conjugate electric field operators: $\Big(E^a_+(n;\hat i), U(n;\hat i),  E_{-}^a(n+\hat i;\hat i)\Big)$  
%on the links $(n,i)$ 
satisfying  (\ref{ccr11}) and shown in Figure \ref{flef}-a. 
We construct  an iterative  series of canonical transformations  to transform them into: 
\begin{itemize} 
\item a set of ${\cal P}$  ``physical" SU(N) plaquette loop flux operators and their conjugate loop electric fields: 
$$\Big({\cal E}^a_{-}(n),~{\cal W}(n),~{\cal E}^a_{+}(n)\Big),~~~a=1,2,\cdots,N^2-1,$$
%aslo satisfying canonical commutation relations 
\item  a set of independent $({\cal N}-1)$  ``unphysical" SU(N) string flux operators and their conjugate string electric fields: $$\Big({\sf E }^a_{-}(n), ~{ \sf T}(n),~{\sf E }^a_{+}(n)\Big),~~~a=1,2,\cdots,N^2-1.$$
%The string ${\sf T}(n)$  connect the origin to the lattice site n. 
\end{itemize}
These new loop \& string flux operators
\footnote{    
The  canonical transformations and  hence the loop \& string operators  depend on the paths chosen for loops \&
strings. We have made a particular choice, shown in Figure \ref{fpls}, which lead to the simplest plaquette magnetic field term as well as a simple  Hamiltonian in the continuum $(g^2 \rightarrow 0)$ limit.} and the location of their electric fields are shown in Figure \ref{fpls}. 
As is clear from this figure, the convention chosen for loop \& string electric fields is that ${\cal E}_-^a(n), {\sf E}_-^a(n) 
~\left({\cal E}_+^a(n), {\sf E}_+^a(n)\right)$ are located
at the initial (final) points of the loop \& string flux lines.  
They satisfy canonical commutation relations amongst themselves. 
% and  create-annihilate SU(N) fluxes along  loops and strings attached to every  lattice site $n$ in $d=2$. 
The degrees of freedom exactly match as $ {\cal L}= {\cal P} + ({\cal N}-1)$. 
We will show that 
the right electric field operators 
%${\sf E}^a_+(n)$ 
of the string  attached to a site $n$  are the 
Gauss law generators (\ref{su2gln}) at n: 
\begin{eqnarray} 
{\sf E }^a_+(n) = {\cal G}^a(n).
\label{cycvar} 
\end{eqnarray}  
%In (\ref{cycvar}) ${\cal G}^a(n)$ are the SU(N) Gauss 
%law generators (\ref{su2gln}) at lattice site $n$. 
Therefore, all $({\cal N}-1)$  string flux operators ${\sf T}(n)$ create unphysical states $\notin {\cal H}^p$  and hence can be  ignored without any loss of physics.  
The traces of ${\cal P}$ plaquette loop flux operators of the form $$Tr \Big(({\cal W}(n_1))^{q_1}({\cal W}(n_2))^{q_2} \cdots ({\cal W}(n_p))^{q_{\cal P}}\Big)$$ create-annihilate all possible  physical loop states in ${\cal H}^p$. Above $(q_1,q_2,\cdots ,q_{{}_{\cal P}})$ are 
sets of ${\cal P}$ integers.
%As mentioned earlier, the transition from links to loops is obtained through a series of iterative canonical %transformations. 
We now discuss the canonical transformations.
To keep the discussion simple, we 
start with a single plaquette case before dealing with 
the entire lattices in $d=2, 3$. Some of the issues in this section were covered briefly  
in \cite{ms1}. 
% We now discuss them with sufficient  details.   
%along with new issues like  the origin and the role of SO(4,2) in  the context of SU(2) 
%loop dynamics.
%As we will see, all concepts and techniques can be  illustrated in this simple single plaquette example \cite{mstbp}.
%\section{\small Canonical Transformations, loop operators \& states, Hydrogen atoms} 

%\subsection{A single plaquette case}
\subsection{\bf Canonical transformations on a single plaquette}  
\label{ctsp} 
\begin{figure}[]
\begin{center}
\includegraphics[scale=0.85]{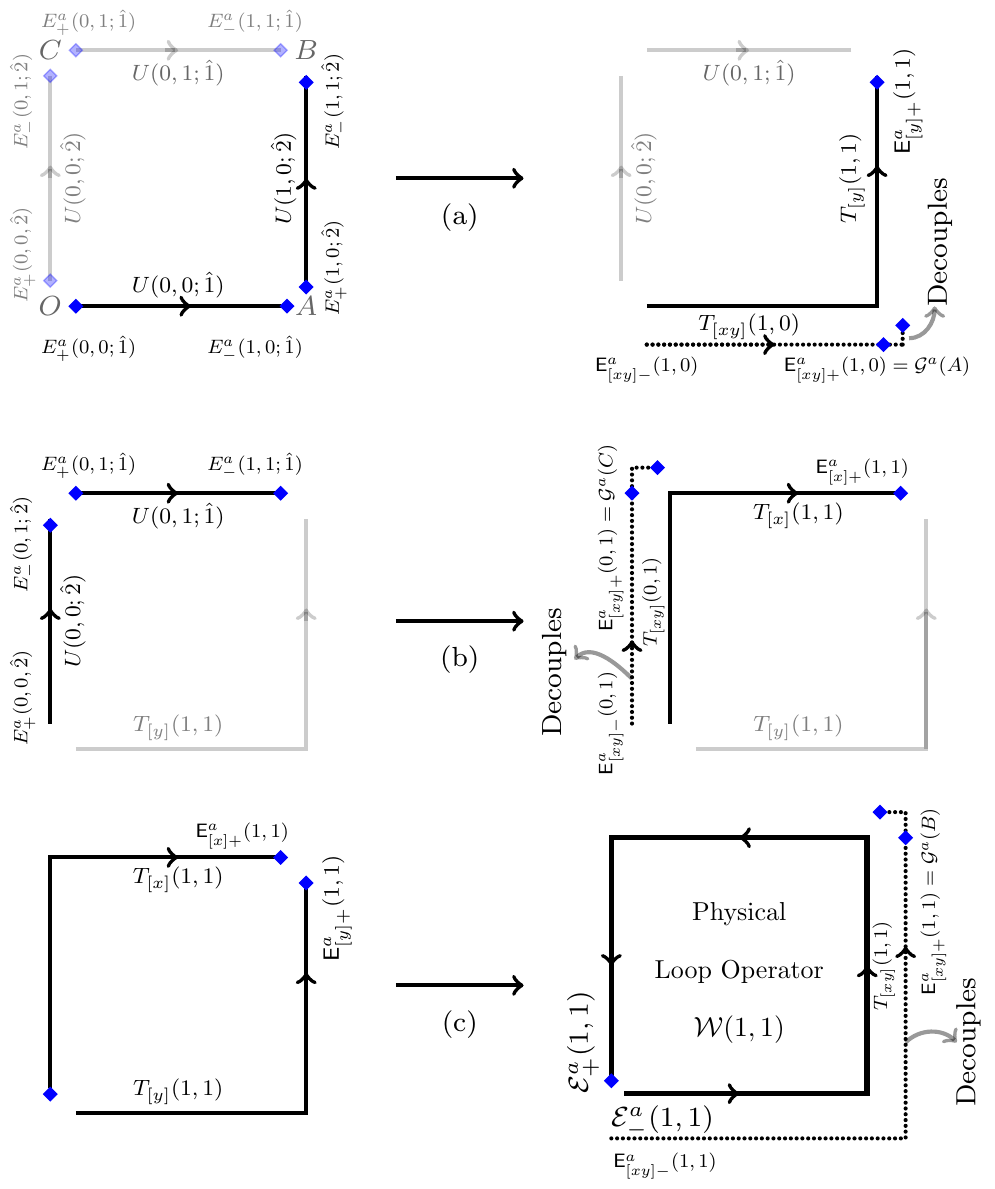} 
\end{center}
\caption{Three canonical transformations on the four link flux operators of a plaquette OABC leading to a single physical plaquette loop flux operator ${\cal W}_{\alpha\beta}(1,1)$ in (c).  The three right 
electric fields $\sf E^a_{[xy]+}(1,0),~ \sf E^a_{[xy]+}(1,1),~ \sf E^a_{[xy]+}(0,1)$ of the three string flux operators ending at A, B and C respectively are the Gauss law generators ${\cal G}^a(A), ~{\cal G}^a(B)$ and ${\cal G}^a(C)$ respectively. The Gauss law at the origin is: ${\cal G}^a(O) = E^a_+(0,0;\hat 1)+ E^a_+(0,0;\hat 2) = {\cal E}_-^a(1,1)+{\cal E}_+^a(1,1) =0$.}
\label{foabc} 
\end{figure}
We start with a plaquette OABC with the following 
Kogut-Susskind SU(N) link flux operators \cite{kogut}: $\left(E^a_+(0,0;\hat 1), ~U(0,0;\hat 1), ~E_-^a (1,0;\hat 1)\right)$ on  link OA, ~$\left(E^a_+(1,0; \hat 2), U(1,0; \hat 2), E^a_-(1,1; \hat 2)\right)$ on  link AB, ~$\left(E^a_+(0,1;\hat 1),~ U(0,1;\hat 1),~E^a_-(1,1;\hat 1)\right)$ on  link CB and finally  the operators $(E^a_+(0,0;\hat 2),~U(0,0;\hat 2),~E^a_-(0,1;\hat 2))$  on the link OC. These link operators and their locations are 
clearly illustrated on the left hand side of  Figure \ref{foabc}-a. 
%The left, right electric fields of $\Big(U_1,U_2,U_3,U_4\Big)$ are denoted by: $\Big((E_+(O,1),E_-(A,1)), ~(E_+(A,2),E_-(B,2)),~ (E_-(B,1),E_+(C,1)),~(E_-(C,2),E_+(O,2)\Big)$ respectively.   
%The basic quantization rules (\ref{ccr11}) show that the left (right) rotations on $U_l$ are generated by the left (right) electric fields 
%$E^a_L(l) ~(E^a_R(l))$ with $a=1,2, \cdots ,(N^2-1)$. 
As is clear from this figure, the SU(N) Gauss laws at four corners O, A, B and C are:
%{\footnotesize 
\begin{eqnarray} 
{\cal G}^a(0,0) & = & E^a_+(0,0;\hat 1) + E^a_+(0,0;\hat 2)=0; \nonumber  \\ 
{\cal G}^a(1,0) & = & E^a_-(1,0;\hat 1) + E^a_+(1,0;\hat 2)=0; \nonumber \\
{\cal G}^a(1,1) & = & E^a_-(1,1;\hat 2) + E^a_-(1,1;\hat 1)=0; \nonumber \\ 
{\cal G}^a(0,1) &=& E^a_+(0,1;\hat 1) + E^a_-(0,1;\hat 2)=0. 
%,~~&&
%{\cal G}^a(1,0) = E^a_-(1,0;\hat 1) + E^a_+(1,0;\hat 2)=0, \nonumber  \\
%{\cal G}^a_C = E^a_+(C,1) + E^a_-(C,2)=0,~~&& 
%${\cal G}^a_O = E^a_+(O,1)+ E^a_+(O,2) =0.
\label{4gl}
\end{eqnarray}
%}
%\begin{eqnarray}
%\left[E^a_+,U_{\alpha\beta}\right] = \left(U %\frac{\sigma^a}{2}\right)_{\alpha\beta} =>   
%\left[E^a_+,E^b_+\right] = i \epsilon^{abc} E^c_+ %\nonumber \\  
%\left[E^a_-,U_{\alpha\beta}\right] = - %\left(\frac{\sigma^a}{2} U\right)_{\alpha\beta}  =>   
%\left[E^a_-,E^b_-\right] = i \epsilon^{abc} E^c_-      
%\label{ccr} 
%\end{eqnarray} 
%As operators on different links commute, we have suppressed the link index $l$ in (\ref{ccr}). 
%One can check that $E_+^a = - R_{ab}(U^\dagger) E_-^b$ where $R_{ab}(U) = \frac{1}{2} Tr\left(\sigma^a U \sigma^bU^\dagger\right)$ is 
%SO(3) rotation matrix implying $ (\vec{E}_-)^2 = (\vec{E}_+)^2$ and $\left[E^a_-,E^b_+\right] = 0$ on every links. 
We now make canonical transformations  to fuse 
%4 link operators 
${\cal L}$ (=4) Kogut Susskind SU(N) flux operators 
$U(0,0;\hat 1), ~U(1,0;\hat 2), ~U(0,1;\hat 1)$ and $U(0,0;\hat 2)$     
%$[U_1,U_2,U_3,U_4]$
%,U(2),U(3),U(4)$ 
into ${\cal N}-1$  (= 3) unphysical string flux operators \footnote{The notations used here are as follows. The subscripts $[xy]$ on the three unphysical  flux operators
$[{\sf T}_{[xy]}(1,0), {\sf T}_{[xy]}(1,1), {\sf T}_{[xy]}(0,1)]$ are used to encode the structure of their right electric fields in (\ref{ct1}), (\ref{ct2}) and (\ref{ct3}). These are  sums of 
 the Kogut-Susskind electric fields in $x,y$ directions (denoted by subscript $[x,y]$) or equivalently Gauss law operators at corners A, B and  C respectively. During qualitative discussions  we will often  suppress these subscripts. The plaquette loop operator ${\cal W}(1,1)$ is defined at $(1,1)$ (and not at $(0,0)$) because of later convenience when we deal with canonical transformations on a finite lattice.} $\Big[{\sf T}_{[xy]}(1,0), {\sf T}_{[xy]}(1,1), {\sf T}_{[xy]}(0,1)\Big]$  
and ${\cal P} ~(=1)$ physical SU(N) plaquette loop  flux operator ${\cal W}(1,1)$ around the plaquette OABC as shown in Figure \ref{foabc}.
 The corresponding  right string and loop
electric fields are denoted by $\Big[{\sf  E}_{[xy]+}^a(1,0), {\sf  E}_{[xy]+}^a(1,1), {\sf  E}_{[xy]+}^a(1,0)\Big]$ and ${\cal E}_{+}^a(1,1)$ respectively. All left electric field operators are defined using (\ref{elerrel}).  

The canonical transformations are performed in 3 sequential steps as shown in Figure (\ref{foabc}-a), (\ref{foabc}-b)
and  (\ref{foabc}-c) respectively. 
%We also note that the Gauss laws at A,B,C,D state  
%${\cal G}^a(A)=E^a_+(1)+E^a_-(2)=0,    
%{\cal G}^a(B)=E^a_+(2)+E^a_-(3)=0  
%{\cal G}^a(C)=E^a_+(3)+E^a_-(4)=0  
%{\cal G}^a(D)=E^a_+(4)+E^a_-(1)=0$ respectively.    
The first canonical transformation  fuses $U(0,0;\hat 1)$ with $U(1,0;\hat 2)$ into ${\sf T}_{[xy]}(1,0)$ and ${{\sf T}}_{[y]}(1,1)$ as follows:
%\begin{align}
%{\footnotesize
\begin{eqnarray}
\label{ct1}  
% U(0,0;\hat 1) \rightarrow 
&{\sf T}_{[xy]}(1,0) \equiv U(0,0;\hat 1), 
%~U(1,0;\hat 2) \rightarrow 
~~{{\sf T}}_{[y]}(1,1) \equiv U(0,0;\hat 1) U(1,0;\hat 2), \nonumber  \\ 
%\nonumber  
%\nonumber  \\ %\label{ct1}  
& ~~{{\sf E}}_{[xy]+}^a(1,0)  =  E^a_-(1,0;\hat 1) + {E}^a_+(1,0;\hat 2) \equiv {\cal G}^a(1,0), \nonumber  \\ 
&{{\sf E}}^a_{[y]+}(1,1)  = {E}^a_-(1,1;\hat 2).
%%\label{ct11} 
%%\end{align}
\end{eqnarray}
%}
 All steps in (\ref{ct1}) are also illustrated in Figure \ref{foabc}-a.
 Note that  the resulting new canonical  pairs $\left({\sf T}_{[xy]}(1,0), {\sf E }_{[xy]+}(1,0)\right)$ and $\left({\sf T}_{[y]}(1,1), {\sf E^a}_{[y]+}(1,1)\right)$  satisfy the standard 
  canonical commutation relations simply by construction in (\ref{ct1}): 
 \begin{eqnarray}
\left[\sf E^a_{[xy]+}(1,0),{\sf T}_{[xy]}(1,0)\right]
=\sf T_{[xy]}(1,0)\left(\frac{\lambda^a}{2}\right), \nonumber \\ 
\left[{\sf E}_{[y]+}^a(1,1),T_{[y]}(1,1)\right]
={\sf T}_{[y]}(1,1)\left(\frac{\lambda^a}{2}\right). 
\label{fgt}
\end{eqnarray}  
%The corresponding electric fields are also independent: 
They are also mutually independent:
\begin{eqnarray} 
 \left[\sf E^a_{[xy]+}(1,0),{\sf T}_{[y]}(1,1)\right]
 = 0, ~~
 \left[{\sf E}^a_{[y]}(1,1),{\sf T}_{[xy]}(1,0)\right]
 = 0. \nonumber \\
 \left[\sf E^a_{[xy]+}(1,0),{\sf E}_{[y]}(1,1)\right]
  = 0, ~~
  \left[{\sf T}_{[y]}(1,1),{\sf T}_{[xy]}(1,0)\right]
  = 0. \nonumber 
%\label{fgt2} 
\end{eqnarray}
%From (\ref{ct1}) the two right string electric fields 
%${\sf E}^a_{[xy]+}(1,0)$ and ${\sf E}^a_{[xy]+}(1,1)$ 
%also commute with each other.
Therefore the resulting new canonical pairs  
should be treated exactly on the same footing as 
the initial Kogut-Susskind canonical pairs on links. 
%pairs on links and satisfy:
% They gauge transform as: 
% \begin{eqnarray}
% T_1 \rightarrow \Lambda_0 ~T_1~\Lambda_A^\dagger; ~~ %T_{12} \rightarrow \Lambda_o~ T_{12} ~\Lambda_B^\dagger 
% \end{eqnarray}   
The left electric fields are given by 
%defined using the standard relations (\ref{elerrel}):
\begin{eqnarray}
\sf E^a_{[xy]-}(1,0)  & \equiv & -R_{ab}\left({\sf T}_{[xy]}(1,0)\right) {\sf E}^b_{[xy]+}(1,0), \nonumber \\
\sf E^a_{[y]-}(1,1) &\equiv&  -~R_{ab}(\sf T_{[y]}(1,1))~ ~\sf E^b_{[y]+}(1,1). 
\label{lrefc}  
\end{eqnarray}
From the third equation in (\ref{ct1}) and ${\cal G}^a(1,0) =0$, it is clear that the string flux operator  $\sf T_{[xy]}(1,0)$ is unphysical as its action on any state takes that 
state  out of ${\cal H}^p$.  Therefore,  we ignore it henceforth. We now iterate the above canonical transformations with $U(0,0;\hat 1), ~U(1,0;\hat 2)$ in (\ref{ct1})  replaced by ${U}(0,0;\hat 2),~ U(0,1;\hat 1)$ respectively. We  define: 
\begin{eqnarray}
& {\sf T}_{[xy]}(0,1)  \equiv  U(0,1;\hat 2),~~ {\sf T}_{[x]}(1,1) \equiv U(0,0;\hat 2) ~U(0,1;\hat 1),  \nonumber  \\  
& {\sf E}^a_{[xy]+}(0,1)  =  {E}^a_-(0,1; \hat 2) + {E}^a_+(0,1;\hat 1) ={\cal G}^a(0,1), \nonumber \\
%= E_+^a(2)+E^a_-(3) = 0
& {\sf E}^a_{[x]+}(1,1)= E^a_-(1,1;\hat 1).
\label{ct2}
\end{eqnarray} 
Again, the canonical transformations (\ref{ct2}) are  illustrated in Figure \ref{foabc}-b. 
% and relations similar to (\ref{fgt}), (\ref{lrefc}) are valid with this new sets.    
 The resulting two new canonical pairs  of string operators $\left(\sf T_{[xy]}(0,1),\sf E^a_{[xy]+}(0,1)\right)$ and $
\left(\sf T_{[x]}(1,1),\sf E^a_{[x]+}(1,1)\right)$ are canonical as well as mutually  independent like the previous two sets in (\ref{fgt}).  
The  left electric fields ${\sf E}^a_{[xy]-}(0,1), ~{\sf E}^a_{[x]-}(1,1)$ are again 
defined through parallel transports as in  (\ref{elerrel})  or  (\ref{lrefc}). As a consequence of Gauss law at C  the string operator ${\sf T}_{[xy]}(0,1)$ (like $\sf T_{[xy]}(1,0)$) becomes unphysical. 
The last sets of canonical transformations fuse the remaining two strings ${\sf T}_{[y]}(1,1)$ and ${\sf T}_{[x]}(1,1)$ to define the final physical plaquette loop conjugate operators $\left({\cal W}(1,1), {\cal E}_+(1,1)\right)$: 
\begin{eqnarray} 
{\sf T}_{[xy]}(1,1)  \equiv  \sf T_{[y]}(1,1),~{\cal W}(1,1) \equiv  \sf T_{[y]}(1,1) ~{\sf T}^\dagger_{[x]}(1,1),  \nonumber \\ 
{\sf E}_{[xy]+}(1,1) =  {\sf E}^a_{[y]+}(1,1) + {\sf E}^a_{[x]+}(1,1)={\cal G}^a(1,1) 
%= E^a_+(3) +E^a_-(4) 
= 0, \nonumber \\
{\cal E}^a_+(1,1) = {\sf  E}^a_{[x]-}(1,1).~~~~~~~~~~~~~~~        
\label{ct3}
\end{eqnarray} 
The above canonical transformations are illustrated in Figure \ref{foabc}-c. In the third equation in (\ref{ct3}), the right electric fields 
${\sf E}^a_{[y]+}(1,1)$ and  ${\sf E}^a_{[x]+}(1,1)$ 
have been substituted in terms of the Kogut-Susskind electric fields using (\ref{ct1}) and (\ref{ct2}) to get the SU(N) Gauss laws:  ${\cal G}^a(1,1)=0$ at lattice site B. Now ${\sf T}_{[xy]}(1,1)$ decouples and 
\begin{eqnarray} 
%{\cal W}_{\alpha\beta} \equiv 
\hspace{-0.19cm} {\cal W}(1,1) \equiv U(0,0;\hat 1)~U(1,0;\hat 2)U^\dagger(0,1;\hat 1)U^\dagger(0,0; \hat 2)
\label{fpt}
\end{eqnarray} 
emerges as the final physical plaquette loop flux operator. Its left and  right electric fields are \footnote{ 
Defining $U_1=U(0,0;\hat 1),~ U_2=U(1,0;\hat 2),~ U_3= U(0,1;\hat 1), ~U_4 =U(0,0;\hat 2),~ {\cal W} = U_1U_2U_3^\dagger U_4^\dagger$ we get:
%{\tiny
\unexpanded{ 
{\footnotesize
\begin{align}
&~~~~~{\cal E}^a_-(1,1)  \equiv  
-R_{ab}\left({\cal W}\right) {\cal E}^b_+(1,1) = -R_{ab}\left({\cal W}\right)E^b_+(0,0;\hat 2)\nonumber\\
&~~~~ = R_{ab}\left({\cal W}U_4\right)E^b_-(0,1;\hat 2)  
= -R_{ab}({\cal W}U_4) E^b_+(0,1;\hat 1) \nonumber\\
 &~~~~= R_{ab}({\cal W}U_4U_3) E^b_-(1,1;\hat 1) 
= - R_{ab}\left({\cal W}U_4U_3\right) E^b_-(1,1;\hat 2)\nonumber\\
&~~~~= R_{ab}\left({\cal W}U_4U_3U_2^\dagger\right) E^b_+(1,0;\hat 2)
=-R_{ab}({\cal W}U_4U_3U_2^\dagger) E^b_-(1,0;\hat 1)\nonumber\\
 &~~~~=R_{ab}\left({\cal W}U_4U_3U_2^\dagger U_1^\dagger\right) E^b_+(0,0;\hat 1) =E^a_+(0,0;\hat 1).\nonumber
 \end{align}
}}
%}}
}: 
%\begin{align} 
%{\cal E}^a_-(1,1) = E^a_+(0,0;\hat 1).  
%\label{lefw} 
%\end{align} 
 \begin{align} 
{\cal E}^a_-(1,1) = E^a_+(0,0;\hat 1), ~~{\cal E}^a_+(1,1) = E^a_+(0,0;\hat 2). 
%&{\cal E}^a_+(1,1)  =  {\sf E}^a_{[x]-}(1,1) \equiv 
%-R_{ab}\left({\sf T}_{[x]}(1,1)\right) {\sf E}^b_{[x]+}(1,1) =\nonumber \\
%& -R_{ab}\left({\sf T}_{[x]}(1,1)\right)E^b_-(1,1;\hat 1)   = R_{ab}(U(0,0;\hat 2) E^b_+(0,1;\hat 1)  \nonumber \\
%& ~~~~= -R_{ab}(U(0,0;\hat 2)) E^b_-(0,1;\hat 2) = E^a_+(0,0;\hat 2).
\label{refw}
\end{align} 
%In (\ref{refw}) we have used the Gauss law at C: ${\sf T}^a_+(0,1) = E^a_+(0,1;\hat 1)+ E^a_-(0,1;\hat 2) =0.$ Similarly,

Thus we have converted all  link operators  into  string \& loop operators. Note that by construction the canonical structures are rigidly maintained at all three steps ((\ref{ct1}), (\ref{ct2}) and (\ref{ct3})). 
%  we first deal with the string sector. It is clear from the canonical relatons  that 
The string flux operators and their conjugate electric fields satisfy
\begin{eqnarray}
\left[{\sf E}^{a}_{[xy]+}(x,y),{\sf T}(x',y')\right]  & = &   \delta_{x,x'}\delta_{y,y'} \left({\sf T(x,y)}\frac{\lambda^a}{2}\right), \nonumber \\   %\Rightarrow  
\left[{\sf E}^{a}_{+}(x,y),{\sf E}^{b}_{+}(x',y')\right]  & = & i \delta_{x,x'}\delta_{y,y'}f^{abc} {\sf E}^{c}_{+}(x,y). ~~~~~  
%\nonumber 
% && \left[{\sf E}^{a}_{-}(x,y),{\sf T}(x',y')\right] = ~~ \delta_{x,x'}\delta_{y,y'}\left({\sf T}(x,y)\frac{\lambda^a}{2}\right); %~  \Rightarrow   
 %\left[{\cal E}^{a}_{-},{\cal E}^{b}_{-}\right] = if^{abc} {\cal E}^{c}_{-}  \nonumber
\label{ccrs12} 
\end{eqnarray} 
Above $(x,y), ~(x',y')=(1,0),(1,1),(0,1)$. The string electric fields ${\sf E}_{[xy]+}^a(x,y)$ at $(x,y)$ satisfy SU(N) algebra and commute if they are at different lattice sites. 
Under SU(N) gauge transformations, these string operators  transform as: 
\begin{eqnarray}
{\sf T}_{[xy]}(x,\nonumber y) & \rightarrow &  \Lambda(0,0) ~{\sf T}_{[xy]}(x,y) \Lambda^\dagger(x,y), \nonumber \\ 
{\sf E}_{[xy]+}(x,y) & \rightarrow & \Lambda(x,y) {\sf E}_{[xy]+}(x,y)  \Lambda^\dagger(x,y). %~~~~~~~(x,y)=(1,0),(1,1),(0,1). 
\label{sttr} 
\end{eqnarray}
Therefore, none of the three strings can form any gauge invariant operators at their end points $(0,1), (1,1), (0,1)$.   The SU(N) Gauss laws at A, B, C state 
this simple fact. Having removed the three unphysical strings, we now focus on the plaquette loop operators $\left({\cal E}_{-}^a(1,1), ~{\cal W}(1,1), ~{\cal E}_{+}^a(1,1)\right) \equiv \left({\cal E}_{-}^a, ~{\cal W}, ~{\cal E}_{+}^a\right)$. 
%Like in Kogut Susskind link formulation \cite{kogut}, loop electric fields ${\cal E}_\mp$ and flux operators ${\cal W}_{\alpha\beta}$ are canonically conjugate and 
Again by construction,  they satisfy the canonical quantization relations:
% like Kogut-Susskind link operators in (\ref{ccr11}) and string operators 
%in (\ref{ccrs12}):
\begin{eqnarray}
 \left[{\cal E}^{a}_{+},{\cal W}\right] & =&   - \left(\frac{\lambda^a}{2} {\cal W}\right)  \Rightarrow  
 \left[{\cal E}^{a}_{+},{\cal E}^{b}_{+}\right]  =  if^{abc} {\cal E}^{c}_{+}, \nonumber ~~~~~~ \\
\left[{\cal E}^{a}_{-},{\cal W}\right] &= &~~ \left({\cal W}\frac{\lambda^a}{2}\right)~  \Rightarrow   
 \left[{\cal E}^{a}_{-},{\cal E}^{b}_{-}\right] = if^{abc} {\cal E}^{c}_{-}.
  \label{ccr12}
 \end{eqnarray} 
Above ${\cal E}^a_- \equiv - R_{ab}({\cal W}) ~ {\cal E}^b_+$  implying $(\vec{\cal E}_-)^2 = (\vec{\cal E}_+)^2 \equiv (\vec{\cal E})^2$  
%or equivalently  $\hat N_a= \hat N_b \equiv \hat N$ 
% N not yet defined.
and $[{\cal E}^a_-,{\cal E}^b_+] =0$.
They  gauge transform at the origin as:
%remaining Gauss law at the origin.  
\begin{eqnarray} 
%\label{gtl} 
{\cal E}_{\mp} \rightarrow \Lambda ~{\cal E}_{\mp}~ \Lambda^\dagger,~~~~~~{\cal W}  \rightarrow \Lambda ~{\cal W} ~ \Lambda^\dagger.   
%\\  
%{\sf L}^a  \equiv   {\cal E}_-^a + {\cal E}_+^a \approxeq E^a_-(1) + E^a_+(4) = 0.  
\label{gl}
\end{eqnarray}%unphysical strings.
We have defined ${\cal E}_{\mp} \equiv \sum_{a} {\cal E}_{\mp}^a ~\lambda^a$ and $\Lambda \equiv \Lambda(0,0)$ denotes the gauge rotation 
at the origin. The corresponding Gauss law at the origin is: %)\footnote{   
%In (\ref{abc}) we have used ${\cal W}_+^a = E^a_+(4)$, (\ref{elerrel}) and the Gauss laws at A,B,C to write:
%\begin{eqnarray}
%E^a_-(1) &= &-R_{ab}(U_1)~E^b_+(1) 
%= R_{ab}(U_1) E^b_-(2)
%=-R_{ab}(U_1U_2)~E^b_+(2)
% =R_{ab}(U_1U_2)E^b_-(3) 
% = -R_{ab}(U_1U_2U_3)~E^b_+(3) \nonumber \\ 
%&=&R_{ab}(U_1U_2U_3) E^b_+(3)=
%R_{ab}(U_1U_2U_3) E_-^b(4) = 
%=-R_{ab}(U_1U_2U_3U_4)~E^b_+(4) = -R_{ab}({\cal W}) ~{\cal E}_+^b \equiv {\cal E}_-^a. \nonumber 
%\end{eqnarray}}: 
\begin{eqnarray} 
{\cal G}^a(0,0) = {\cal E}_-^a+{\cal E}_+^a = E^a_+(0,0; \hat 1) + E^a_+(0,0;\hat 2) =0.  ~~~
%-R_{ab}(U_1)E^b_+(1) + {\cal W}^a_+ = R_{ab}(U_1) R_{bc}(U_2)E_+(2) +{\cal W}_+
\label{abcd}
\end{eqnarray}
%We now show that the Gauss law (\ref{abcd}) is eactly equivalent to $E^a_+(0,0;\hat 1) + E^a_+(0,0;\hat 2) =0$. F
%In (\ref{abcd}) the sign $\approxeq$ is used to imply that  the relationship is valid acting on any physical state $ \in {\cal H}^p$ where are all states are annihor this we need to inverse ilated  
The relations (\ref{abcd}) are valid within ${\cal H}^p$ because we have ignored all string electric fields because of the Gauss laws: ${\sf E}_{[xy]}^a(x,y) =0,~ (x,y)=(1,0),(1,1),(0,1)$.    
%In (\ref{abc}) we have used ${\cal W}_+^a = E^a_+(4)$ and the Gauss laws at A,B,C to write:
%\begin{eqnarray}
%E^a_-(1) &=& -R_{ab}(U_1)~E^b_+(1) 
%= R_{ab}(U_1) E^b_-(2)
%=-R_{ab}(U_1U_2)~E^b_+(2)
% =R_{ab}(U_1U_2)E^b_-(3) = 
%\nonumber \\
% = -R_{ab}(U_1U_2U_3)~E^b_+(3) \nonumber \\
%=R_{ab}(U_1U_2U_3) E^b_+(3)=
%=R_{ab}(U_1U_2U_3) E_-^b(4) = 
%&=&-R_{ab}(U_1U_2U_3U_4)~E^b_+(4) = -R_{ab}({\cal W}) ~{\cal W}_+^b \equiv {\cal W}_-^a. \nonumber 
%\end{eqnarray}  
%They are covariant under gauge transformation 
%(\ref{gtl}) at O. The SU(2) Gauss laws at O,A,B,C reduce %to a global constraint (\ref{gl}) at O:
%At O the gauge transformations are:  
 % % % % % % % % % % % % % % % % % % %
 % % % % % % % % % % % % % % % % % % % % % % %
%The (\ref{ct3}) implies $Tr(U_1U_2U_3U_4) = Tr {\cal W}$. 
%We ignore the 3 unphysical strings  
\subsubsection{\bf Inverse relations}
It is  instructive and useful 
to invert the canonical transformations (\ref{ct1}), (\ref{ct2}) and (\ref{ct3}) to write Kogut Susskind link operators in terms of strings and loop variables. These relations also enable us to write  the Kogut Susskind  Hamiltonian (\ref{ksham}) in terms of loop operators (see (\ref{loopham})). 
 It is clear from Figure \ref{foabc}-a,b,c that
 %{\footnotesize
 \begin{eqnarray} 
 \label{uinvr}
U(0,0;\hat 1)= \sf T(1,0), ~~U(1,0;\hat 2) = \sf T^\dagger(1,0)\sf T(1,1) ~ \\  
U(0,0;\hat 2) = \sf T(0,1), ~~
U(0,1; \hat 1) = \sf T^\dagger(0,1) {\cal W}^\dagger \sf T(1,1). \nonumber 
 \end{eqnarray}
 %} 
 Above we have ignored subscript $[xy]$ and used ${\sf T}(x,y) \equiv {\sf T}_{[xy]}(x,y)$.
  Similarly, the electric field relations in (\ref{ct1}), (\ref{ct2}) and (\ref{ct3}) can also be inverted to write (see appendix B for details): 
%{\tbifootnotesize   
\begin{eqnarray}
\label{invr} 
E_+^a(0,0;\hat 1) & = &{\sf  E}^a_{[xy]-}(1,0)+{\sf E}^a_{[xy]-}(1,1)+{\cal E}^a_-,\nonumber \\  
E_+^a(1,0;\hat 2) &= & R_{ab}\left({\sf T}^\dagger_{[xy]}(1,0)\right)\Big({\sf E}^b_{[xy]-}(1,1)+{\cal E}^b_-\Big), \nonumber \\ 
E_+^a(0,0;\hat 2) &= &{\cal E}_+^a + {\sf E}^a_{[xy]-}(0,1),  \nonumber  \\ 
E_+^a(0,1;\hat 1) & =& R_{ab}\left({\sf T}^\dagger_{[xy]}(0,1)\right)~{\cal E}_+^b. 
\end{eqnarray}
%}
%\begin{eqnarray}
%&&E_+^a(0,0;\hat 1) = ~~~{\sf  E}^a_{[xy]-}(1,0)+{\sf E}^a_{[xy]-}(1,1)+{\cal E}^a_- \nonumber \\
%&& E_+^a(1,0;\hat 2) = ~~~R_{ab}\left({\sf T}^\dagger_{[xy]}(1,0)\right)\Big({\sf E}^b_{[xy]-}(1,1)+{\cal E}^b_-\Big), \nonumber \\
%&&E_+^a(0,1;\hat 1) = ~~~R_{ab}\left({\sf T}^\dagger_{[xy]}(0,1)\right)~{\cal E}_+^b \nonumber \\
%&& E_+^a(0,0;\hat 2) = ~~~~{\cal E}_+^a + {\sf E}^a_{[xy]-}(0,1).
% \label{invr} 
% \end{eqnarray}
 These  canonical relations between links \& loops have the following interesting features: 
 \begin{itemize} 
 \item They are consistent with
  gauge transformations (\ref{gt1n}), (\ref{sttr}) and 
  (\ref{gl}) as well as with SU(N) algebras 
   of link, string and loop electric fields given in (\ref{ccr11}) and (\ref{ccrs12}), (\ref{ccr12}).
   \item %The inverse relations (\ref{uinvr}) and (\ref{invr}) are mutually compatable.  
   The canonical commutation relations between SU(N)  link flux operators and their link electric fields  also remain intact under the links to loops \& stings mappings (\ref{uinvr}) and (\ref{invr}).  %from string, loop to link operators.
   %As an example, it is easy to see that $E^a_+(1,0;\hat 2)$ leaves $U(0,0;\hat 1), U(0,0; \hat 2), U(0,1;\hat 1)$ unchanged and rotates $U(1,0;\hat 2)$ from the left: 
   %{\footnotesize
   %\begin{align}
   %{[E^a_+(1,0;\hat 2), U(1,0;\hat 2)]  = 
   %R_{ab}({\sf T}^\dagger_{{[xy]}}(1,0)) 
   %\underbrace{{\sf %T}^\dagger_{{[xy]}}(1,0)\frac{\lambda^b}{2}
   %{\sf T}_{{[xy]}}(1,0)}_{=R_{bc}\left({\sf T}_{[xy]}(1,0)\right)\frac{\lambda^c}{2}}U(1,0;\hat 2) 
   %\frac{\lambda^a}{2} U(1,0;\hat 2)}. \nonumber 
   %\end{align}
   %}
   %All other commutation relations can be directly read off from 
   %(\ref{uinvr}) and (\ref{invr}).
 \item 
   No strings (${\sf T}(x,y)$ or ${\sf E}^a(x,y)$) can appear in a gauge invariant operator in ${\cal H}^p$. As an example, the gauge invariant electric field terms in the Kogut Susskind Hamiltonian are:   
 {\footnotesize 
 \begin{align}
 \left(\vec E_+(0,0;\hat 1)\right)^2 = \left(\vec E_+(1,0;\hat 2)\right)^2 = 
 %~~~\nonumber \\
 \left(\vec E_+(0,1;\hat 1)\right)^2 \nonumber \\  = \left(\vec E_+(0,0;\hat 2)\right)^2  =  \left(\vec {\cal E}_-\right)^2.
 %,~~~~~~  &&~~~
 %E_+^a(B,2) = ~R_{ab}({T^\dagger(A)})~ {\cal E}^b_-, \\
 %E_-^a(C,1) = ~R_{ab}({ T(C)})~{\cal E}^b_-,  && ~~~
 %E_{\sf L}^a(4) = ~R_{ab}({T(3)}){\cal E}^b_-. \nonumber
 \label{invrr} 
 \end{align}
 }
We have used the Gauss laws ${\sf E}^a_{[xy]+}(x,y) =0$ in (\ref{invr}) within ${\cal H}^p$. In other words, while expressing 
 Kogut-Susskind  link electric fields in terms of loop 
 electric fields, the strings can appear only  
 in the  overall parallel transport factors. This is also required for the consistency with SU(N) gauge transformations in (\ref{invr}). 
 % incan appear only as  overall parallel transport factors. % Thus  the Kogut-Susskind  electric field terms can be simply described in terms  of the  loop electric fields.
 \end{itemize} 
%:$\left[E^a_L(I),E^b_L(J)\right] =
  %i \delta_{{I,J}} f^{abc}~E^c_L(I);~ I,J=1,2,3,4$.    
 % Having defined new string and loop operators, it is also interesting to convert (\ref{ct1}), (\ref{ct2}) and (\ref{ct3}) and write Kogut Susskind link operators 
 %in terms of strings and loop variables. These relations 
 %will also lead us from  link Hamiltonian (\ref{ksham}) to  loop Hamiltonian (\ref{loopham}). 
 %It is clear from Figure xxx that $U_1= \mathbb T(1), ~ U_2= \mathbb T^\dagger(1)~\mathbb T(2), ~ U_3 = \mathbb T^\dagger(2)~\mathbb T(3),~ U_4= \mathb T^\dagger(3)~{\cal W}$. Similarly, the Figure xxx also transcribes the Kogut Susskind electric fields in terms 
 %of string and loop electric fields.   
% \begin{eqnarray} 
 %&&E_-^a(1) = -R(U_1)E_+^a(1) = -R({\mathbb T}(1))\left[{\mathbb E}_+^a(1) - E^a_-(2)\right]= {\mathbb E}_-^a(1) - R_{ab}({\mathbb T}(2))E^b_+(2) \nonumber \\ 
% && = {\mathbb E}^a_-(1)+R_{ab}({\mathbb T}(2))\left[{\mathbb E}^b_+(2)- E_-^b(3)\right] = {\mathbb E}^a_-(1) + {\mathbb E}_-^a(2) + {\mathbb E}^a_-(3) + {\cal W}_-.
 %\label{invr}  
 %\end{eqnarray}
 %Note that the relations (\ref{invr}) are consistent
 %gauge transformations as well as SU(N) lie algebras 
 %of electric fields:$\left[E^a_-(I),E^b_-(J)\right] =
 %\delta_{IJ} i f^{abc}E^c_-(I)$.

\subsubsection{\bf Loop prepotential operators}

The  physical loop electric fields ${\cal E}_{\pm}^a$ discussed in the previous section can be conveniently  described in terms of the  prepotential creation-annihilation  
%creation-annihilation operators for the fundamental and physical SU(2) flux loop operator ${\cal W}_{\alpha\beta}$ are the prepotential operators \cite{manu,rmi} defined through the electric fields
operators:
%can be described terms of prepotentials \cite{manu}: 
\begin{eqnarray}   
{\cal E}_-^a =  a^\dagger ~\left(\frac{\sigma^a}{2}\right) ~a,~~~~~~~~~~{\cal E}_+^a = b^\dagger ~\left(\frac{\sigma^a}{2}\right)~  b.
\label{su2sb} 
\end{eqnarray} 
In (\ref{su2sb}) ${\cal E}_\pm^a$ are SU(2) electric fields and $(a^\dagger_\alpha,b^\dagger_\beta)$ and  $\left(a_\alpha,b_\beta\right)$ are the SU(2) prepotential creation and  annihilation  SU(2) 
doublets
\footnote{The $SU(N), ~N \ge 3$ case can be similarly analyzed  by replacing SU(2) prepotentials with SU(N) irreducible prepotentials  discussed in the context of SU(N) lattice gauge theories in \cite{rmi2}.} 
with $\alpha,\beta =1,2$. We also define the total number operators $\hat N_a = a^\dagger \cdot a \equiv a^\dagger_1a_1 + a^\dagger_2a_2$, ~$\hat N_b = b^\dagger \cdot b \equiv b^\dagger_1b_1 + b^\dagger_2b_2$.  The constraint ${\cal E}_+^2 = {\cal E}_-^2$ implies $$\hat N_a = \hat N_b \equiv \hat N.$$  
Under SU(2) gauge transformations (\ref{gl}): 
\begin{eqnarray} 
a_{\alpha} \rightarrow \Lambda_{\alpha\beta}~ a_\beta, ~~~~
b_{\alpha} \rightarrow \Lambda_{\alpha\beta} ~b_\beta.
\label{sbgto}
\end{eqnarray}
The prepotential formulation also has an important additional U(1) invariance \cite{manujp,manuplb}: 
\begin{eqnarray} 
a_\alpha \rightarrow e^{i\theta} ~a_\alpha, ~~~ b_\alpha \rightarrow e^{-i\theta} ~b_\alpha. 
\label{abgi} 
\end{eqnarray}
The prepotential operators defining relations (\ref{su2sb}) are invariant under (\ref{abgi}).
The gauge invariant strong coupling vacuum $|0\rangle ~(\equiv |0\rangle_a 
\otimes |0\rangle_b)$ is also the prepotential harmonic oscillator vacuum satisfying: $a_\alpha|0\rangle =0, ~~ b_\alpha|0\rangle=0.$ 
The  quantization 
rules (\ref{ccr11}) and the gauge transformations (\ref{gl}), (\ref{abgi}) imply \cite{manujp,manuplb}:
\begin{eqnarray} 
{\cal W}_{\alpha\beta} & = & \frac{1}{\sqrt{(\hat N+1)}} \left(a_{\alpha} \tilde b_{\beta}-\tilde a^{\dagger}_{\alpha}  
b^{\dagger}_{\beta}   \right) 
\frac{1}{\sqrt{(\hat N+1)}} \nonumber ~~~~~~~~\\
& \equiv &
\frac{1}{\sqrt{(\hat N+1)}}\left({\cal W}^{(-)}_{\alpha\beta} + {\cal W}^{(+)}_{\alpha\beta}\right)\frac{1}{\sqrt{(\hat N+1)}}.  
\label{holo}
\end{eqnarray}
It is easy to check that  (\ref{su2sb}) and (\ref{holo}) satisfy the canonical commutation relations (\ref{ccr12}).  Further, the above prepotential representation also maintains the non-trivial relations: ${\cal W} ~{\cal W}^\dagger = {\cal W}^\dagger ~{\cal W}= {\cal I},~  |{\cal W}| = +1$  as well as  the canonical commutation relations:  
%\begin{eqnarray} 
$\left[{\cal W}_{\alpha\beta}, {\cal W}_{\gamma\delta}\right]
=0, ~~\left[{\cal W}_{\alpha\beta}, {\cal W}^\dagger_{\gamma\delta}\right]=0.$ 

 We now construct a complete  orthonormal loop basis in ${\cal H}^p$ with  the prepotential operators in a straightforward manner in the next section. We further  
show that ${\cal H}^p$ can be exactly identified with all possible spherically symmetric ``s-states of a hydrogen atom" \cite{ms1}.  
% Before discussing this loop dynamics,     
%we first construct the physical Hilbert space ${\cal H}^p$ in terms of  loop prepotential creation operators $a^\dagger_\alpha$ and $b^\dagger_\alpha$ defined in (\ref{su2sb}). 

\subsubsection{\bf Physical loop Hilbert space and Hydrogen atom}

In the standard approach all four link flux operators in
Figure \ref{foabc}-a are  fundamental with each of them  gauge transforming differently. Therefore, the construction of gauge invariant states is more involved
compared to working with a single loop flux operator 
${\cal W} $.  In this section we exploit this simple fact 
and show that the physical or loop Hilbert space can be completely realized in terms of a hydrogen atom Hilbert space.
%Having extracted the fundamental gauge covariant plaquette loop operators and its conjugate electric fields,
%%k+,k_,k0 not defined yet
%% gauge invariant 
%%loop creation $k_+$, annihilation $k_-$ and flux 
%%counting $k_0$ operators,
% we now construct 
%a basis in ${\cal H}^p$ before considering loop 
%dynamics. 
%As shown in \cite{ms1}, a complete  
%basis  of a single plaquette lattice gauge theory 
%can be exactly realized in terms of the energy eigenstates $|n,l,m\rangle$ of a hydrogen atom. These states transform covariantly under gauge transformatons at the origin (see (\ref{nlmtp})). The Gauss law in this basis is trivially solved by putting $l=0,~ m=0$. 
This correspondence is achieved by identifying the loop electric fields $\vec {\cal E}_{\mp}$ of SU(2) lattice gauge theory with the  angular momentum $\vec {\sf L}$ and Laplace Runge Lenz 
 vector $\vec {\sf A}$ of the hydrogen atom. More precisely: 
 %More precisely, in the single plaquette case 
 %considered so far, we identify the plaquette loop electric fields with $\vec {\sf L}$ and $\vec {\sf A}$ 
 %of hydrogen atom:
 \begin{eqnarray} 
  \vec {\cal E}_{\mp} \equiv \frac{1}{2} \left(\vec {\sf L} \mp \vec {\sf A}\right).
  % ~~~~~~{\cal E}_+ \equiv   \frac{1}{2} \left( \vec {\sf L} - \vec {\sf A}\right).
  \label{lrl}
\end{eqnarray} 
In the above identification, the identity  $\vec {\cal E}_-^{~2} = \vec {\cal E}_+^{~2} \equiv \vec {\cal E}^{~2}$ in (\ref{consn}) holds naturally as $\vec {\sf L} \cdot \vec {\sf A}=0$ \cite{wyb}.   %With the above correspondence the Gauss law (\ref{abcd}) states that the ``s-states" of hydrogen atom with vanishing angular momenta also describe a complete orthonormal loop basis in ${\cal H}^p$. 
%angular momentum $\vec {\sf L} ~(\equiv  \vec {\cal E}_-+\vec {\cal E}_+)$ of hydrogen atom states must vanish. 
We  can also have three separate identifications like (\ref{lrl}) for  the three string electric fields  ${\sf T}_{[xy]\mp}(1,0),~{\sf T}_{[xy]\mp}(1,1), ~{\sf T}_{[xy]\mp}(0,1)$. But these identifications will be in the unphysical sector in the case of pure gauge theories
and hence we ignore them in this work.

We first construct the  eigenstates of the complete set of commuting operators (CSCO-I) consisting of $(\vec {\cal E}_-^{~2} \equiv \vec {\cal E}^{~2},~{\cal E}_-^{a=3})$ and  $( \vec {\cal E}_+^{~2} \equiv \vec {\cal E}^{~2}, ~{\cal E}_+^{a=3})$ which form $SU(2)\otimes SU(2)$ representations: 
%\begin{eqnarray}
%\label{csco1} 
%\begin{split}
%|j,m,\bar m\rangle  =  
%|j,m\rangle_a \otimes |j,\bar m\rangle_b \equiv   \left(\frac{(a^{\dagger}_1)^{(j+m)}~ (a^{\dagger}_2)^{(j-m)}}{\sqrt{(j+m)!~(j-m)!}}\right) ~|0\rangle_a \otimes  \left(\frac{(b^{\dagger}_1)^{(j+\bar m)}~ (b^{\dagger}_2)^{(j-\bar m)}}{\sqrt{(j+\bar m)!~(j-\bar m)!}}\right) |0\rangle_b.       
% \end{split}
% \label{jmm}
%\end{eqnarray}
%\begin{eqnarray}
%\label{csco1} 
%\begin{split}
%\hspace{-0.4cm}|j,m,\bar m\rangle  =  
%|j,m\rangle_a \otimes |j,\bar m\rangle_b \equiv   %\left(\frac{(a^{\dagger}_1)^{(j+m)} %(a^{\dagger}_2)^{(j-m)}}{\sqrt{(j+m)!(j-m)!}}\right)  %\left(\frac{(b^{\dagger}_1)^{(j+\bar m)} %(b^{\dagger}_2)^{(j-\bar m)}}{\sqrt{(j+\bar m)!(j-\bar %m)!}}\right) |0\rangle.~~~       
% \end{split}
% \label{jmm}
%\end{eqnarray}
\begin{eqnarray}
%\label{csco1} 
%\begin{split}
|j,m_-, m_+\rangle  =  
 %~\otimes ~ \nonumber \\ \equiv   
 \underbrace{T^{j}_{m_-}(a^\dagger)~|0\rangle_a}_{|j,m_-\rangle_a} ~\otimes~ 
\underbrace{T^{j}_{m_+}(b^\dagger) ~|0\rangle_b}_{|j, m_+\rangle_b}.       
% \end{split}
 \label{jmm}
\end{eqnarray}
In (\ref{jmm}), the tensor operator $T^j_m$  are defined as
%\begin{eqnarray}
{\footnotesize
$$T^{j}_{m}(a^\dagger) \equiv \sqrt{(2j)!}
\left(\frac{(a^{\dagger}_1)^{(j+m)} (a^{\dagger}_2)^{(j-m)}}{\sqrt{(j+m)!(j-m)!}}\right).$$
}
%\nonumber  
%T^j_{m_+}(b^\dagger) \equiv \sqrt{(2j)!} \left(\frac{(b^{\dagger}_1)^{(j+ m_+)} (b^{\dagger}_2)^{(j-m_+)}}{\sqrt{(j+m_+)!(j-m_+)!}}\right). \nonumber 
%\end{eqnarray}  
%\begin{eqnarray} 
%T^{j}_{m_-}(a^\dagger) \equiv \sqrt{(2j)!}
%\left(\frac{(a^{\dagger}_1)^{(j+m_-)} (a^{\dagger}_2)^{(j-m_-)}}{\sqrt{(j+m_-)!(j-m_-)!}}\right), 
%T^j_{m_+}(b^\dagger) \equiv \sqrt{(2j)!} \left(\frac{(b^{\dagger}_1)^{(j+ m_+)} (b^{\dagger}_2)^{(j-m_+)}}{\sqrt{(j+m_+)!(j-m_+)!}}\right). \nonumber 
%\end{eqnarray} 
 The states (\ref{jmm}) are invariant under U(1) gauge transformations 
(\ref{sbgto}). They are eigenstates of the above CSCO-I: 
\begin{align} 
\vec {\cal E}^{~2} |j,m_-,m_+ \rangle =  j(j+1) 
|j,m_-,m_+\rangle, \nonumber \\
\vec {\cal E}_\mp^{a=3} |j,m_-,m_+ \rangle  = m_\mp  |j,m_-,m_+\rangle.
%\vec {\cal E}_+^{a=3} ~|j,m,\bar m \rangle &=& \bar m~ 
%|j,m,\bar m\rangle \nonumber. 
\label{meqnj} 
\end{align}
%These $SU(2) \otimes SU(2)$ representation states occur 
%both in hydrogen atom \cite{wyb} as well as SU(2) lattice %gauge theory  \cite{manu} and also in loop quantum 
%gravity \cite{thiemannrovelli,mh}. 
In the context of hydrogen atom, the states (\ref{jmm}) 
are the energy eigen states   with energy 
\cite{wyb} $E_{\sf n} \sim \frac{1}{{\sf n}^2}$ with ${\sf n} \equiv 2j+1$. The  two magnetic quantum numbers $m_\mp$ describe their degeneracies. On the other hand,  in the  gauge theory context the states $|j~m_- ~ m_+\rangle $ in (\ref{meqnj}) describe loops    
carrying non-abelian quantized SU(2) loop electric fluxes
 \footnote{
% on  \cite{sharat1,robson,manu} and over a single plaquette 
%in the present loop formulation (see also next section). 
Similar SU(2) spin network basis in terms of harmonic oscillators or prepotentials  have also been discussed in the context of loop quantum gravity \cite{levhm}.  They define the quantum states of space geometry as spin networks \cite{rovbook}.}. 
 Further, 
as ${\cal E}_-^a + {\cal E}_+^a \equiv {\sf L}^a$ the gauge rotations  at the origin of the flux states in (\ref{jmm}) 
%${\cal W}(1,1)$ 
correspond to the spatial rotations of the hydrogen 
atom.
%We now solve the Gauss law in this simplest one plaquette problem by rotating the basis so that the angular momentum operators 
%$\left(\vec {\sf L}^2, {\sf L}^{a=3}\right)$ are diagonal. 
% to get the loop basis states or spin network  states which can be identified with hydrogen atom energy eigenstates $|n,l,m\rangle$ with vanishing angular momentum. 
%in terms of 
%the energy eigenstates of hydrogen atom.
%The states (\ref)on neighbouring links (plaquettes) are intertwined together to give a gauge invariant basis in ${\cal H}^p$. 
Under these gauge transformations:
\begin{eqnarray} 
|jm_-{m}_+\rangle \rightarrow 
%\sum_{m_-'  m'_+} 
\sum_{m_\mp'} |jm'_-{m}'_+\rangle  D^{~~j}_{m'_-m_-}(\Lambda)~ D^{~~j}_{{m}'_+ {m}_+}(\Lambda).  
\label{jmmtrans} 
\end{eqnarray}
In (\ref{jmmtrans}), 
%We now redefine $|j,m,\bar m\rangle \rightarrow 
%(-)^{j-m}|j,-m,\bar m \rangle \equiv |j,m,\bar m\rangle$ 
%so that: 
%\begin{eqnarray} 
%\ket{j,m,\bar{m}} \rightarrow  \ket{j,m',\bar{m}'}  D^j_{mm^\prime}(\Lambda^\dagger) D^j_{\bar m'\bar m}(\Lambda). 
%\end{eqnarray} 
$D^{~~j}_{mm^\prime}(\Lambda)$ are the Wigner  matrices,
$\Lambda \equiv \Lambda(0,0)$ denotes the gauge parameters at the origin. We have used the gauge transformations (\ref{sbgto}) 
 and the definition (\ref{jmm})  to get (\ref{jmmtrans}). In order to solve Gauss law systematically,  
we construct a coupled basis from  (\ref{jmm}) so that  the following coupled and complete set of commuting operators (CSCO II) are diagonal:
{\footnotesize
$$\left\{\vec {\cal E}_-^{~2} = \vec {\cal E}_+^{~2} = \vec {\cal E}^{~2}, %{\cal E}_+^2 = {\cal E}^2, 
~~~(\vec {\cal E}_- + \vec {\cal E}_+)^2, ~~~(\vec {\cal E}_-+ \vec {\cal E}_+)^{a=3}\right\}
$$ 
$$ 
\equiv \left\{  \vec {\cal E}^2,~ (\vec {\sf L})^2,~ (\vec {\sf L})^{a=3}\right\}.$$
} 
% Note that the defining equation (\ref{lrl}) states ${\cal E}_-^a+{\cal E}^a_+ \equiv {\sf L}^a$. 
The eigenbasis states of CSCO-I and CSCO-II are related by Clebsch-Gordan coefficients: 
\begin{eqnarray}
\vert {\sf n}~l~m \rangle  \equiv  \sum_{m_-, m_+} C_{jm_-,j m_+}^{~~l,m}~ |j~m_- m_+\rangle \nonumber \\
 =  \sum_{m_-,m_+} C_{jm_-,j m_+}^{~~l,m}~ |j,m_-\rangle_a~ |j, m_+\rangle_b. 
\label{nlmm} 
\end{eqnarray}
Above  ${\sf n} \equiv 2j+1 =1,2,\cdots ; ~~~l= 0,1,\cdots ,2j ~(\equiv {\sf n}-1); ~~~ m=-l,-(l-1), \cdots , (l-1), l$. 
The states in (\ref{nlmm}) are eigenstates of CSCO II:
%{\footnotesize
\begin{align} 
\vec {\cal E}^{~2} ~|{\sf n} ~l ~ m \rangle & ~ =~   \frac{({\sf n}^2-1)}{4} ~~|{\sf n} ~l ~ m\rangle, \nonumber \\ 
\vec {\sf L}^{~2} ~|{\sf n} ~l  ~m \rangle & ~=~ l(l+1) ~ ~
|{\sf  n} ~l~ m\rangle,\nonumber \\
 {\sf L}^{a=3}~|{\sf n}~l~m \rangle & ~=  ~~~m~
~|{\sf n}~l~ m\rangle.
\label{midone}    
\end{align}
%} 
%\begin{eqnarray} 
%&& ~~~~\left(\vec {\cal E}\right)^2 ~~~~~~~~|{\tt n} ~l ~ m \rangle~~ =  \frac{({\tt n}^2-1)}{4} ~|{\tt n} ~ l ~ m\rangle, \nonumber \\
%\label{midone}
%&& \left(\vec {\cal E}_-+ \vec {\cal E}_+\right)^2 ~~~~|{\tt n}~ l ~ m \rangle ~=~ l(l+1)  ~ 
%|{\tt n}~l~ m\rangle,  \\
%&& \left(\vec {\cal E}_- +\vec {\cal E}_+\right)^{a=3} ~|{\tt n}~l~m \rangle ~=  ~~m~
%|{\tt n}~l ~m\rangle.  \nonumber 
%\end{eqnarray} 
 Note that the states $|{\sf n}~l~m\rangle$ in (\ref{nlmm}) are also the standard hydrogen atom energy eigenstates  \cite{wyb} characterized by the principal, angular momentum and magnetic quantum numbers ${\tt n}, ~l$ and $m$ respectively.    
 Under gauge transformations, the coupled states (\ref{nlmm}) have  much simpler transformation property as compared to the states in (\ref{jmmtrans}): 
  \begin{eqnarray}
  |{\sf n}~l~ m\rangle \rightarrow \sum_{\bar m} D^{~l}_{m \bar m}(\Lambda) ~|{\sf n} ~l ~\bar{m}\rangle.
  \label{nlmtp}
  \end{eqnarray} 
 Thus the principal and angular momentum quantum numbers are  gauge invariant.  The Gauss law in this single plaquette case (\ref{abcd}) states that  ${\cal E}^a_- + {\cal E}^a_+ \equiv  {\sf L}^a =0$. Therefore, all possible orthonormal solutions are  the s-states $|{\sf n},l=0,m=0\rangle$ of hydrogen atom. 
  This gauge invariant hydrogen atom loop 
 basis  can be easily  constructed in 
terms of the prepotential operators.  There are three possible $SU(2) \times U(1)$ gauge invariant operators: 
\begin{eqnarray} 
k_-  \equiv a \cdot \tilde b,~~
k_+  \equiv a^\dagger \cdot \tilde b^\dagger,~~ 
k_0 \equiv \frac{1}{2}~ \underbrace{(\hat N_a+ \hat N_b+2)}_{2(\hat N+1)}
\label{su11} 
\end{eqnarray} 
In (\ref{su11}) $a\cdot \tilde b  = a_\alpha \tilde b_\alpha =  a_\alpha \epsilon_{\alpha\beta}b_\beta \equiv (a_1b_2-a_2b_1)$ and $\hat N_a = \hat N_b = \hat N$.
% as $\epsilon_{11} = \epsilon_{22} =0, \epsilon_{12}=-\epsilon_{21}=1$. 
They are  gauge invariant  loop creation-annihilation operators  $k_\pm =  {\rm Tr} {\cal W}^{(\pm)}$. On the other hand, gauge invariant $k_0$ has the interpretation of loop flux counting operator. They satisfy SU(1,1) algebra: 
\begin{eqnarray}
\left[k_-,k_+\right] = 2k_0,~~~~~ ~~~~\left[k_0,k_\pm\right] = \pm k_\pm. 
\label{su11a} 
\end{eqnarray}
They are also invariant 
 \footnote{The  three  operators: $\kappa_+ = a^\dagger \cdot b, ~ \kappa _- \equiv b^\dagger \cdot a,~ \kappa_0 \equiv \frac{1}{2}\left(N_a-N_b\right)$ are also invariant under SU(2) and follow SU(1,1) algebra 
%which commutes with  satisfy satisfying  
$\left[\kappa_+,\kappa_-\right] = 2 \kappa_0,~ \left[\kappa_0,\kappa_\pm\right] = \pm \kappa_\pm$. They commute with $SU(1,1)$ generators $\left[k_0,k_\pm\right]$ and ${\sf L}^a$. However,   they 
are  not invariant under abelian gauge transformations (\ref{abgi}) and hence irrelevant.}
under U(1) transformations (\ref{abgi}).
%\footnote{The other three SU(1,1) operators, also invariant under SU(2) transformation (\ref{sbgto}), are: $$ \kappa_- \equiv a^\dagger \cdot b, ~~~~\kappa_+ \equiv b^\dagger \cdot a, ~~~~\kappa_0 \equiv  \frac{1}{2} \left(N_a-N_b -2\right).$$ However, they are not invariant under U(1) transformations (\ref{abgi}) (also see (\ref{sbgtop} and (\ref{abgip}) for general ${\cal P}$ plaquette case). Hence they are irrelevant and ignored.}
 The SU(1,1) Casimir operator is defined as: 
\begin{eqnarray} 
{\cal C} \equiv k_0^2 - \frac{1}{2} \left(k_-k_+ + k_+k_-\right).
\label{su11cas} 
\end{eqnarray} 
%FFFFFFFFFFFFFFFFFFFFFFFFFF
%  The loop states (\ref{nlmm}) diagonalizing the electric field term in the Hamiltonian now takes a simple form: 
%For a single plaquette considered in this section, the 
%Gauss laws (\ref{abc}) implies $l =m =0$ and the states 
%in (\ref{nlmm}) take the simple form (\ref{lb}). 
% We call this basis the tadpole basis. %(fig \ref{f:tadpole}). %For a single plaquette lattice 
%In constructing this basis, we are guided by the need to construct a gauge invariant basis and also to simplify the matrix elements of $TrW$. 
%\begin{eqnarray} 
%|j,m,n\rangle = |j,m\rangle 
% % % % % % % % % % % % % % % % % % % % %
%A basis in the physical Hilbert space ${\cal H}^p$  %diagonalizing electric field term is: 
All possible orthonormal hydrogen atom loop states can be easily constructed using SU(1,1) or loop creation operators $k_+$: 
\begin{eqnarray} 
|{\sf n} \rangle ~\equiv ~\frac{1}{\sqrt{ ({\sf n}-1)!~{\sf n}!}} ~\left(k_+\right)^{ {\sf n}-1}|0\rangle; ~~~{\sf n} = 1,2\cdots. 
%~~~~~~ H_E ~| j\rangle  = g^2~ j\left(j+1\right) ~| j\rangle.
\label{lb} 
\end{eqnarray} 
%In (\ref{lb}), ${\mathtt n}  =2j+1$ where ${\tt n}$ is the principal quantum number of hydrogen atom states in (\ref{midone}). 
 %SU(1,1) creation operator  $k_+$  creates all possible  s-states of hydrogen atom in (\ref{midone}): $\ket{{\sf n}, l=0, m=0}$ 
%in (\ref{lb}). 
%$|{\mathtt n},l=m=0\rangle = \frac{1}{\sqrt{{\mathtt %n}!({\mathtt n}-1)!}}k_+^{{\mathtt n}-1}\ket{0}$. 
 %The loop basis (\ref{lb}) is in the electric representation: 
% \begin{eqnarray} 
%(\vec{\cal E})^2~|n \rangle = \frac{\sf n}{2}\left(\frac{\sf n}{2} +1\right)~~|n\rangle. 
%\label{eev}
%\end{eqnarray} 
The single plaquette loop states in (\ref{lb}) 
%are orthonormal as well as complete in ${\cal H}^p$ and
%\begin{eqnarray}
%$\langle n^\prime|n\rangle = \delta_{n^\prime,n}, ~~ \sum_{n\in Z_+} |n\rangle \langle n| = 1.$ 
%\label{con} 
%\end{eqnarray}
% These loop states 
 %are annihilated by the SU(1,1) Casimir ${\cal C}$ as  ${\cal C} | 0 \rangle = 0$ and $\left[{\cal C},k_+\right] = 0$. Further, they 
 form a discrete representation of SU(1,1) with Bargmann index \footnote{Discrete SU(1,1) representations are characterized by: ~~$ {\cal C} \ket{{\mathtt k}, {\tt k}+m}= {\tt k}({\tt k}-1) \ket{{\mathtt k}, {\tt k}+m}, ~ 
k_0\ket{{\mathtt k}, {\tt k}+m} =({\tt k}+m) \ket{ {\mathtt k}, {\tt k}+m},~ k_+\ket{{\mathtt k}, {\tt k}+m} =\sqrt{(2{\tt k}+m)(m+1)} \ket{{\mathtt k}, {\tt k}+m+1}$ and $k_-\ket{ {\mathtt k}, {\tt k}+m} =\sqrt{(2{\tt k}+m-1)m} \ket{{\mathtt k}, {\tt k}+m-1}$. We have  ${\tt k}=1$,  $m=({\tt n}-1) =2j$ and $\ket{{\mathtt k}=1, 1+m} \equiv \ket{\tt n}$. }
%In (\ref{su11r}) $\ket{n} =\ket{m+1, {\mathtt k}=1}$ 
${\mathtt k} =1$:
 \begin{eqnarray}
 \label{su11r} 
 {\cal C} \ket{ \tt n} = 0,~~ && ~~k_0\ket{ \tt n} = \tt n \ket{\tt n}  \\
 k_+\ket{ \tt n} = \sqrt{\tt n(\tt n +1)}~ \ket{\tt n+ {1}}, &&
k_-\ket{\tt n} = \sqrt{\tt n (\tt n -1)}\ket{\tt n-{1}} \nonumber 
 \end{eqnarray}  
% and $\left[C, k_+\right] = 0$.  
These gauge invariant fundamental loop flux 
creation-annihilation and counting operators govern the
loop dynamics which we discuss in the next section.
Note that in the hydrogen atom loop basis all topological effects of the compactness of SU(2) gauge group are contained in the discreteness of the principal quantum numbers ${\tt n}$ of hydrogen atom.    
 
% As we will see in the next section, the loop or spin network basis in ${\cal H}^p$ on the entire lattice is also made up of these basic hydrogen atom states
 
\subsubsection{\bf Loop dynamics and $SU(1,1) \subset SO(4,2)$}
\label{ldsp}

We consider SU(N)  
Kogut-Susskind Hamiltonian \cite{kogut}:  
\begin{eqnarray}
H  & = & {g^{2}} \sum_{l=1}^{4} \vec{E}^2(l) 
+ \frac{K}{g^2}  \left[2N - {\textrm Tr} \left(  U_1~U_2~U_3^\dagger ~U_4^\dagger ~ +~ h.c~  \right)\right]   \nonumber \\
& &  \equiv ~~~~~H_E ~~+ ~~H_B.  
\label{ksham}   
\end{eqnarray} 
In (\ref{ksham}), 
%$K$ is a constant and 
%$U_{\square} \equiv U_1U_2U_3U_4$ is the SU(N) plaquette operator and 
K is a constant and $U_1 \equiv U(0,0;\hat{1}), U_2 \equiv U(1,0;\hat{2}), U_3 \equiv U(0,1;\hat{1}), U_4 \equiv U(0,0;\hat{2})$. Using links to loop relations  (\ref{fpt}) and (\ref{invrr}), the SU(N) 
loop Hamiltonian for the single plaquette is:
\begin{eqnarray}
H =  4 g^{2}  \vec{\cal E}^2 
+ \frac{K}{g^2}  \left[2N-{\textrm Tr} \left({\cal W}  + {\cal W}^\dagger\right)\right].  
\label{loopham} 
\end{eqnarray}  
%and $K$ is a constant. 
At this stage we specialize to SU(2)  case \footnote{Similar construction is also possible for 
SU(N) and involves SU(N) irreducible prepotential operators discussed in the context of SU(N) lattice gauge theories in \cite{rmi2}.}. 
The Hamiltonian (\ref{ksham}) can be completely 
rewritten in terms of loop creation, annihilation and counting operators forming SU(1,1) algebra. The electric field term is:  
\begin{eqnarray} 
H_{E} = g^2 \sum_{l=1}^{4} \vec{E}^2(l)
= 4 g^2 \vec{\cal E}^2 = g^2\left(k_0^2 -1\right). 
\end{eqnarray} 
The  four link magnetic field term takes its simplest  possible form: 
\begin{align} 
H_{B}  & =
% \frac{K}{g^2} ~ Tr U_{\square} = 
%\frac{1}{g^2} ~
\frac{1}{g^2}~ ~Tr (U_1~U_2~U_3^\dagger ~U_4^\dagger)
  = 
%\frac{1}{g^2} ~
\frac{1}{g^2}~~Tr~{\cal W} \nonumber \\
& =
%\frac{1}{g^2} ~
\frac{1}{g^2}~~\frac{1}{\sqrt{k_0}}~\Big(k_- + k_+\Big)~
\frac{1}{\sqrt k_0}.
\label{sf} 
\end{align} 
%As a result, the magnetic field term (\ref{sf}) also takes a  simple form. 
The magnetic field term, important in the weak coupling continuum limit, simply  
creates and annihilates the fluxes on the plaquette loop:  
\begin{align} 
H_B ~ \ket{ \sf n }~ & = \frac{1}{g^2}~Tr\left(U_1~U_2~U_3^\dagger ~U_4^\dagger\right)\ket{ \sf n}~=\frac{1}{g^2}~ Tr~ {\cal W} ~|\sf n \rangle \nonumber \\
& =\frac{1}{g^2}
~\Big[{\sf n} +{1}\rangle + \ket{{\sf n} -1} \Big]. 
\label{mft}
\end{align} 
 %Using Wigner Eckart theorem \cite{manu}, the matrix elements of the magnetic field term in this basis can also be expressed as: 
%\begin{eqnarray} 
%\langle n^\prime = 2 j^\prime | H_B|n = 2j\rangle = \frac{1}{g^2}~ 
%\Bigg\{\begin{array}{ccc}
%j^\prime & j & \frac{1}{2}\\
%j & j^\prime &  0\\
%\end{array} 
%\Bigg\} 
%\label{s6j}
%\end{eqnarray} 
Note that the magnetic field term which was product of 
four (link) flux operators reduces to a single (loop) flux operator. This is the simplest possible form 
of the important $(1/g^2)$ magnetic field term.
% which, to begin with,  was product of 4 flux creation-annhilation operators and therefore contained 16 terms \cite{manu}. 
In the Appendix \ref{matheqn} we show that the loop Schr\"odinger equation easily reduces to Mathieu equation in the magnetic basis. 
% % % % % % % % % % % % % % % %

In the case of finite lattice, considered in the next sections,  the states (\ref{midone}) of  hydrogen atoms are associated with every plaquette. Like in single plaquette case, they  describe the electric fluxes flowing around the corresponding plaquettes.  The Gauss law is solved by Wigner coupling all the hydrogen atom  states and demanding that the three components of the total angular momenta vanish. Further, 
the role of SU(1,1) in this section gets generalized to  the dynamical symmetry group
SO(4,2) of hydrogen atoms (see section \ref{sso42}).
%The prepotential operators also 
% enable us to construct $SU(2) \times U(1)$ gauge invariant operators following SU(1,1) algebra. We define
% \begin{eqnarray} 
% k_+ \equiv a^\dagger \cdot \tilde b^\dagger,~~~~ k_- \equiv a \cdot \tilde b, ~~~~
% k_0 = \frac{1}{2} \left(\hat N_a+ \hat N_b+2\right) = (\hat N+1). 
% \label{su11}  
% \end{eqnarray}  
% $k_\pm = Tr {\cal W}_\pm$.
% 
%
\subsection{\bf Canonical transformations on a finite lattice}

On a finite $d=2$ lattice  we  canonically transform    the $3 {\cal L}$ Kogut-Susskind conjugate operators  
$\left[U(x,y; \hat i), ~E^a_\mp(x,y;\hat i)\right]$ 
satisfying (\ref{ccr11}) on every link into
\begin{enumerate} 
\item  $3 ({\cal N}-1)$ unphysical  string conjugate  operators \footnote{In the appendix the string operators are denoted by $\left[{\sf T}_{[xxyy]}(x,y), {\sf E}^a_{[xxyy]\mp}(x,y)\right]$.
 The subscript $[xxyy]$ encodes the  Gauss law structures of the string electric field  at $(x,y)$. In this section, for the sake of notational convenience, we have ignored the subscripts and simply denoted them by ${\sf T}(x,y)$ and  ${\sf E}^a_\mp(x,y)$.} $\left[{\sf T}(x,y), {\sf E}^a_{\mp}(x,y)\right]$ satisfying (\ref{ccrs12}) at every  site. These operators  are shown in Figure \ref{fpls}-a.   The string ${\sf T}(x,y)$ start at $(0,0)$
and end at $(x,y)$ following the path $(0,0) \rightarrow (x,0) \rightarrow (x,y)$.
\item  $3 {\cal P}$ physical loop conjugate operators   $\Big[{\cal W}(x,y),~{\cal E}^a_\mp(x,y)\Big]$ satisfying (\ref{ccr12}) on every plaquette or 
equivalently at every  dual site. These operators are shown in Figure \ref{fpls}-b. The plaquette loop 
flux operator ${\cal W}(x,y)$ is along the path: $(0,0)\rightarrow (x-1,0) \rightarrow (x-1,y-1) 
\rightarrow (x,y-1) \rightarrow (x,y) \rightarrow (x-1,y)
\rightarrow (x-1,0) \rightarrow (0,0)$. 
  \end{enumerate} 
  
The above two sets are mutually independent. 
 As mentioned earlier, the total degrees of freedom match because   ${\cal L} = {\cal P} + ({\cal N} - 1)$. 
 %The canonical transformations 
 %leading to the above new string \& loop flux operators and their conjugate electric fields are explicitly constructed in the appendix A. 
  
 \subsubsection{\bf Canonical relations}
 \begin{figure}
 \centering
 \includegraphics[scale=0.9]{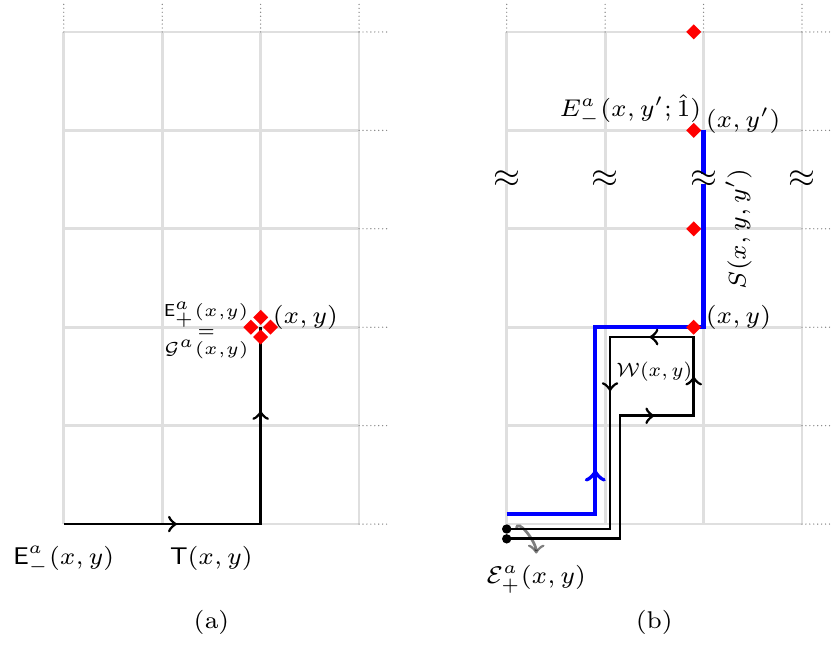} 
 \caption{  Graphical representation of the canonical relations (\ref{slef}). The Kogut Susskind electric fields are denoted by {\tiny {\color{red}$\blacklozenge$}} and the plaquette loop electric fields are denoted by $\bullet$. We show
 a) string electric field in terms of Kogut-Susskind 
 electric fields %in (\ref{slef}) 
 and (b) plaquette loop electric fields ${\cal E}_+^a(x,y)$ in terms of the original Kogut-Susskind link electric fields. In (a) the 4 {\tiny{\color{red}$\blacklozenge$}} at $(x,y)$ denotes the Gauss law operator at ${\cal G}^a(x,y)$. In (b) Kogut Susskind link electric fields $E_-^a(x,y'; \hat{1}) ; y'=y,y+1\cdots{\sf N}$ are parallel transported by $S(x,y,y')$ (denoted by thick line) to give the loop electric field ${\cal E}_+^a(x,y)$ .
% Graphical representation of (a) plaquette loop electric fields in terms of the original Kogut-Susskind link electric fields (see eqn. (\ref{slef})); (b) link electric field $E_-^a(x,y; \hat{1})$ , (c) $E_-^a(x,0; \hat{1})$ and (d) $E_-(x,y; \hat{2})$ in terms of plaquette loop operators and electric field. (see eqn. ($\ref{kstowele}$)). $\bullet$ represents plaquette electric fields and $\color{red}\times$ represents link electric fields.
}
 \label{flooplink}
 \end{figure}
 \noindent The  $({\cal N}-1)$ string in Figure \ref{fpls}-a and $({\cal P})$ plaquette loop flux operators in Figure \ref{fpls}-b are related to the initial $({\cal L})$ Kogut-Susskind link operators as (see appendix A for details): 
 \begin{eqnarray} 
{\sf T}(x,y)~~ = ~~\prod_{x^\prime=0}^{x}U(x^\prime,0; \hat{1}) \prod_{y^\prime=0}^{y} U(x,y^\prime; \hat{2}), \nonumber \\
  {\mathcal W}(x+1,y+1) = ~~{\sf T}(x,y)~U_p(x,y) ~{\sf T}^\dagger(x,y).
  \label{splo}
  \end{eqnarray} 
  In (\ref{splo}), the strings ${\sf T}(x,y)$ are defined at all lattice sites away from the origin 
  and the loop operators ${\cal W}(x,y)$ are located  at  $x,y=1,2,\cdots ,\sf N$. The Kogut-Susskind plaquette operators are defined as: $U_p(x,y)=U(x,y;\hat{1})~U(x+1,y;\hat{2})~
U^\dagger(x+1,y+1;\hat{1})~U^\dagger(x,y+1;\hat{2})$. The conjugate string and plaquette loop electric fields in terms of the initial Kogut-Susskind link electric fields are (see appendix A for details): 
\begin{eqnarray}
{\sf E}^a_+(x,y) &= &\sum_{i=1}^{2} \left[ E^a_-(x,y;\hat i)+ E^a_+(x,y;\hat i)\right] = \underbrace{{\cal G}^a(x,y)}_{{\bf {=0}}}, \nonumber \\
{\mathcal E}_+^a(x,y) &=& -\sum\limits_{y'=y}^{\sf N} R_{ab}(S(x,y,y^\prime))E_-^b(x,y'; \hat{1}).
\label{slef} 
 \end{eqnarray}
In (\ref{slef}), we have defined:   
%\begin{eqnarray}
$S(x,y,y^\prime)  \equiv {\sf T}(x-1,y)~U(x-1,y;\hat{1})~\prod_{y''=y}^{y'}~U(x,y''; \hat{2})$ and $x\neq 0;y\neq 0$. %\nonumber 
%\end{eqnarray}
%On a single plaquette lattice, eqn. (\ref{slef}) reduces to eqn.(\ref{refw}) as expected.
 The relations (\ref{slef}) between the new string and loop electric fields and old Kogut-Susskind electric fields are derived in appendix A (see (\ref{cycle}) and (\ref{looptolinkr})). They are illustrated in Figure \ref{flooplink}-a and  Figure \ref{flooplink}-b respectively.
Because of the SU(N) Gauss laws all string operators, containing gauge degrees of freedom away from the origin, naturally decouple from the theory. The remaining physical plaquette loop operators can be thought of as a set of collective coordinates which  describe the theory without any redundant loop or local gauge degrees of freedom. %These operators, being related to the initial Kogut-Susskind link operators through canonical transformations, constitute a set  of fundamental  conjugate loop operators in ${\cal H}^p$. 
These ${\cal P}$ SU(N) loop flux operators are all mutually independent (no SU(N) Mandelstam constraints) and obey the canonical quantization conditions with their loop electric fields exactly like the original Kogut-Susskind link operators in (\ref{ccr11}). Note that in the special single plaquette case the 
the relations  (\ref{slef}) reduce  to the relations already derived in the section \ref{ctsp}. As an example the second relation in (\ref{slef}) states
${\cal E}_+^a(1,1) = -R_{ab}({\sf T}_{[x]}(1,1)) E_-^b(1,1;\hat 1)$ which is included in (\ref{refw}).

\subsubsection{\bf Inverse relations} 
%\begin{figure}
%    \centering
%    \includegraphics[scale=0.8]{new2.pdf} 
% %   \includegraphics{new2.pdf}
%    \caption{ Graphical representation of the inverse canonical relations (\ref{kstowele}): a) link electric field $E_+^a(x,y=0; \hat{1})$, (b) $E_+^a(x,y \neq 0; \hat{1})$ and (c) $E_+(x,y; \hat{2})$ in terms of plaquette loop operators and loop electric field.
%        The $\bullet$ represents plaquette loop electric fields and {\tiny{$\color{red}\blacklozenge$}} represents Kogut-Susskind link electric fields. All loop electric fields $\bullet$ are parallel transported along blue lines to give Kogut Susskind link operator $E^a_+(x,y;\hat i)$ or  {\tiny{$\color{red}\blacklozenge$}} in (\ref{kstowele}). In (a)  $\sum_p {\sf L}^b(p)$ gives  $\Delta_X^b(x,y=0)$ in (\ref{kstowele}),  the summations is over the plaquettes in the dotted region. In (c)  we show $\Delta_Y(x,y)$ where the summation is again over the plaquettes in the dotted region. The shaded region in (c) represents ${\cal W}_{xy}(x,y)$ in the second equation in  (\ref{kstowele}). }
%    \label{flinkloop}
%    \end{figure} 
\begin{figure}
    \centering
    \includegraphics[scale=0.82]{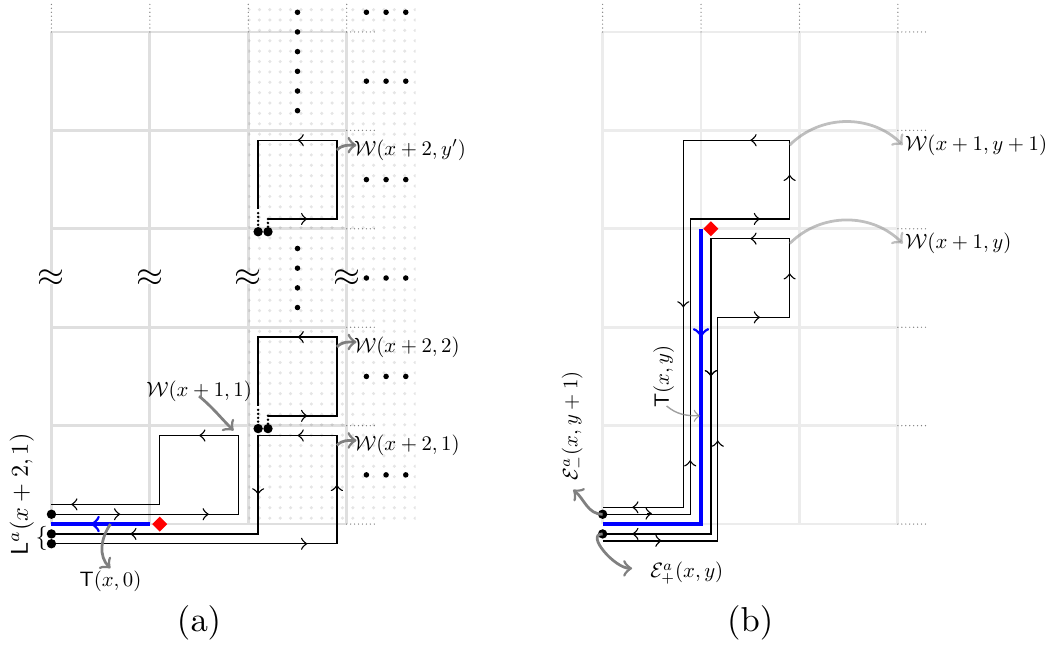} 
    \caption{ Graphical representation of the inverse canonical relations (\ref{kstowele}): a) link electric field $E_+^a(x,y=0; \hat{1})$, (b) $E_+^a(x,y \neq 0; \hat{1})$ in terms of plaquette loop operators and loop electric field.
        The $\bullet$ represents plaquette loop electric fields and {\tiny{$\color{red}\blacklozenge$}} represents Kogut-Susskind link electric fields. All loop electric fields $\bullet$ are parallel transported along thick lines to give Kogut Susskind link operator $E^a_+(x,y;\hat i)$ or  {\tiny{$\color{red}\blacklozenge$}} in (\ref{kstowele}). In (a)  $\sum_p {\sf L}^b(p)$ gives  $\Delta_X^b(x,y=0)$ in (\ref{kstowele}),  the summations is over the plaquettes in the dotted region. }
    \label{flinkloop}
    \end{figure} 
\noindent  The  Kogut Susskind link flux operators in terms of the string \& loop flux operators are: 
\begin{eqnarray}
U(x,y; \hat{1}) &  = & ~~{\sf T}^\dagger(x,y)~{\cal W}(x+1,y)~{\cal W}(x+1,y-1) \nonumber \\
&& \cdots\cdots~~{\cal W}(x+1,1)~{\sf T}(x+1,y)\nonumber\\
U(x,y; \hat{2})& = &~~~~~~{\sf T}(x,y+1)~{\sf T}^\dagger(x,y).
\label{aabb}
\end{eqnarray}
The relations (\ref{aabb}) are clear from Figure \ref{fpls}-a,b. 
%In (\ref{aabb}) and below we denote $\sf T_{[yyxx]}(x,y)$ simply by $\sf T(x,y)$ for notational convenience.
The Kogut-Susskind link electric fields in terms of the loop electric fields  are (see appendix B for details):
%{\footnotesize 
\begin{eqnarray}
\label{kstowele}
& E_+^a(x,y; \hat{1})~ = ~ R_{ab}({\sf T}^\dagger(x,y))\Bigg[{\cal E}_-^b(x+1,y+1)~~+
\nonumber \\  & ~~~~{\cal E}_+^b(x+1,y)  
 + \underbrace{\delta_{y,0}\sum_{\bar x=x+2}^{\sf N}\sum_{\bar y=1}^{\sf N } {\sf L}^b(\bar x,\bar y)}_{\Delta_X^b(x,y)}\Bigg], \nonumber \\  \\
& E_+^a(x,y; \hat{2}) ~ = ~ 
R_{ab}({\sf T}^\dagger(x,y))\bigg[{\cal E}_+^b(x+1,y+1)
~~+ \nonumber \\
& ~~~~ R_{bc}({\cal W}_{xy}){\cal E}_-^c(x,y+1)
+\underbrace{\sum\limits_{\bar y=y+2}^{\sf N} {\sf L}^b(x+1,\bar y)}_{\Delta_Y^b(x,y)} 
\bigg]. \nonumber 
\end{eqnarray}
%}
In (\ref{kstowele}) we have defined the parallel transport:
\begin{eqnarray} 
R_{bc}({\cal W}_{xy}) 
\equiv R_{bc}\big(
%\prod_{\bar y=y}^{1}
{\cal W}(x,1)~{\cal W}(x,2)~\cdots {\cal W}(x,y)
\big). 
\label{partran}
\end{eqnarray}
and  used:~
${\cal E}_\pm^a(x,y=0) \equiv 0,~ ~~~{\sf L}^a(x,y) \equiv \left[{\cal E}^a_-(x,y) +{\cal E}^a_+(x,y)\right].$
The inverse relations (\ref{kstowele}) and (\ref{partran}) for $E^a(x,y=0;\hat 1), E^a(x,y \neq 0;\hat 1)$ and $E^a(x,y;\hat 2)$  are illustrated in Figure \ref{flinkloop}-a,b and Figure \ref{flinkloopc} respectively. 
On a single plaquette lattice (\ref{kstowele}) reduces to (\ref{invr}) as expected.
%In  Figure 
%\ref{flinkloop}-a we graphically represent   
%$E^a(x,y=0;\hat 1)$ in terms of loop electric fields. %The Figure 
%\ref{flinkloop}-b illustrates the same when $y \neq 0$.
%In Figure \ref{flinkloop}-c  we represent 
%$E^a(x,y;\hat 2)$ in (\ref{kstowele}) in terms of loop electric fields.  These inverse relations  are derived in appendix B (see (\ref{e+fin}), (\ref{final1}) and (\ref{final2})).
% and are used to derive SU(N) 
%loop dynamics after discussing  the physical loop Hilbert space ${\cal H}^p$ in the next section. 
 \begin{figure}
      \centering
      \includegraphics[scale=0.8]{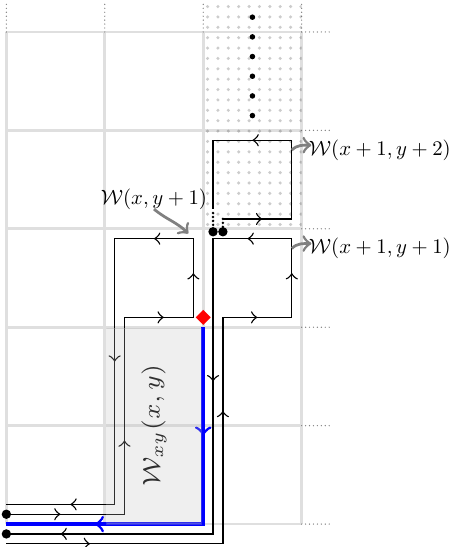} 
      \caption{ Graphical representation of the inverse canonical relations (\ref{kstowele}): $E_+(x,y; \hat{2})$ in terms of plaquette loop operators and loop electric field.
%          The $\bullet$ represents plaquette loop electric fields and {\tiny{$\color{red}\blacklozenge$}} represents Kogut-Susskind link electric fields. All loop electric fields $\bullet$ are parallel transported along blue lines to give Kogut Susskind link operator $E^a_+(x,y;\hat i)$ or  {\tiny{$\color{red}\blacklozenge$}} in (\ref{kstowele}).
 We show $\Delta_Y(x,y)$ where the summation is over the plaquettes in the dotted region. The shaded region represents ${\cal W}_{xy}(x,y)$ in the second equation in  (\ref{kstowele}). }
      \label{flinkloopc}
      \end{figure} 
 \subsubsection{\bf Physical loop Hilbert space  ${\cal H}^p$ and Hydrogen atoms}
 
 Like in the single plaquette case, the SU(N) Gauss law does not permit any string excitation and the $({\cal N}-1)$ string operators  become irrelevant. Therefore, all  possible SU(N) gauge invariant operators are made up  
 of the ${\cal P}$ fundamental plaquette loop operators 
 and their conjugate electric fields. 
 In other words, 
 the non-trivial problem of SU(N) gauge invariance over the entire lattice reduces to the problem of residual SU(N) global invariance of $3{\cal P}$ loop operators, all starting and ending  at the origin. Further, 
 all  
 $3 {\cal P}$ 
 loop operators gauge transform as adjoint matter fields at the origin: 
\begin{align}
{\cal W}(p)   \rightarrow \Lambda ~{\cal W}(p) ~\Lambda^\dagger, ~~~~
{\cal E}_{\pm}(p)   \rightarrow  \Lambda ~{\cal E}_{\pm}(p) ~\Lambda^\dagger.
 %~p=1,2, \cdots ,{\cal P}.
 \label{wgtao}
\end{align}
%\begin{eqnarray}
%{\cal E}_{\pm}(p) & \rightarrow & \Lambda_o ~{\cal E}_{\pm}(p) ~\Lambda^\dagger_o, \nonumber \\
% {\cal W}(p) & \rightarrow & \Lambda_o ~{\cal W}(p) ~\Lambda^\dagger_o, ~~~~~~~~~~p=1,2, \cdots ,{\cal P}.
% \label{wgtao}
%\end{eqnarray}
%\begin{eqnarray} 
%{\cal W}(p) \rightarrow& \Lambda_o ~{\cal W}(p)~ \Lambda^\dagger_o, ~~~~
%{\cal W}_\pm(p) \rightarrow \Lambda_o ~{\cal W}_\pm(p)~ \Lambda^\dagger_o;  ~~~~ \forall~ p =1,2,\cdots , {\cal P}. 
%\label{wgtao}  
%\end{eqnarray} 
%the origin is excluded $x=y \neq 0$ and
In (\ref{wgtao}), $\Lambda = \Lambda(0,0)$ are the gauge 
transformations at the origin.
This global invariance at the origin is fixed  by the residual $(N^2-1)$ SU(N) Gauss 
laws:
%This SU(N) gauge invariance of at the origin results in  the global SU(N) invariance of the final loop Hamiltonian (\ref{sunloopham}). The corresponding Gauss laws in terms of loop electric fields (see below and  Appendix A) is:
\begin{eqnarray} 
{\cal G}^a(0,0) 
%= \Big(E^a_+(0,0;\hat 1) + E^a_+(0,0; \hat 2)  \Big) 
= \sum_{p=1}^{\cal P}\Big[{\cal E}^a_-(p) + 
{\cal E}^a_+(p)\Big] \equiv \sum_{p=1}^{\cal P} ~{\sf L}^a(p) =0. 
\label{lglao}
\end{eqnarray} 
%\subsubsection{The Physical Loop Hilbert Space} 
We now solve the Gauss law (\ref{lglao}). 
%To keep the discussion simple we focus on SU(2) gauge group \cite{msplb}. We  have ${\cal P}$ physical loop operators ${\cal W}_{\alpha\beta}(p)$ and their conjugate electric fields ${\cal E}_\pm(p)$.
%For an $L\times L$ lattice, the total Hilbert space is given by the direct product of the plaquette Hilbert spaces. 
A basis in the full Hilbert space of SU(2) lattice gauge theory on a ${\cal P}$ plaquette lattice is given by  $\ket{j,m_-,{m}_+}_{1} \otimes \ket{j,m_-,{m}_+}_{2} \otimes \cdots \otimes\ket{j,m_-,{m}_+}_{\cal P}$.
We are interested in constructing the physical Hilbert space ${\cal H}^p$ which is the $SU(2)$ invariant subspace of the above direct product Hilbert space. As seen in the single plaquette case, it is convenient to 
define prepotentials for this purpose. 
We generalize (\ref{su2sb}) and write: 
\begin{eqnarray}   
{\cal E}_-^a(p) =  a^\dagger(p) \left(\frac{\sigma^a}{2}\right) a(p), ~{\cal E}_+^a(p) = b^\dagger(p) \left(\frac{\sigma^a}{2}\right)  b(p). %~~~~p=1,2,\cdots ,{\cal P}.
\label{su2sbp} 
\end{eqnarray} 
We define the number operators on every plaquette: 
$\hat N_a(p) \equiv a^\dagger(p) \cdot a(p)$ and 
$\hat N_b(p) \equiv b^\dagger(p) \cdot b(p)$. As the magnitudes of left and right electric field operators are equal we  have the following constraint:
\begin{align} 
\hat N_a(p) = \hat N_b(p) \equiv \hat N(p) 
%~~~~~~~~~~~~~~p=1,2,\cdots,{\cal P}, 
\label{abcons}
\end{align}
on every plaquette p.  
The loop flux operators (\ref{holo}) also generalize: 
%{\footnotesize
\begin{eqnarray} 
{\cal W}_{\alpha\beta}(p) &= &%\frac{1}{\sqrt{(N(p)+1)}} 
\hat F_p~\Big[ a_{\alpha}(p)~ \tilde b_{\beta}(p)~-~\tilde a^{\dagger}_{\alpha}(p)~  
b^{\dagger}_{\beta}(p) \Big] ~\hat F_p
%\frac{1}{\sqrt{(N(p)+1)}} 
\nonumber \\
& \equiv& 
\hat F_p~
%\frac{1}{\sqrt{(N(p)+1)}}
\Big[{\cal W}^{(-)}_{\alpha\beta}(p) ~+~ {\cal W}^{(+)}_{\alpha\beta}(p)\Big] ~\hat F_p.
%\frac{1}{\sqrt{(N(p)+1)}}.  
\label{holop}
\end{eqnarray}
%}
In (\ref{holop}), $\hat F_p \equiv  \frac{1}{\sqrt{(\hat N(p)+1)}}$ are the normalization constants so that 
${\cal W}$ is unitary.    
Under SU(2) (global) gauge transformations (\ref{wgtao}): 
\begin{eqnarray} 
a_{\alpha}(p) \rightarrow \Lambda_{\alpha\beta} ~a_\beta(p), ~~~~
b_{\alpha}(p) \rightarrow \Lambda_{\alpha\beta} ~b_\beta(p). % ~~~~~~~~p=1,2,\cdots ,{\cal P}.
\label{sbgtop}
\end{eqnarray}
In the prepotential representation, we have new U(1) local gauge invariance on each plaquette loop: 
\begin{eqnarray} 
a_\alpha(p) \rightarrow e^{i\theta(p)} ~a_\alpha(p), ~~~ b_\alpha(p) \rightarrow e^{-i\theta(p)} ~b_\alpha(p). %~~~~~~~~ p=1,2,\cdots ,{\cal P}. 
\label{abgip} 
\end{eqnarray}
 %The transformations (\ref{sbgtop}) and (\ref{abgip}) are straightforward generalizations of the single plaquette transformations (\ref{sbgto}) and (\ref{abgi}) respectively. 
 The transformation (\ref{abgip}) is generalization of 
 (\ref{abgi}).  The abelian gauge angle now depends on the location of the plaquette loop. The electric fields (\ref{su2sbp}) and the loop flux operators (\ref{holop}) are invariant under (\ref{abgip}).  This abelian gauge invariance will play a role later in constructing SO(4,2) loop operators in section (\ref{sso42}).
The   hydrogen atom states $\ket{{\sf n}_p~l_p~m_p}$  
for each individual plaquette  p $(=1,2,\cdots,{\cal P})$   can be  constructed exactly like in (\ref{jmm}) and (\ref{nlmm}).  %The  eigenvalue equations (\ref{midone}) are associated with each plaquette now.  
Under gauge transformation $\Lambda$ at the origin, all states transform together as: 
\begin{eqnarray} 
\ket{{\sf n}_p~l_p~m_p} \rightarrow \sum_{\bar m_p = -l_p}^{l_p}D^{~~l_p}_{m_p \bar m_p}\left(\Lambda\right) ~\ket{{\sf n}_p~l_p~\bar m_p}. %~~~~~~~~~~ p=1,2,\cdots ,{\cal P}.
\label{nabgip} 
\end{eqnarray}
\begin{figure}
\centering
\includegraphics[scale=.53]{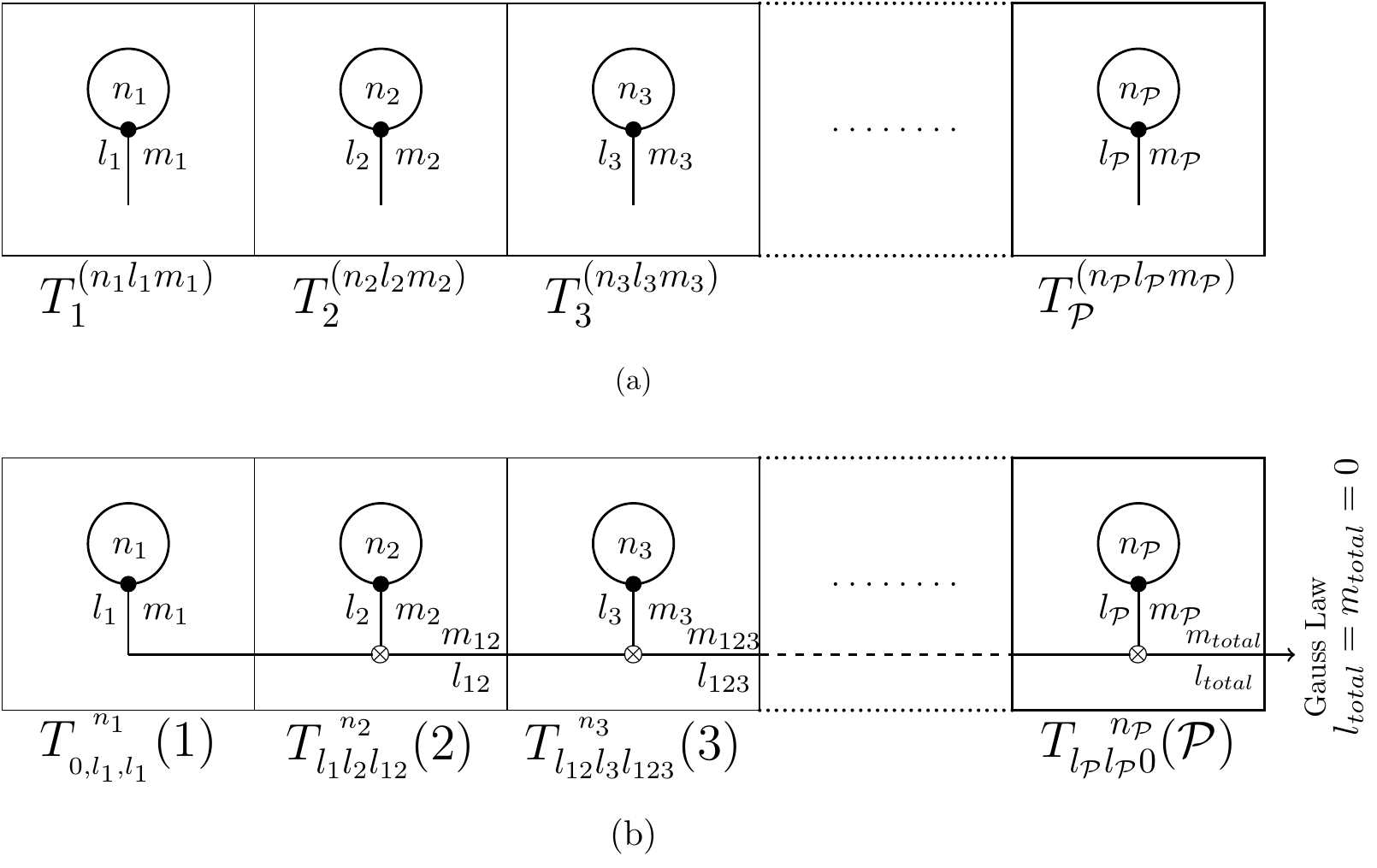}
\caption{a) Uncoupled and b) Coupled hydrogen atom loop basis. The global Gauss law is solved by putting the total angular momentum ${\sf L}_{\textrm total}=0$. The tensors involved in the  matrix product states in section \ref{smps} are also shown at the bottom. In (a) and (b)  $\bullet$ represents the $j$-$j$ coupling or contraction of $j$ flux lines within a plaquette in (\ref{nlmm}) and in (b) $\otimes$ represents $l$-$l$ couplings or contraction of $l$ 
flux lines between neighbouring plaquettes (see eqn.(\ref{entlatt}) and [44]).}
\label{f:finlatlss}
\end{figure}
Therefore, all principal and angular momentum quantum numbers ${\sf n}_p,~ l_p$ are already gauge invariant. To proceed further, we  separate the 
gauge variant part of the hydrogen atom state 
$\ket{{\sf n}~l~m}$ 
in (\ref{nlmm}) from its gauge 
invariant part on each plaquette. 
%For this purpose 
We write it as:
\begin{eqnarray} 
\ket{{\sf n}~l~m}  = 
% {\sf K} 
% \sum_{\{m_1\cdots m_{2l}\}_m} %\underbrace{
%\left\{a^\dagger_{m_1}  \cdots a^\dagger_{m_l}
%b^\dagger_{m_{l+1}}  
% \cdots b^\dagger_{m_{2l}}\right\} 
%~
%\left(k_+\right)^{{\sf n}-l-1}~{\ket 0} 
{\sf K} \hat {\cal S} \hat {\cal A} ~\ket{0}. 
\label{nha}
\end{eqnarray}
In (\ref{nha}), ${\sf K}$ is a normalization constant,
$\hat {\cal S}$ and $\hat {\cal A}$ define the symmetric and anti-symmetric parts as follows: 
$$\hat {\cal S} \equiv \sum_{\{m_1\cdots m_{2l}\}_m} \left\{a^\dagger_{m_1}  
%a^\dagger_{m_2} 
\cdots a^\dagger_{m_l}
b^\dagger_{m_{l+1}}  
%b^\dagger_{m_{l+2}} 
\cdots b^\dagger_{m_{2l}}\right\} $$ 
and 
$$\hat {\cal A} \equiv \left(k_+\right)^{{\sf n}-l-1} 
= \left(a^\dagger_1b^\dagger_2-a^\dagger_2 b^\dagger_1\right)^{{\sf n}-l-1}.$$
All magnetic quantum numbers in ${\cal S}$ $m_1,\cdots ,m_{2l} = \pm  \frac{1}{2}$ are summed over  
such that the condition, $m=m_1+m_2+\cdots +m_{2l}$, is satisfied.  In (\ref{nha}), the anti-symmetric operator ${\cal A} \equiv (k_+)^{{\sf n} -l -1}$  
represents the gauge invariant flux loops in  (\ref{nabgip}) within a plaquette.  On the other hand, the symmetric operator ${\cal S } \equiv \sum_{\{m_1\cdots m_{2l}\}_m} %\underbrace{
\left[a^\dagger_{m_1}  a^\dagger_{m_2} \cdots a^\dagger_{m_l}
b^\dagger_{m_{l+1}}  b^\dagger_{m_{l+2}} \cdots b^\dagger_{m_{2l}}\right]$ represents the uncoupled 
open flux lines coming out of the plaquette and forming the vector part of the state $\ket{n~l~m}$. If $l$ has its minimum value $l=0$ on a plaquette then 
${\cal S}$ is an identity operator.  All $2j ~(=({\sf n}-1))$ plaquette flux lines are mutually contracted like in the single plaquette case  and  
(\ref{nha}) reduces to (\ref{lb}). This is $j$-$j$ coupling in (\ref{nlmm}) within a plaquette. At the other limit, if $l$ has its maximum value $l=({\tt n} -1) =2j$ then all $4j$ plaquette loop prepotential operators  in (\ref{nha}) are symmetrized and  
there  is no anti-symmetrization or self contraction by $k_+$ operator. 
In other words all $4j$ flux lines flow out of the plaquette
and need to be contracted  with similar symmetrized flux lines from other plaquettes to get all possible gauge invariant loop states over the entire lattice.  This is $l$-$l$ coupling (see [44]).  A  hydrogen atom state has 
$0 \le l \le ({\sf n}-1)$. Therefore, it is convenient to represent the hydrogen atom states $\ket{{\sf n}~l~m}$ by tadpoles on every plaquette as shown 
in Figure \ref{f:finlatlss}-a. The tadpole loop at the top represents the flux flowing in a loop within the plaquette. This is the anti-symmetrized part ${\cal A}$ in (\ref{nha}). The vertical stem of the tadpole is the symmetrized part ${\cal S}$, it represents the flux leakage $(l,m)$ through the plaquette. 
We now consider the direct product states of all ${\cal P}$ hydrogen atoms in Figure \ref{f:finlatlss}-a:
%{\footnotesize 
\begin{align}
\left\vert \begin{array}{cccc} {\sf n}_1 & {\sf n}_2 &  ~~\cdots ~~~{\sf n}_{\cal P} ~~\\ 
l_1 & l_2 &  \cdots   ~~~l_{\cal P} \\ 
m_1 & m_{2}& ~~\cdots~~  m_{\cal P} \end{array} \right\rangle & \equiv  
|{\sf n}_1~l_1~m_1\rangle \otimes  |{\sf n}_2~l_2~m_2\rangle  \cdots \nonumber \\
& \cdots\cdots \otimes |{\sf n}_p~l_p~m_p\rangle.
\label{dphals}  
\end{align} 
%}
% already contain    
%$2{\cal P}$ gauge invariant quantum numbers:
%$\left(n_1,n_2,\cdots,n_{\cal P}\right)$ and %$\left(l_1,l_2,\cdots ,l_{\cal P}\right)$.
In order to solve the  Gauss law  (\ref{lglao}) 
we describe the states  (\ref{dphals}) in a coupled basis shown in Figure \ref{f:finlatlss}-b. We couple ${\sf L}^a_1, {\sf L}^a_2, \cdots ,{\sf L}^a_{\cal P}$ and go to a basis where 
%trade off the ${\cal P}$ gauge variant quantum numbers $(m_1,m_2,\cdots ,m_{\cal P})$ in terms of 
%$({\cal P} -3)$ gauge invariant quantum numbers,  we  transform to the  coupled   angular momentum  basis in a sequential manner. 
in addition to the diagonal $(J_1^2, J_2^2,\cdots ,J_{\cal P}^2)$ and $({\sf L}_1^2, {\sf L}_2^2,\cdots ,{\sf L}_{\cal P}^2),$ the following $({\cal P}-3)$ angular momentum operators, commuting with the above two sets,  are diagonal: 
$$\Bigg[({\sf L}_1+{\sf L}_2)^2, ~({\sf L}_1+{\sf L}_2+ {\sf L}_3)^2, \cdots\cdots, 
\underbrace{({\sf L}_1+{\sf L}_2+\cdots {\sf L}_p)}_{=0 ~({\rm Gauss~Law})}{}^2,$$
~~~~~~~~~~~~$\underbrace{({\sf L}_1+{\sf L}_2+ \cdots +{\sf L}_p)}_{=0~({\rm Gauss ~Law})}{}^{a=3}\Bigg].$

Note that the total  angular momentum is zero implying $({\sf L}_1+{\sf L}_2+\cdots +{\sf L}_{{\cal P}-1})^2 = {\sf L}^2_{\cal P}$ (see Figure \ref{f:finlatlss}-b).
Thus we have traded off ${\cal P}$ gauge variant magnetic quantum numbers $(m_1,m_2,\cdots ,m_{\cal P})$
in (\ref{dphals})  in terms of $({\cal P}-3)$ 
gauge invariant eigenvalues of the coupled ${\sf L}$ operators 
shown above. Therefore, in total  
there are $3({\cal P}-1)$ members of the complete set of commuting operators.    
The resulting SU(2) gauge invariant loop basis on a lattice with ${\cal P}$ plaquettes is given by  
\footnote{
More explicitly, the states in (\ref{entlatt}) are:
\unexpanded{
{\footnotesize
\begin{align} 
& 
\left\vert{[{\sf n}]~ [l]~  [ll]}\right\rangle
\hspace{-0.1cm}  \equiv \hspace{-0.4cm} \sum_{\{all ~m\}} 
\Big\{
C_{l_1m_1;l_2m_2}^{~~l_{12}m_{12}} ~C_{l_{12}m_{12};l_3m_3}^{~~l_{123}m_{123}} ~C_{l_{123}m_{123};l_4m_4}^{~~l_{1234}m_{1234}} \nonumber \\ 
& 
\cdots  
C_{l_{12\cdots (p-1)} m_{12\cdots (p-1)};l_p m_p}^{~~~\bf{l_{total} =0, m_{total}=0}}\
\Big\} 
\left\vert \begin{array}{cccc} {\sf n}_1 & {\sf n}_2 &  \cdots {\sf n}_{\cal P} \\
l_1 & l_2 &  \cdots  l_{\cal P} \\ 
m_1 & m_{2}& \cdots  m_{\cal P} \end{array} \right\rangle. \nonumber
%\label{llL} 
\end{align}}}
}:
\begin{align}
&\left\vert \begin{array}{cccc} {\sf n}_1 & {\sf n}_2 &  ~~\cdots ~~~{\sf n}_{\cal P} ~~\\ 
l_1 & l_2 &  \cdots   ~~~l_{\cal P} \\ 
l_{12} & l_{123}& ~~~~~\cdots  l_{total}=0 \end{array} \right\rangle = 
\Bigg\{|{\sf n}_1~l_1~m_1\rangle \otimes  |{\sf n}_2~l_2~m_2\rangle \nonumber \\
&\phantom{xxxxxxxxxxxxxxxx}\cdots \otimes |{\sf n}_p~l_p~m_p\rangle\Bigg\}^{{l_{total}=0}}_{{m_{total}=0}}.
\label{entlatt} 
\end{align}
Note that, like in the single plaquette case, all 
topological effects of the compactness of gauge 
group are now  contained  in the 
principal and angular momentum quantum numbers of hydrogen atom  $\ket{{\tt n}~ l~ m}$.
The above loop basis will be briefly denoted by $\left\vert{\{{\sf n}\}~ \{l\}~ \{ll\} }\right\rangle$. 
 The symbols $\{{\sf n}\}, ~\{l\}$ and $\{ll\}$ stand for the sets $({\sf n}_1,{\sf n}_2, \cdots, {\sf n}_{\cal P}): {\cal P}$  principal quantum numbers; $~ 
(l_1,l_2,\cdots, l_{\cal P}): {\cal P}$ angular momentum 
quantum numbers and $(l_{12},l_{123}, \cdots ,l_{123 \cdots ({\cal P}-1)} =l_{\cal P}, l_{123\cdots {\cal P}}=0): {\cal P}-3$ coupled angular momentum quantum numbers respectively. 
%\subsubsection{The gauge invariant SU(N) loop quantum numbers} 
These $3({\cal P}-1)$ principal, angular momentum quantum numbers characterizing the loop basis are 
gauge invariant as is clear from the gauge transformations (\ref{nabgip}). As expected, this  is also the number of  physical degree 
of freedom in the original Kogut-Susskind formulation. 
In fact, in  SU(N) Kogut-Susskind lattice gauge theory  in terms of link operators,  the total number of physical degrees of freedom is given by the dimension of the quotient space: 
\begin{align} 
{\mathbb N}^d_{SU(N)} =\left[\frac{\otimes_{links} ~SU(N)}{\otimes_{sites} ~SU(N)}\right] = \left(N^2-1\right)\left({\cal L}-{\cal N}\right). 
%= 3\left({\cal P}-1\right)
\label{dqs}
\end{align} 
Above, ${\cal L}$ and ${\cal N}$ are the numbers of links and sites of space lattice in d dimension. In $d=2$ we have ${\cal L} -{\cal N} = {\cal P}  -1$ and if we further choose $N=2$ then, as mentioned above,   ${\mathbb N}_{SU(2)}$ in (\ref{dqs}) is also the number of  gauge invariant principal and angular momentum quantum numbers appearing in the orthonormal hydrogen atom loop basis  (\ref{entlatt}) in ${\cal H}^p$.   

We now discuss pure $SU(N),~ N \ge 3$  lattice gauge theory in two and three space dimension. 
A $SU(N)$ tadpole state over a plaquette, analogous to the $SU(2)\otimes SU(2)$ state $\ket{j~m} \otimes \ket{ j~m^\prime} \sim \ket{n~l~m}$ in (\ref{jmm}) and illustrated in Figure \ref{f:finlatlss}, is  characterized by the representations of $SU(N) \otimes SU(N)$ group. 
These representations or equivalently orthonormal SU(N) tadpole states on each plaquette  are labelled by  $(N^2-1)$ loop quantum numbers
\footnote{A SU(N) irreducible representation is characterized by $(N-1)$ eigenvalues of Casimir operators and $\frac{1}{2}N(N-1)$ ~``SU(N) magnetic quantum numbers". As an example, the three ``SU(3)  magnetic quantum numbers" are the SU(2) isospin,  its third component and the hypercharge. 
 The $SU(2)\otimes SU(2)$ tadpole or hydrogen atom  states $\ket{j,m_-,{m}_+}$ are now replaced by $\ket{p,q,i_-,m_-,y_-, i_+,  m_+, y_+}$ where $p,q$ are the common eigenvalues of the two SU(3) 
Casimir operators and $i_\mp, m_\mp,y_\mp $ represent their isospin, magnetic isospin and hypercharge quantum numbers respectively.  These 8 quantum numbers are associated with a $SU(3)
\otimes SU(3)$  tadpole diagram.
Therefore, all $SU(N) \otimes SU(N)$ representations with equal Casimirs or SU(N) tadpole states are characterized by $(N-1) + N(N-1) = (N^2-1)$ quantum numbers.}.  
Therefore, in $d=2$ where all ${\cal P}$  plaquette loops are fundamental and mutually independent, there are 
%${\cal P}$ tadpole states, each characterized by $(N^2-1)$ loop quantum numbers leading to 
$(N^2-1)~{\cal P}$ loop quantum numbers. Subtracting out global $(N^2-1)$  degrees of freedom (or gauge transformations at the origin),  we again see that there are total $(N^2-1)({\cal P}-1)$
% = (N^2-1)({\cal L}-{\cal N}) = {\mathbb N}^{d=2}_{SU(N)}$ 
gauge invariant  SU(N) loop quantum numbers. This exactly matches with  ${\mathbb N}^{d=2}_{SU(N)}$   
in (\ref{dqs}) as $({\cal P}-1) = ({\cal L}-{\cal N})$ in $d=2$. 
\begin{figure}
\centering
\includegraphics[scale=.3]{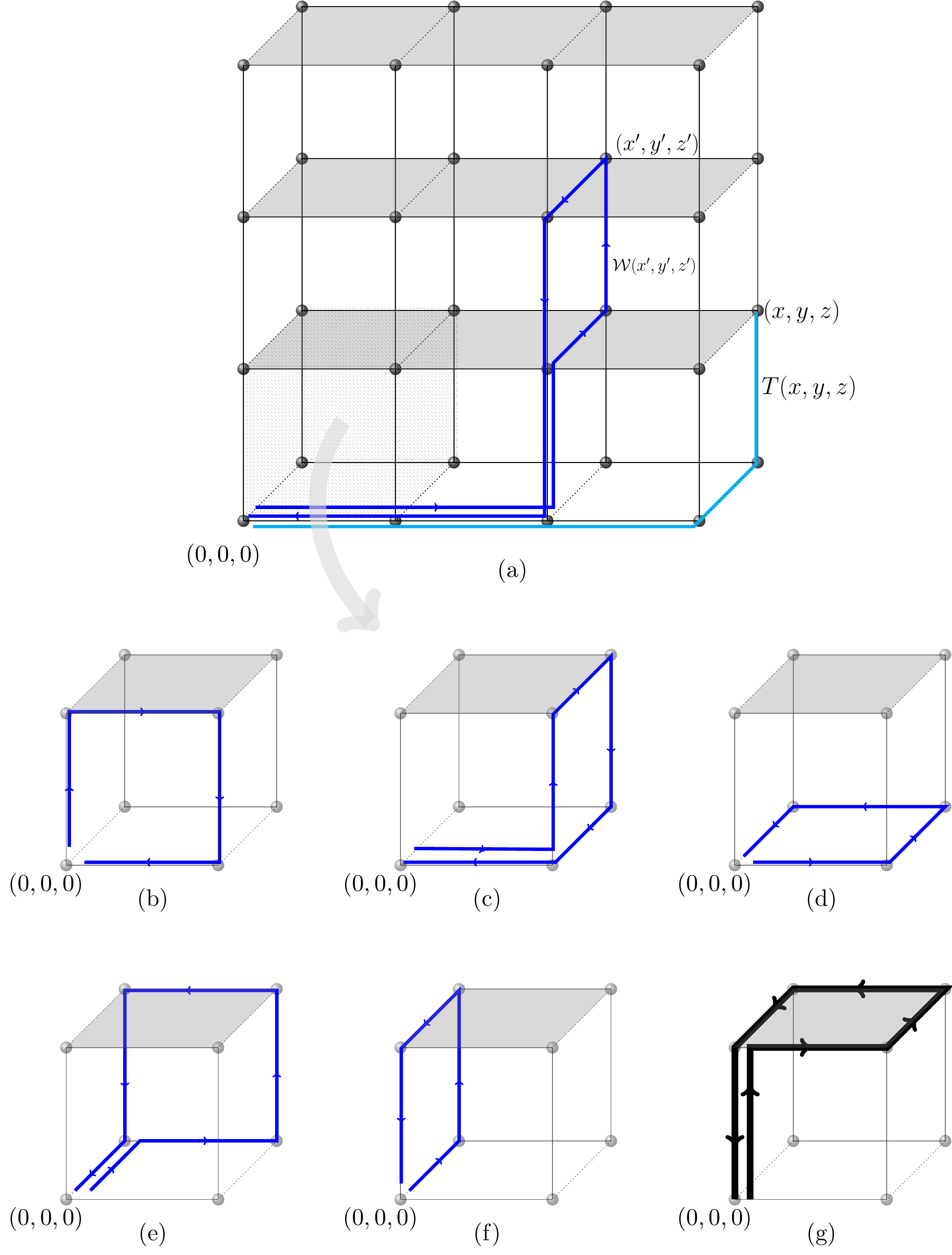}
\caption{(a) Graphical representation of the fundamental plaquette operators obtained by canonical transformations in $d=3$. 
%A string operator $T(n')$ to a site $n'$ is also shown. %There are many unitarily equivalent ways of constructing the plaquette and string operators through canonical transformation along different maximal trees. Our choice of maximal tree to a point $(x,y,z)$ is along the path $(0,0,0)\rightarrow (x,0,0)\rightarrow(x,y,0)\rightarrow(x,y,z)$. 
The shaded horizontal plaquettes  are not obtained by canonical transformations as explained in the text.  They are also not independent: the shaded plaquette  operator in (g) is the product of the fundamental plaquette loop operators in (b),(c),(d),(e),(f) in that order. 
This is just the SU(N) Bianchi identity on lattice. 
%The plaquette loop operators along the shaded region are not obtained through canonical transformations. Therefore, canonical transformations bypasses the issue of Bianchi identities.
}
\label{f:3dls}
\end{figure} 
In 3 dimension we repeat $d=2$ canonical transformations 
on the $z=0$ plane and then extent the string operators 
${\sf T}(x,y,z=0)$ in the z directions to construct plaquette loops in $xz$ and $yz$ planes as shown in Figure \ref{f:3dls}. Thus the canonical transformations already convert all horizontal links on $(xy)$ planes at $z \neq 0$ %(shown as shaded planes in Figure \ref{f:3dls}-a)
%$z=1,2,\cdots {\sf N}$ planes 
in forming plaquette loops in the perpendicular  $(xz)$ and $(yz)$ planes. Therefore, there are no fundamental $xy$ plaquette loops on $z=1,2, \cdots {\sf N}$ surfaces. These  surfaces are shown as shaded planes in Figure \ref{f:3dls}. In fact, the $(xy)$ plaquette loops at 
$z \neq 0$ can be  written in terms of the fundamental 
plaquette loops in $(xz)$ and $(yz)$ planes as shown in Figure \ref{f:3dls}-b,c,d,e,f,g.
This way the canonical transformations also  bypass the problem of SU(N) Bianchi identity constraints  confronted in the loop formulation  of SU(N) lattice gauge theories \cite{bat} in any dimension $d \ge 3$. In $d=3$, we have ${\cal N} = \left({\sf N}+1\right)^3, ~ {\cal L} = 3 {\sf N}\left({\sf N}+1\right)^2$ and ${\cal P}= 3 {\sf N}^2\left({\sf N} +1\right)$.  The total number of 
$(xy)$ plaquettes is ${\cal P}_{xy} \equiv \frac{\cal P}{3} = {\sf N}^2(\sf N+1)$. The number of $(xy)$ plaquette at $z=0$ plane is ${\cal P}_{xy}(z=0)
= \frac{{\cal P}_{xy}}{\sf N+1} = \sf N^2$. Therefore,  
the number of dependent $(xy)$ plaquettes 
${\cal P}_{xy}(z\neq 0) = {\cal P}_{xy} - {\cal P}_{(xy)}(z= 0) ={\sf N}^3 \equiv $ the number of Bianchi identities.  Hence the number of independent SU(N) loop quantum numbers after subtracting $(N^2-1)$ gauge degrees of freedom at the origin $= \left(N^2-1\right) ~ \left({\cal P} - {\cal P}_{xy}(z\neq 0) -1\right) =  \left(N^2-1\right) ~ \left({\cal L} - {\cal N}\right) = {\mathbb N}^{d=3}_{SU(N)} $. 
%The horizontal plaquettes lead to do not lead to independent loops transformations do not convert the $(x,y)$ plaquettes at above $z=0$. 
%  iteratively converting links into strings and finally  independent links not all plaquette loop operators are mutually independent. the canonical transformations lead to ${\cal P}=(2{\sf N}+3){\sf N}^2$  independent SU(N) plaquette loop flux operators as shown in figure \ref{f:3dls}-a. The shaded plaquettes in the Figure \ref{f:3dls}-a are not fundamental. Therefore, in $d=3$ the total number of angular momentum quantum numbers appearing in the loop basis in ${\cal H}^p$ (like (\ref{entlatt}) in 
%$d=2$)  
%is $(N^2-1)~({\cal P}-1)$. We have  subtracted $(N^2-1)$ degrees of freedom associated with the Gauss law at the origin. On the other hand, ${\mathbb N}_{SU(N)} =\left[\frac{\otimes_{links}SU(N)}{\otimes_{sites}SU(N)}\right] = (N^2-1) ({\cal L} -{\cal N})$. As  the number of links:~${\cal L} = 3{\sf N}({\sf N}+1)^2$ and the number of sites: 
%${\cal N} = ({\sf N}+1)^3$ in $d=3$, we get 
%$${\mathbb N}_{SU(N)} = (N^2-1)({\cal P}-1)$$
%as expecected. 
%The fusion of links into loops removes all spurious gauge degrees of freedom associated with strings without  introducing  any redundant loops  degrees of freedom. 
This is again an expected result  because the canonical transformations used for converting links into (physical) loops \& (unphysical) strings  can not introduce any spurious degrees of freedom in any dimension. 
%For the 
%same reason, we also bypass all  Bianchi identity constraints \cite{bat} in any dimension. 
Therefore, the SU(N) plaquette loop operators are mutually independent and  contain complete physical information. The 
corresponding SU(N) coupled tadpole basis is  orthonormal as well as complete in ${\cal H}^p$ bypassing
\footnote{To the best of author's knowledge, solving Mandelstam constraints in $SU(N)|_{N\ge 3}$ lattice gauge theories 
is an open problem. The degree of difficulty and the number of Mandelstam constraints  increases  with increasing N \cite{migdal}. 
In  $SU(2)$ case, the solutions are the 
spin networks discussed earlier.}
 all  non-trivial and notorious 
SU(N) Mandelstam or  Bianchi identity constraints 
which have been extensively discussed in the past \cite{migdal,bishop,loll,rest5,manuplb}. 
% Their explicit construction analogous to hydrogen atom states in (\ref{jmm}) now involves 
%irreducible SU(N) prepotentials \cite{mmra} which take care of the multiplicity problems of SU(N) represntations for $N\ge3$. The coupling of these SU(N) states within a plaquette leading to  
%SU(N) tadpole diagrams will involve SU(N) Clebsch-Gordan coefficients. These issues of technical nature will be 
%discussed elsewhere.  
% % % % % % % % % % % %HHHHHHHHHHHHHHHHHHHHHHHHHHHHHHHHHH
%\begin{figure}
%\centering
%\includegraphics[scale=.4]{BI.pdf}
%\caption{The graphical representation of fundamental plaquette loop operators obtained by canonical transformations on a 3 dimensional lattice. Plaquette operators along the shaded planes are not independent and are not obtained through the canonical transformations.Therefore, canonical transformations bypasses the issue of Bianchi identities. }
%\label{f:biden}
%\end{figure} 
 
\subsubsection{\bf Dynamical symmetry group SO(4,2) of hydrogen atom}
\label{sso42}

Having constructed  the gauge invariant  loop basis in terms of the new plaquette loop operators or in terms of hydrogen atom states  
%in (\ref{jmm}), (\ref{nlmm}) and (\ref{entlatt}) 
in the  previous sections, we now discuss the  structure of a general gauge invariant operator in ${\cal H}^p$.  
We again illustrate these structures using SU(2) gauge group. 
In the simplest single plaquette case, we have already seen that the basic $SU(2)$ gauge  invariant  operators are 
$\left[k_0, ~k_{\pm}\right]$. 
They
(a) are  invariant under U(1) gauge transformations (\ref{abgip}),~
(b) form $SU(1,1)$ algebra and  
(c) generate  transitions $\ket{\sf n} \rightarrow \ket{\bar {\sf n}}$ within  the hydrogen atom basis
(\ref{lb}) in ${\cal H}^p$. 
We now generalize these three  results to the entire lattice in this section. 
%We show that any gauge invariant 
%operator is built out of 15 SO(4,2) generators associated with every plaquette. 
%Under SU(2) gauge transformation at the origin, they all transform together as tensors. 
We note that all  $4{\cal P}$ loop prepotential operators $\left(a_\alpha^\dagger(p),~a_\beta(p)\right)$ and $\left(b^\dagger_\alpha(p),~b_\beta(p)\right)$ of the theory transform 
 as matter doublets under SU(2) gauge transformations 
 (\ref{sbgtop}).  Therefore, the basic 
SU(2) tensor operators which are also invariant under U(1) gauge transformations (\ref{abgip}) can be classified 
%\footnote{Like single plaquette case, the remaining quadratic operators:  
%$[a^\dagger_\alpha(p)a^\dagger_\beta(p), b^\dagger_\alpha(p)b^\dagger_\beta(p),  %a_\alpha(p)a_\beta(p),~b_\alpha(p)b_\beta(p),
%~a^\dagger_\alpha(p) b_\beta(p) %a_\alpha(p)b^\dagger_\beta(p)
%]$ and their hermitian conjugates are not invariant under U(1) gauge transformations (\ref{abgip}) and therefore ignored.} 
into the following four classes:
\begin{align} 
\hspace*{-.3cm}\left[a^\dagger_\alpha(p)b^\dagger_\beta(p);~ a_\alpha(p) b_\beta(p); ~a^\dagger_\alpha(p)a_\beta(p); 
 ~b^\dagger_\alpha(p) b_\beta(p)
\right].
 % ~~~ p=1,2,\cdots,{\cal P}.
\label{set}
\end{align}
%\begin{widetext}
%\begin{center}
%\begingroup
%\squeezetable 
%\squeezetable 
\begin{table*}[] 
\begin{center}
\begin{tabular}{ |c|c|c|c| } 
\hline 
 ${\sf L}_{ab} = \epsilon_{abc}\left({\cal E}_-^c + {\cal E}_+^c\right)$  
&   $L_{45} = -\frac{i}{2}\big(k_+ - k_-\big)$       &
 ${\sf L}_{a5} = \frac{1}{2} Tr{\sigma_a \left({\cal W}^{(+)} - {\cal W}^{(-)} \right)}$ 
& 
\\ 
&  
&  &
${\sf L}_{56} = k_0$ % \frac{(N_a+N_b+2)}{2}}$ 
 \\ 
${\sf L}_{a4} =   {\left({\cal E}_-^a - {\cal E}_+^a\right)}$ 
&     $L_{46}= \frac{1}{2} \big(k_++k_-\big)$     &
${\sf L}_{a6} =  \frac{i}{2} Tr{ \sigma_a \left({\cal W}^{(+)} + {\cal W}^{(-)}\right)}$
& 
\\ 
 \hline
\end{tabular}
\end{center}
\caption{All possible  (15)  SU(2) covariant operators on a plaquette which are also U(1) gauge invariant. They form  SO(4,2) algebra. We have defined (\ref{holop}) ~${\cal W}^{(+)}_{\alpha\beta} \equiv -\tilde{a}^\dagger_{\alpha}b^\dagger_\beta$ and ${\cal W}^{(-)}_{\alpha\beta} \equiv a_{\alpha}\tilde b_\beta$.
} 
\label{tso42} 
\end{table*} 
%\endgroup 
%\end{center}
%\end{widetext}
%\endgroup
These are 16  SU(2) gauge covariant and U(1) gauge invariant  operators on every plaquette of the lattice. 
 The magnitude of the left and the right electric fields 
on every plaquette being  equal (\ref{abcons}), the number operators on each plaquette satisfy $a^\dagger(p)\cdot a(p) = b^\dagger(p)\cdot b(p)= \hat N(p)$. Thus their number reduces to 15. These 15 operators on every plaquette, arranged as in Table {\ref{tso42}},  form SO(4,2) algebra 
%(see \cite{wyb,gilmore} in the context of a hydrogen atom)
: 
\begin{eqnarray}
\left[L_{AB},L_{CD}\right]=-i~\Big(g_{AC}~L_{BD}~+~g_{AD}~
L_{CB} \nonumber \\ 
+~g_{BC}~L_{DA} ~+~g_{BD}~L_{AC}~\Big). 
\label{so42} 
\end{eqnarray}
 Above, $A,B =1,\cdots,6$ and $g_{AB}$ is the metric $(----++)$.  The  algebra (\ref{so42}) can be explicitly checked using the prepotential representations of ${\cal E}^a_\mp$ and 
${\cal W}_\mp$ in  (\ref{su2sb}) and (\ref{holo}) respectively. Note that the fundamental loop quantization relations (\ref{ccr12}) are also 
contained in (\ref{so42}). 

In fact, the emergence of SO(4,2) group in SU(2) loop dynamics in the present loop formulation 
is again an expected result. This can be seen as follows. Let  $\ket{\psi}$ be a physical  state and $\hat {\cal O}$ be any gauge invariant operator. Then the state $\ket{\psi'} \equiv \hat {\cal O} \ket{\psi}$ is also a physical state. As $\ket{\psi}, \ket{\psi'} \in {\cal H}^p$, both  can be expanded in the ``hydrogen atom loop basis". We, therefore,  conclude that 
%any SU(2) 
any gauge invariant operator $\hat {\cal O}$ will 
generate a transition: 
$$\ket{{\sf n}~l~m} ~~\overset{\hat {\cal O}}{\longrightarrow} ~~
 \sum_{\bar {\sf n},\bar l, \bar m}~ O_{{\sf n}~l~m}^{{}^{\bar {\sf n}~\bar l~\bar m}} ~\ket{\bar {\sf n}~ \bar l~ \bar m}.$$ 
 Above, $O_{{\sf n}~l~m}^{{}^{\bar {\sf n}~\bar l\bar~ m}}$ are some coefficients depending on the operator $\hat {\cal O}$. On the other hand, any transition  $\ket{{\sf n} ~l~m} \rightarrow \ket{\bar {\sf n} ~\bar l ~\bar m}$  can be  generated by SO(4,2) generators. This is  a very old and well known result in the hydrogen atom literature \cite{wyb}. 
Therefore, all gauge invariant operators (including the Hamiltonian in the next section) are SU(2) invariant combinations of these SU(2) covariant and U(1) invariant SO(4,2) generators on different plaquettes of the lattice. These results can also be appropriately generalized to higher SU(N) group by replacing SU(2) prepotential operators by SU(N) irreducible prepotential operators discussed in \cite{rmi2}.

\section{\bf SU(N) Loop Dynamics} 
\label{sldyn}

In this section we discuss dynamical issues associated with  the SU(N) Kogut-Susskind Hamiltonian after rewriting it in terms of the new fundamental plaquette loop operators.  We show that in terms of these plaquette loop operators  the initial SU(N) local gauge invariance reduces to global SU(N) invariance and the loop Hamiltonian has a simple weak coupling $g^2\rightarrow 0$ continuum limit. 
The Kogut Susskind Hamiltonian \cite{kogut} is: 
\begin{align}
H = g^2\sum_{l}  \vec E^{2}_l + \frac{K}{g^2} 
%H = g^2\sum_{(x,y) \in \Lambda} \sum_{i=1,2} E^2(x,y; \hat i) \frac{K}{g^2}
\sum_{p} \left(2N - {\textrm Tr} \left(~U_{p}+ U_p^\dagger\right)\right). 
%H  = g^2\sum_{I=1}^{4} \vec E^2(I) + \frac{K}{g^2} \Big(2 - %Tr (U(1)U(2)U(3)U(4)\Big).  \nonumber   
\label{ks} 
\end{align}
In (\ref{ks}) K is a constant, $l \equiv (x,y; \hat i)$ 
denotes a link in $\hat i$ direction, $p$ denotes a plaquette. The plaquette operator: $U_p(x,y)=U(x,y;\hat{1})~U(x+1,y;\hat{2})~
U^\dagger(x+1,y+1;\hat{1})~U^\dagger(x,y+1;\hat{2})$
% and 
%$Tr~ U_p \equiv Tr~\left(U_1U_2U_3^\dagger U_4^\dagger\right)$
defines  the  magnetic field term on  a plaquette  $p$. As mentioned earlier, we choose space dimension $d=2$.
 Substituting the Kogut Susskind electric fields 
in terms of the loop electric fields given in (\ref{kstowele}), 
we get: 
%The Kogut Susskind Hamiltonian in terms of  SU(N) loop 
%operators  is:
\begin{eqnarray}
H  & = &\sum_{x,y \in \Lambda} \Bigg\{g^2\Big[\vec{\cal E}_-(x,y) + \vec{\cal E}_+(x,y-1) + {\Delta}_{X}(x,y)\Big]^2   \nonumber  \\ 
&+& g^2 \Big[\vec{\cal E}_+(x,y) +  R\left({\cal W}_{xy}\right) \vec{\cal E}_-(x-1,y)  +  \Delta_Y(x,y)\Big]^2   
\nonumber \\ 
%&+& 
& + & \frac{K}{g^2} \Big[2N - {\textrm Tr}\Big({\cal W}(x,y)+ h.c  
%{\cal W}^\dagger (x,y)
\Big)\Big]\Bigg\}.  
\label{loophamp}
 \end{eqnarray}
In  (\ref{loophamp}) all operators vanish when x,y are negative or  %$x,y= 0 %{\sf N}+1$ 
 zero as plaquette loop operators are labelled by top right corner (see Figure \ref{fpls}-a).  The operators $\Delta_{X,Y}$ are defined as:
%The electric field, flux operators %${W}_{\mp}(m,n),{W}(m,n)$ are located at (m,n) as shown in Figure 2-
\begin{align}
{\Delta}^a_{X}(x,y) & \equiv  \delta_{y,0} \sum\limits_{\bar x=x+1}^{\sf N}\sum\limits_{\bar y=1}^{\sf N} ~{\sf L}(\bar x, \bar y), \nonumber \\ 
{\Delta}^a_{Y}(x,y)   & \equiv ~~~~ \sum\limits_{\bar y=y+1}^{\sf N} {\sf L}^a(x,\bar y)
%W_-^b(n,p)+W_+^b(n,p),~ R({W}) \equiv  \prod_{l=0}^{m-1}R_{ab}(W(n,l))W_-^b(n-1,m) \nonumber 
\label{nlt} 
\end{align} 
%In (\ref{loophamp}) 
We have also used the  relations: $Tr ~U_p(x,y) = Tr\left({\sf T}^\dagger(x,y) ~{\cal W}(x+1,y+1) 
~{\sf T}(x,y)\right) = Tr ~{\cal W}(x+1,y+1)$.
The Hamiltonian ({\ref{loophamp}) describes gauge invariant dynamics directly in terms of the bare essential,  fundamental plaquette loop creation and annihilation operators without any gauge fields.  
As expected, the unphysical strings do not appear in the 
loop dynamics.
There are many interesting and novel features of the Kogut-Susskind Hamiltonian (\ref{ks}) rewritten in terms of loop operators (\ref{loopham}):
\begin{itemize}
\item 
There are no local SU(N) gauge degrees of freedom and at the same time there are no redundant loop operators. The 
$(N^2-1)$ residual  SU(N) gauge degrees of freedom in (\ref{wgtao}) can be removed by working in the coupled hydrogen atom basis (\ref{entlatt}).   %There are no  redundant loop operators in(\ref{loophamp}). % thus bypassing all non-trivial SU(N) Mandelstam constraints. 
\item In going from links to loops ((\ref{ks}) to (\ref{loophamp})), 
all interactions have shifted from the magnetic field part to  the electric field part.  Therefore, the interaction strength now is $g^2$ and not $\frac{1}{g^2}$. Therefore, the loop Hamiltonian (\ref{loophamp}) 
can be used to develop a weak coupling 
%($g^2\rightarrow 0$) 
gauge invariant loop perturbation theory near the continuum limit. 
\item  The magnetic field term, dominating in the weak coupling continuum ($g^2 \rightarrow 0$) 
limit, acquires its simplest possible form.
It  creates and annihilates single electric plaquette flux loops exactly like in the  single plaquette case (\ref{sf}):
$Tr{\cal W} \sim (k_+ + k_-)$ with $k_{\pm} \in SU(1,1) \subset SO(4,2)$. 
% \footnote{Defining $SU(1,1)$ 
 %(\subset SO(4,2))$ 
% operators $k_{+}(k_-) \equiv a^\dagger\cdot \tilde{b}^\dagger (a\cdot\tilde b)$,   $k_0 \equiv 1/2(N_a+N_b+2)$ with every plaquette loop flux (\ref{holo})
% on the lattice,   $Tr{\cal W} =  (1/\sqrt{k_0})(k_+ +k_-)(1/\sqrt{k_0})$.}
 %$g^2 \rightarrow 0$ 
 %Note that in going from links to loops ((\ref{ks}) to (\ref{loopham})),  
 %Therefore, as opposed to strong coupling  ($g^2 \rightarrow \infty$) expansion with simple $g^2 E^2$ term in (\ref{ks}), the 
%loop formulation (\ref{loopham}) with the simple $1/g^2~ Tr{\cal W}$ term provides an opportune framework 
%to develop gauge invariant weak coupling ($g^2\rightarrow 0$) expansion near the continuum limit\footnote{In strong coupling ($g^2 \rightarrow \infty$) expansion non-interacting terms $g^2E^2_{link}$ are trivially diagonalized and 4 flux interaction terms $1/g^2~ Tr(U_1U_2U^\dagger_3U^\dagger_4)$ are  treated in perturbation. However, one is far away from continuum.}. 
% continuum limit.
%\item For $N=2$, the loop Hamiltonian is constructed out of $SO(4,2)$ generators on every plaquette given in Table 1.
\item  In the hydrogen atom or tadpole  basis  (\ref{entlatt}):
 %{\footnotesize 
 \begin{align}   
 & \langle  {\sf n}^\prime ~ l^\prime ~{m}^\prime |H_B| {\sf n}~l~{m}\rangle  \equiv
 \frac{K}{g^2}\langle {\sf n}^\prime~l^\prime {m}^\prime | Tr {\cal W}| {\sf n}~l~{m}\rangle  \nonumber \\ 
 & ~~~~~~~~= 
 \frac{K}{g^2}\delta_{l,l^\prime}\delta_{m,m^\prime}
 \left[\delta_{{\sf n}',{\sf n}+1} + \delta_{{\sf n}',{\sf n}-1}\right]. 
 %\Bigg\{\begin{array}{ccc}
 %j^\prime & j & \frac{1}{2}\\
 %j & j^\prime &  l\\
 %\end{array} 
 %\Bigg\} 
 \label{s6jj}
\end{align}
%}
In (\ref{s6jj}) ${\sf n}=2j+1$ and ${\sf n}'=2j'+1.$
We have ignored the constant and taken $H_B \equiv Tr {\cal W}$. 
% We have considered  magnetic field term over a single plaquette and its matrix elements 
%in  the corresponding hydrogen atom/tadpole  basis to avoid multiple delta functions over other plaquette loop indices. 
If we put $l=0$ in (\ref{s6jj}), we recover the single plaquette result (\ref{mft}). 
In fact, the matrix elements (\ref{s6jj}) in the hydrogen atom loop basis are valid in arbitrary d dimensions.  This is in a  sharp  contrast  to the magnetic field term  in the standard SU(2) spin network basis
\footnote{{Just for the sake of comparison, we draw 
attention to the same Kogut-Susskind magnetic field term in $d=2$ in the standard SU(2) spin network basis \cite{robson,kolawa,pietri,ani,manuplb}:
\unexpanded{\footnotesize \begin{align}
&~~~~~~~\frac{K}{g^2}\langle \bar{j}_{abcd}|H_B| {j}_{abcd} \rangle ~ = ~\frac{K}{g^2}~\langle \bar{j}_{abcd}|{\textrm Tr}U_{abcd}| {j}_{abcd} \rangle ~= \nonumber\\ &    
 \frac{N_{abcd}}{g^2}\underbrace{ \left[ \begin{array}{ccccccccccccc}
j_1   &    & j_4   &   &  {j}^d_{12}  &   &  j_3  &    &  j_2  &   &  {j}^b_{12}  &   \\
   & {j}^a_{12}    &    & j_3^d  &   & j_2^d  &    & {j}^c_{12}   &   & j_1^b  &   & j_4^b  \\
\bar{j}_1   &    &  \bar{j}_4  &   & \bar{j}^d_{12}  &     & \bar{j}_3  &    & \bar{j}_2   &   &  \bar{j}^b_{12}  &    \\
\end{array} \right]}_{\rm 18j~ coefficient~ of~ the ~ second~ kind} \nonumber\\&
~~~~~~~~~~~~~~\times \prod_{i=1}^{4} 
\left(
\delta_{\bar{j}_i,j_i \pm \frac{1}{2}}. 
+ \delta_{\bar{j}_i,j_i - \frac{1}{2}}\right) \nonumber
%\label{18j} 
\end{align}}} 
\noindent The angular momentum quantum numbers ($j_1,j_2,j_3,\cdots$ $,j_{12}^a,j_{12}^b\cdots )$ 
%in (\ref{18j}) 
above are analogues of hydrogen atom quantum numbers $(n,l,m)$ in (\ref{s6jj}) and specify the spin network loop states 
$\ket{j_{abcd}}$ in Kogut-Susskind formulation.  The details can be found in \cite{manuplb}. However,
the comparison of (\ref{s6jj}) and the above 18j symbol makes it amply clear that hydrogen atom loop basis is much simpler than the  spin network basis 
for any practical calculation especially in the weak coupling $(g^2\rightarrow 0)$ continuum limit.} 
leading to 
(18-j) Wigner coefficients in $d=2$ and (30-j) Wigner coefficient in $d=3$ \cite{manuplb}. 
%transformations of all loop symmetry: 
%electric, flux operators: 
%\begin{eqnarray}
%{\cal E}_{\pm}(m,n) & \rightarrow & \Lambda_0 {\cal %E}_{\pm}(m,n) \Lambda^\dagger_0, \nonumber \\
% {\cal W}(m,n) & \rightarrow & \Lambda_0 {\cal W}(m,n) \Lambda^\dagger_0.
%\end{eqnarray}
%This global invariance is fixed by the Gauss law (\ref{e:globgauss}) at the origin.
\item The non local terms in the Hamiltonian, ${\Delta}^a_{X}(x,y), ~{\Delta}^a_{Y}(x,y)$ and $R({\cal W}_{xy})$ get tamed in the weak coupling limit.
In this $g^2\rightarrow 0$ limit, the relations (\ref{elerrel}) imply: 
 $${\sf L}^a(x,y)= {\cal E}^a_-(x,y)+{\cal E}^a_+(x,y) \sim 0.$$   
 Therefore, ${\Delta}^a_{X}(x,y) \sim 0,~{\Delta}^a_Y(x,y) \sim 0$. Further,  $R_{ab}({\cal W}_{xy}) \sim \delta_{ab}.$ The Hamiltonian, in this weak coupling limit, takes a simple form: 
 {\footnotesize
 \begin{align}
 \label{weakh}
 & ~~~~~~H =  \sum\limits_{p=1}^{\cal P} \left\{4g^2\vec{\cal E}^{~2}(p)+ \frac{1}{g^2} Tr \left({\cal W}(p)+h.c\right)\right\} ~ +  \\
& ~~~~ {g^2} \sum_{<pp'>}\left\{\vec{\cal E}_-(p)\cdot \vec{\cal E}_+(p')  +  \vec{\cal E}_+(p)\cdot \vec{\cal E}_-(p')  
 %\vec{\cal E}_+(p)\cdot \vec{\cal E}_-(p') 
 \right\} 
 + g^3 \delta H. \nonumber 
 %\sum_{r=0}^{\infty} g^r {H}_r.
 \end{align}}
%%{\footnotesize
%\begin{align}
%\label{weakh}
%H &=  \sum\limits_{p=1}^{\cal P} \left\{4g^2\vec{\cal E}^{~2}(p)-
%\frac{1}{g^2} Tr \left({\cal W}(p)+h.c\right)\right\}+
%{g^2} \sum_{<pp'>}\left\{\vec{\cal E}_-(p)\cdot \vec{\cal E}_+(p')  +  \vec{\cal E}_+(p)\cdot \vec{\cal E}_-(p')  
%%\vec{\cal E}_+(p)\cdot \vec{\cal E}_-(p') 
%\right\}  \nonumber \\ 
%&+
%g^3 \delta H.
%%\sum_{r=0}^{\infty} g^r {H}_r.
%\end{align}
%%}
\noindent Above   $\sum_{<pp'>}$ denotes summation over nearest neighbour plaquette loop electric fields. 
%These terms  describe nearest neighbour plaquette loop-plaquette loop interactions. 
The non-localities occur in the higher order terms in the coupling. %These terms are denoted by $\delta H$ and  are at least of the order $g^3$. %Larger non-localities come with higher and higher powers of the coupling g. 
Therefore, these terms, collectively denoted by $g^3~\delta H$ in (\ref{weakh}), can be ignored
in the weak coupling limit as a  first approximation.  
%  we focus on the  nearest neighbour loop -loop interacton Hamiltonian ${\mathbb H}$. 
The SU(N) gauge theory Hamiltonian in the loop  picture now reduces to SU(N) spin model Hamiltonian with nearest neighbour interactions. This simple spin Hamiltonian has the same global SU(N) symmetry, dynamical variables as the Hamiltonian in (\ref{loophamp})  or (\ref{weakh}). 
In fact,  this  is an interesting model in its own right  to explore  confinement and the spectrum in the weak coupling continuum limit.
Note that the elementary  but dominant   $1/g^2$ magnetic field terms (see (\ref{s6jj})) are left untouched by this approximation. They need to be treated exactly in the $g^2\rightarrow 0$ limit 
and  should be part of unperturbed Hamiltonian along with contributions from the electric field terms.  
As an example, in the simplest case of single plaquette 
SU(2) lattice gauge theory, the dominant magnetic 
field term can be easily diagonalized using SU(2) characters \cite{robson,kolawa,pietri} (also see Appendix C). However, it has continuous spectrum 
(\ref{mev}). Therefore the magnetic field term alone can not be used as unpertubed Hamiltonian even in the weak coupling $(g^2\rightarrow 0)$ limit. One has 
to include contributions from $(g^2)$ electric field terms in the unperturbed Hamiltonian. 
These issues are currently under investigation  and will be addressed elsewhere. 
\end{itemize}

\subsection{\bf The Schr\"odinger equation in hydrogen atom loop basis} 
\label{seqn} 
In this section we explore the ground  state $\ket{\Psi_0}$ and the first excited state $\ket{\Psi_1}$ of SU(2) lattice gauge theory 
in terms of the SO(4,2) fundamental plaquette loop operators discussed in section \ref{sso42} and 
given in Table \ref{tso42}. 

\subsubsection{\bf A variational ansatz}
\label{va} 
 An easy, intuitive and old approach is the variational or coupled cluster method \cite{greensite}.  The simple ansatzes are:
\begin{eqnarray} 
\ket{\Psi_0} & = & e^{\Gamma} \ket{0}, ~~~~~~~~\bra{\Psi_0}\Psi_0\rangle =1, \nonumber \\
\ket{\Psi_1} & = & \Sigma^+ ~\ket{\Psi_0},  ~~~~\bra{\Psi_0}\Psi_1\rangle =0. 
\label{vargs} 
\end{eqnarray} 
%\begin{figure}
%\centering
%\label{lgroundstate}
%\includegraphics[scale=.5]{lgroundstate.eps}
%\caption{A typical ground state configuration.}
%\end{figure}
In (\ref{vargs}) $\Gamma$ and $\Sigma$ are the $SU(2) \otimes U(1)$  gauge invariant operators  constructed out of SO(4,2) generators in the Table 1.  
It is convenient to write  $\Gamma= \Gamma^+-\Gamma^-$
where ~$\Gamma^- \equiv \left(\Gamma^+\right)^\dagger$
 %$\Sigma=\Sigma^+-\Sigma^-$ 
and  $\Gamma^+,~ \Sigma^+$ have the structures:
\begin{eqnarray} 
\hspace{-0.2cm}\Gamma^+\hspace{-0.2cm} &\equiv & G_1 ~\sum_{p=1}^{\cal P} k_+(p) + \hspace{-0.2cm} \sum_{p_1,p_2=1}^{\cal P}\hspace{-0.2cm} G_2(|p_1-p_2|)  \vec{k}_+(p_1)\cdot\vec{k}_+(p_2) \nonumber \\ 
& +& ~~~~~\cdots \cdots,  \nonumber \\
\Sigma^+  \hspace{-0.2cm} & \equiv & F_1 \sum_{p=1}^{\cal P} k_+(p) + \hspace{-0.2cm} \sum_{p_1,p_2=1}^{\cal P} F_2(|p_1-p_2|) \vec{k}_+(p_1)\cdot\vec{k}_+(p_2) \nonumber \\
 %+\sum_{p,p^\prime,p^{\prime\prime}=1}^{\cal P} 
& +& ~~~~~\cdots\cdots.
\label{cropex}
\end{eqnarray} 
In the first term above  $k_+(p)$ is  the gauge invariant  $SU(1,1) \in SO(4,2)$ plaquette loop creation operator. 
In the second term, we have defined  SU(2) adjoint loop flux creation operator $\vec k_+(p)$ on every plaquette $p$ using SO(4,2) generators in Table 1: 
$$k^a_+(p) \equiv L_{a5}(p)- i L_{a6}(p) = Tr\left(\sigma^a {\cal W}^{(+)}(p)\right),~ a=1,2,3.$$
% ie, k^
%Above, we define $\sigma^0=\mathcal{I}$. Therefore, $k^0_+(p)=k_+(p)$.
%We have also used the fact that any Wilson loop operator can be constructed out of the fundamental plaquette loop operaors. 
\begin{figure}
\centering
\includegraphics[scale=.3]{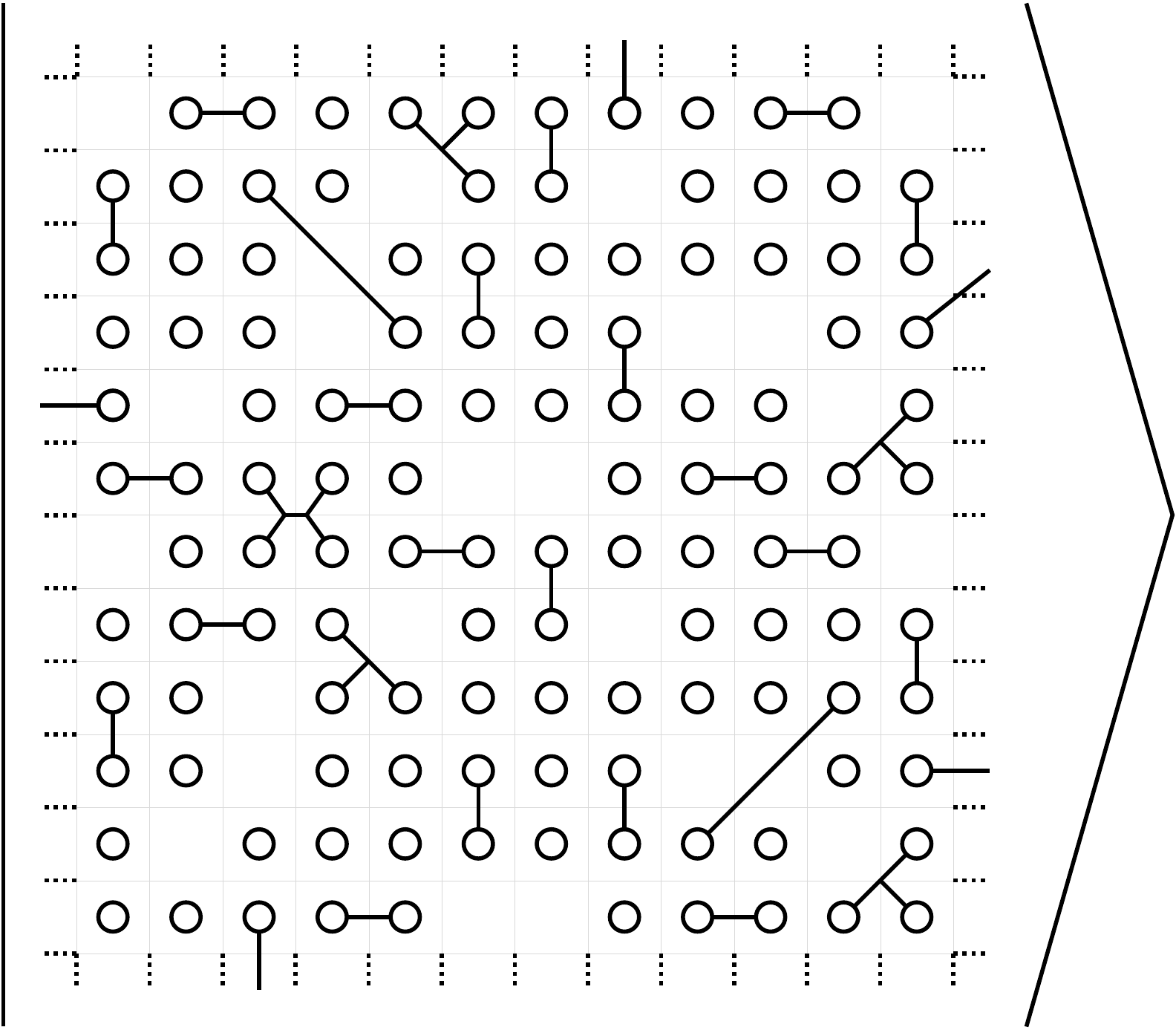}
\caption{The SU(2) ground state picture  in the hydrogen atom basis (\ref{entlatt}).}
\label{gtgs}
\end{figure}
Note that the expansion (\ref{cropex}) is in terms of number of fundamental loops and not in terms of coupling constant.  In fact,  $g^2$ dependence of the 
structure functions $G_1,~G_2,\cdots $ and $F_1,~F_2,\cdots $ have been completely suppressed. 
The physical interpretations of (\ref{vargs}) and (\ref{cropex}) are extremely simple. The operator $e^\Gamma$  acting on the strong coupling vacuum in (\ref{vargs})  creates loops of all 
shapes and sizes in terms of the fundamental loop operators to produce the ground state $\ket{\Psi_0}$. 
%All these  loops of different sizes and shapes are created by fusing ${\cal P}$ fundamental SO(4,2) plaquette loop operators given in Table 1 with structure functions $G_1,~G_2,\cdots$. 
The first term $k_+(p)$ in (\ref{cropex}) 
%is $SU(1,1) \in SO(4,2)$  creation operator.  It 
creates hydrogen atom s-states on plaquette $p$ 
or simple one plaquette loops. These are shown as small circles (tadpoles without legs) in Figure \ref{gtgs}. 
The second term describes  doublets  of 
hydrogen atoms  with vanishing 
total angular momentum. These are shown as two tadpoles joined together  in Figure \ref{gtgs}. The three hydrogen atom or three tadpole states over three plaquettes ($p_1,p_2,p_3$) can be created by including  a term  of the form $\left(\vec k_+(p_1)\times \vec k_+(p_2)\right) \cdot \vec k_+(p_3)$ in $\Gamma^+$ and so on 
and so forth. As shown in Figure \ref{gtgs}, the ground state is a soup of all such coupled tadpoles or coupled hydrogen atom clusters,  each with vanishing angular momentum. 
%The loop creation operator  $\Gamma$ itself is written in terms of the ${\cal P}$ fundamental plaquette loop creation operators $\in SOte (4,2)$. 
  The first excited state in (\ref{cropex}) is obtained by exciting loops in this ground state by a creation operator $\Sigma^+$.    
  %The correlation functions $G_1, G_2, \cdots $ and $F_1,~F_2, \cdots $ are fixed by the Schr\"odinger equation.
The sizes of the ``hydrogen atom clusters" and their importance depend on the structure functions $G$ and $F$ which in turn  are fixed by the loop Schr\"odinger equation with Hamiltonian (\ref{weakh}). 
 These qualitative features can be made more precise by 
putting the ansatz (\ref{vargs}) in (\ref{weakh}). 
 The resulting  Schr\"odinger equation can be analyzed for the 
structure constants\footnote{A reasonable assumption is $G_1 >> G_2>> \cdot\cdot, ~F_1>> F_2>> \cdot\cdot $ in 
(\ref{cropex}). In this case 
the matrix elements of $\delta H$  in (\ref{weakh}) are small in the states in (\ref{vargs}): $\bra{\Psi_0} \delta H \ket{\Psi_0} \approx 0$ and $\bra{\Psi_1} \delta H \ket{\Psi_1}  \approx 0$ as $\left[{\sf L}^a(p),k_\mp(p)\right]=0$ and ${\sf L}^a\ket{0}=0.$} 
 $(G_1,~G_2,\cdots )$ and $(F_1,~F_2,\cdots )$ in the complete, orthonormal  hydrogen atom loop basis (\ref{entlatt}) using its dynamical symmetry group  SO(4,2) algebra  in (\ref{so42}). 
 We postpone quantitative analysis in this direction   to a later publication.   

\subsubsection{\bf A tensor networks ansatz}
\label{smps}
 
The present loop formulation is  tailor-made for  tensor network \cite{tn6} and matrix product state (MPS) ansatzes to explore the interesting and physically relevant part of ${\cal H}^p$ for low energy states. This is 
due to the following two reasons:
\begin{itemize} 
\item  The absence of non-abelian Gauss law constraints at every lattice site. 
\item  The  presence of (spin type) local hydrogen atom orthonormal basis  on  every plaquette.
\end{itemize}
We first briefly discuss matrix product state approach 
in  a simple example of spin chain with spin $s=1$ before directly generalizing it to pure SU(2) lattice gauge theory on a one dimensional chain of plaquettes.  
In the case of spin chain with $s_x =-1, 0, +1$  at every lattice site $x=0,1,\cdots ,N$, any state can be written as: 
\begin{eqnarray} 
\ket{\Psi} = \hspace{-0.4cm}\sum_{s_1,s_2\cdots s_N=0, \pm{1}}
\Psi(s_1,s_2,\cdots s_N) \ket{s_1,s_2,\cdots s_N}.
\label{spinstate}
\end{eqnarray}    
%\begin{enumerate} 
%\item   the absence of local Gauss laws at every lattice %site, 
%\item  the  ``spin type" structures 
%of hydrogen atom orthonormal basis on every plaquette.
%\end{enumerate}  
The  matrix product state method consists of replacing the wave functional by 
\begin{eqnarray} 
\Psi(s_1,s_2,\cdots s_N) = Tr \left(T^{(s_1)}_1T^{(s_2)}_2\cdots T^{(s_N)}_N\right). 
\label{mpsspin}
\end{eqnarray}
In (\ref{mpsspin}) $T^s$ are $D\times D$ matrices where D is the bond length.  
The matrix elements of $T^s$ are fixed by minimizing the spin Hamiltonian. In the hydrogen atoms loop basis 
we have a  similar structure where the three dimensional spin states are replaced by infinite dimensional quantum 
states  of hydrogen atoms: $|s\rangle  \rightarrow |n~l~m\rangle$.  
The most general state in the  hydrogen atom loop basis
can be written as: 
%\begin{eqnarray} 
%\ket{\Psi} = \sum_{\{n\}} ~\sum_{\{l\}}~\sum_{\{ll\}} %\Psi_{\{n\}\{l\}\{ll\}}~~|\{n\}\{l\}\{ll\}\rangle 
%\end{eqnarray}
\begin{eqnarray}
\ket{\Psi} \hspace{-0.1cm} = \hspace{-0.2cm}\sum_{\{n\}\{l\}\{m\}} \hspace{-0.3cm} \Psi{\tiny{\left[ \begin{array}{cccc} n_1 & n_2 &  \cdots n_p \\ 
l_1 & l_2 &  \cdots   ~l_p \\ 
m_{1} & m_{2}& ~\cdots  m_{\cal P} \end{array} \right]}} \hspace{-0.1cm}
\left\vert \begin{array}{cccc} n_1 & n_2 &  \cdots ~n_p ~\\ 
l_1 & l_2 &  \cdots   ~l_p \\ 
m_{1} & m_{2}& ~\cdots  m_{\cal P} \end{array} \right\rangle.
\label{aabbcc}
\end{eqnarray}
%\begin{eqnarray}
%\ket{\Psi}  = \sum \Psi_{\tiny{\left[ \begin{array}{cccc} n_1 & n_2 &  ~~\cdots ~~~n_p ~~\\ 
%l_1 & l_2 &  \cdots   ~~~l_p \\ 
%l_{12} & l_{123}& ~~~~~\cdots  l_{total}=0 \end{array} \right]}} 
%\left\vert \begin{array}{cccc} n_1 & n_2 &  ~~\cdots ~~~n_p ~~\\ 
%l_1 & l_2 &  \cdots   ~~~l_p \\ 
%l_{12} & l_{123}& ~~~~~\cdots  l_{total}=0 \end{array} \right\rangle
%\end{eqnarray}
We now consider  SU(2) lattice gauge theory on a chain of ${\cal P}$ plaquettes as shown in Figure \ref{f:finlatlss}. A  simple  tensor network ansatz, like (\ref{mpsspin} for spins, for the ground state wave function in (\ref{aabbcc}) is
%the %${\cal P}$ hydrogen atoms basis is:  
\begin{eqnarray} 
\Psi_0{\tiny{\left[\begin{array}{cccc} n_1 & n_2 &  ~~\cdots ~~~n_p ~~\\ 
l_1 & l_2 &  \cdots   ~~~l_p \\ 
m_1 & m_2& ~~~~~\cdots  m_{\cal P} \end{array} \right]}} & \equiv &  Tr \Big[T_1^{(n_1l_1m_1)}~T_2^{(n_2l_2m_2)} \cdots \nonumber \\ && ~~~\cdots 
T_{\cal P}^{(n_{\cal P}~l_{\cal P}~m_{\cal P})}\Big]. 
\label{tna}
\end{eqnarray}
In (\ref{tna}) $T_x^{(n_xl_xm_x)}; ~ x=1,2\cdots ,{\cal P}$ are ${\cal P}$
matrices of dimension $D\times D$ where $D $ is the bond 
length describing correlations between hydrogen atoms.
Assuming a bound on the principal quantum number (e.g.,
${\sf n} = 1,2$) and  minimizing the energy of the spin model Hamiltonian within spherically symmetric s-sector should give a good idea of ground state  at least in the strong coupling region. The method can then be extrapolated systematically towards weak coupling by extending the range of hydrogen atom principal quantum number on each plaquette. The global SU(2) Gauss law can also be explicitly implemented through the following ansatz: 
\begin{align}
& \ket{\Psi}  =  \sum_{\{n\}\{l\}\{ll\}} \Psi{\tiny{\left[ \begin{array}{cccc} n_1 & n_2 &  ~~\cdots ~~~n_{\cal P} ~~\\ 
l_1 & l_2 &  \cdots   ~~~l_{\cal P} \\ 
l_{12} & l_{123}& ~~~~~\cdots  l_{12\cdots {\cal P}-2} \end{array} \right]}} ~~~~~~~~~~~~~~~~~~~~\nonumber \\ 
&~~~~~~~~~~~~~~~~~~~~~~~~~~~~
\left\vert \begin{array}{cccc} n_1 & n_2 &  ~~\cdots ~~~n_p ~~\\ 
l_1 & l_2 &  \cdots   ~~~l_p \\ 
l_{1} & l_{12}& ~~~~~\cdots  l_{12\cdots {\cal P}-2} \end{array} \right\rangle.
\label{aabbcc2}
\end{align}
We can now make an explicitly gauge invariant MPS ansatz for the ground state:
\begin{eqnarray} 
\Psi_0{\tiny{\left[ \begin{array}{cccc} n_1 & n_2 &  ~~\cdots ~~~n_p ~~\\ 
l_1 & l_2 &  \cdots   ~~~l_p \\ 
l_{12} & l_{123}& ~~~~~\cdots  l_{12\cdots{\cal P}-2} \end{array} \right]}} \equiv  Tr \bigg[T^{{}^{~n_1}}_{{}_{0,l_1,l_1}}(1)~T^{{}^{~n_2}}_{l_1l_2l_{12}}(2) \nonumber \\
T^{{}^{~n_3}}_{l_{12}l_3l_{123}}(3) \cdots\cdots
T_{l_{\cal P}l_{\cal P}0}^{~~n_{\cal P}}({\cal P})\bigg]. ~~~~
\label{tna2}
\end{eqnarray}
This ansatz is illustrated in Figure \ref{f:finlatlss}-b. Much more work is required to implement these ideas on a computer.  We will discuss these computational issues in a future publication.

\section{Summary and Discussion} 

In this work we have constructed a series of iterative  canonical transformations in pure SU(N) lattice gauge theories to get to a most economical loop formulation without any local 
spurious degrees of freedom. 
%The local SU(N) gauge invariance in  the initial Kogut Susskind link formulation reduces to the global SU(N) invariance in the final loop formulation. 
The  canonical transformations ensure that the total degrees of freedom remain intact at every stage. At the end, as a consequence of SU(N) Gauss laws, 
all local SU(N) gauge degrees of freedom carried by string operators drop out. %We thus obtain  a loop formulation without any extra loops or local gauge degrees of freedom.  In other words, 
The loop operators obtained this way are fundamental and the loop formulation is free of difficult SU(N) Mandelstam as well as Bianchi identity ($d \ge 3$) constraints. The resulting  
SU(N) loop Hamiltonian in two dimension reduces to SU(N) spin Hamiltonian. %It has many interesting features discussed in the text. 
%he SU(N) magnetic field term over a plaquette, important in the weak coupling limit, is simply the sum of  fundamental plaquette loop creation and destruction operators.  
 In the special SU(2) case,  the canonical transformations map the physical loop Hilbert space to the space of Wigner coupled hydrogen atoms 
 and the loop dynamics can  be  completely described in terms of  the generators of the  dynamical symmetry groups SO(4,2) of hydrogen atoms.  Within this loop approach all non-abelian topological effects are contained in the discrete nature of the hydrogen atom energy eigenstates. 
  %We have also proposed 
%a simple nearest neighbour plaquette loop-plaquette loop  interaction model to study confinement and the spectrum near the weak coupling $g^2 \rightarrow 0$ 
%limit.  

We now briefly discuss some new future directions.
The absence of SU(N) Gauss laws should   help us in defining the entanglement entropy in lattice gauge theories. The entanglement entropy of two complimentary regions in a gauge invariant state suffers from the serious  obstacles \cite{eent} created by SU(N) Gauss laws  at the boundary. In the present formulation the two regions can have mutually independent hydrogen atom/tadpole basis which are coupled together  across the boundary through a single flux line at the end. 
The present loop approach may also be interesting 
in the context of cold atom experiments \cite{uca}. 
The  hydrogen atom interpretation of
%physical gauge theory Hilbert space 
${\cal H}^P$ and absence of local gauge invariance 
should bypass the challenging task of imposing non-trivial and exotic non-abelian Gauss law constraints  at every lattice site in 
the laboratory. 

%\vspace{1cm} 

\acknowledgments

\noindent {\it Acknowledgments: We thank  Ramesh Anishetty for useful discussions and comments on 
the manuscript. MM would like to 
acknowledge  H S Sharatchandra 
for introducing him to the canonical transformations 
which led to this work.  
TPS thanks CSIR for financial support.}
\appendix

\section{From links to loops \& strings} %SU(N) canonical transformations on a finite lattice}

In this appendix we generalize the three canonical transformations
(\ref{ct1}), (\ref{ct2}) and (\ref{ct3}) in the single plaquette case to the entire  lattice in two dimension.  
%${\cal N} \equiv (L+1)^2$ lattice sites, ${\cal L} \equiv %2L(L+1)$ links and ${\cal P} \equiv {\sf L}^2$ plaquettes with ${\cal N}+{\cal P}-1 = {\cal L}$. 
%A point on this lattice is denoted by $n =(x,y); ~x,y=0,1,2,\cdots,\sf N$. 
%The link operators 
%and their electric fields in $\hat X$  and $\hat Y$ directions are denoted by $U(x,y; \hat i), ~E_\mp^a(x,y; \hat i); i =1,2$ respectively.  
%The corresponding electric fields are denoted by $E^a_\mp(x,y;\hat 1)$ and $E^a_\mp(x,y; \hat 2)$ respectively.
We define a comb shaped maximal tree with its base along the $X$ axis  and  make a series of canonical transformations along the maximal tree to construct the string operators ${\sf T}_{[xxyy]}(x,y)$ attached to each lattice site $(x,y)$ away from the origin.  This is similar to the construction of 
string operators ${\sf T}_{[xy]}(x,y)$  
% \equiv {\sf T}(1,0), {\sf T}(B) \equiv \sf T(1,1)$ and $\sf T(C)\equiv \sf T(0,1)$ 
attached to the points $A \equiv (1,0),  B \equiv (1,1)$ and $C \equiv (0,1)$  in the simple single plaquette  example illustrated in  Figure \ref{foabc}-a,b,c. The gauge covariant loop operators ${\cal W}(x,y)$ are constructed by fusing the string operators  
%${\sf T}(x,y)$ and ${\sf T}^\dagger(x+1,y)$ 
with the horizontal link operators $U(x,y; \hat 1)$ again through canonical transformations.  As expected, all string operators ${\sf T}_{[xxyy]}(x,y)$ decouple as a consequence of SU(N) Gauss laws ${\cal G}^a(x,y)=0$. Thus only  the fundamental physical loop operators are left at the end.
The iterative canonical transformations are performed in 6 steps.  These 6 steps are also illustrated graphically in Figures 10-15 for the sake of clarity.  
%We are left with  ${\cal P}$ physical  fundamental plaquette loop operators ${\cal W}(x,y)$, like ${\cal W}(1,1)$ in the single plaquette case. 

\subsection{Strings along x axis}
We start by 
defining iterative canonical transformation along the x axis. They transform the  $\sf N$ 
link operators $U(x,0;\hat 1)$ 
%at ${x =0,1,\cdots \sf N}$  
into $\sf N$ string operators  ${{\sf T}}_{[xx]}(x,0)$. These string operators  start at the origin and end at $x=1,2,\cdots {\sf N}$ along the x axis as shown in the Figure \ref{f:can1d}. The canonical transformations 
are defined iteratively as: 
%, denoted by $T_{[xx]}(x,0)$ and 
%These string operators ${ {\sf T}}_{[xx]}(x,0)$    
%start from the origin and end at $(x,0)$. 
%We start with  the 
%initial string operator ${\sf T}_{[x]}(0,0) \equiv {\cal  iteratively define the string operators ${\sf T}_{[x]}(x+1)$ and ${\sf T}_{[xx]}(x,0)$ as: 
%(x,y;\hat 1)$ iteratively as follows:
%{\footnotesize 
\begin{align}
 {\sf T}_{[x]}(x+1,y=0)~ &\equiv~ {\sf T}_{[x]}(x,0) ~~U(x,0; \hat 1), \nonumber\\
{\sf T}_{[xx]}(x,0)  ~&\equiv~  {\sf T}_{[x]}(x,0),  \nonumber \\ 
{\sf E}^a_{[x]+}(x+1,0) ~&= ~E^a_-(x+1,0;\hat 1),\nonumber\\ {\sf E}^a_{[xx]+}(x,0) ~&=~ E^a_-(x,0; \hat 1)+E^a_+(x,0; \hat 1). 
\label{xax1} 
\end{align}
%\label{xax1} }
\begin{figure}
\centering
\includegraphics[scale=0.55]{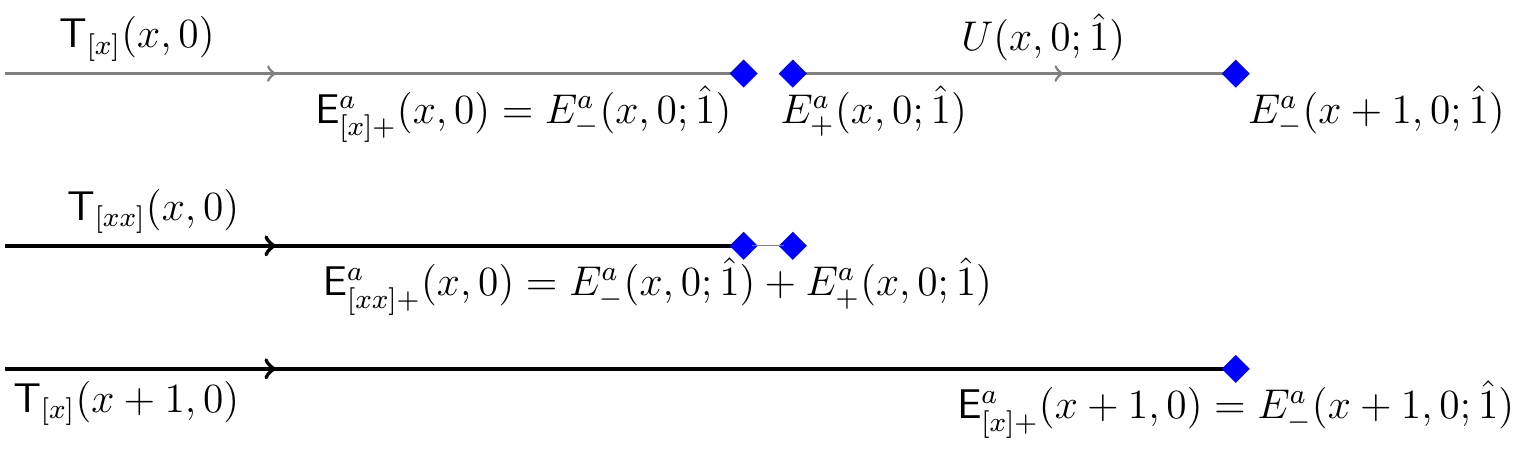}
\caption{Graphical representation of the iterative canonical transformations (\ref{xax1}). The initial 
${\sf T}_{[x]}(x,0)$ and the final ${\sf T}_{[xx]}(x,0)$ string operators at $(x,0)$ are shown. The string operator  ${\sf T}_{[x]}(x+1,0)$ in the third row replaces ${\sf T}_{[x]}(x,0)$ in the first row in the next iterative step. All electric fields involved in 
(\ref{xax1}) are also shown at their locations.} 
\label{f:can1d}
%\caption{Canonical transformations on a 1D lattice. {\sf T}he conjugate electric fields $T_\pm(n)$ corresponding to the string operators $T(n)$ are 0 by the gauss law at $n$ and hence decouples from the physical theory.}
\end{figure}
Above $x=1,\cdots,{\sf N}$ and the starting input for the first equation in (\ref{xax1}) is 
$T_{[x]}(1,0) \equiv U(1,0;\hat 1)$. 
The canonical transformations (\ref{xax1}) iteratively transform the
flux operators $\Big[{\sf T}_{[x]}(x,0),~ U(x,0; \hat 1)\Big]$ and their electric fields into $\Big[{\sf T}_{[xx]}(x,0), ~{\sf T}_{[x]}(x+1,0)\Big]$ and their electric fields as shown in Figure \ref{f:can1d}.     
At the boundary $x = {\sf N}$, we  define 
${\sf T}_{[xx]}({\sf N},0) \equiv {\sf T}_{[x]}(\sf N,0)$ for later convenience.
% $E^a_+(\sf N,0; \hat 1)=0$. 
As is also clear from Figure \ref{f:can1d}, the subscript $[xx]$ on the string flux operator ${\sf T}_{[xx]}(x,0)$ %~(T_{[x]}(x,0))$ 
encodes the structure of its right  electric field  
${\sf E }^a_{[xx]+}(x,0)$  
%~ ({T}^a_{[x]+}(x,0))$ 
in (\ref{xax1}). More explicitly, the last equation in (\ref{xax1}) states that ${\sf E}_{[xx]+}(x,0)$  is  
the sum of 
two  adjacent Kogut Susskind electric fields in x 
direction. 
%For convenience, we define $T_{[xx]}(L+1,0) 
%\equiv {\sf T}(L+1,0)$ at the boundary. 
% and $E^a_+(L,0;\hat 1) \equiv 0$. 
Note that if we were in one dimension with open boundary conditions, the Gauss law   (\ref{su2gln}) would imply ${\cal G}^a(x) \equiv {{\sf T}}^a_{[xx]+}(x,0) = 0; ~\forall x$ making 
all string operators ${\sf T}_{[xx]}(x,0)$ unphysical and irrelevant as expected.
\subsection{Strings along y axis} 

We now iterate the above canonical transformations 
to extend ${\sf T}_{[xx]}(x,0)$ in the y direction to get ${\sf T}_{[y]}(x,y=1)$ and the final unphysical and ignorable  string operators  ${\sf T}_{[xxy]}(x,0)$ along the x axis as illustrated in Figure \ref{f:can2}:
\begin{align}
{\sf T}_{[y]}(x,1) ~& \equiv~  {\sf T}_{[xx]}(x,0) ~U(x,0;\hat{2}), \nonumber\\   
{\sf T}_{[xxy]}(x,0)  ~ &\equiv ~ {\sf T}_{[xx]}(x,0) \nonumber\\
{\sf E }^a_{[y]+}(x,1)~ & =~  E^a_-(x,1; \hat 2), \nonumber\\{\sf E}^a_{[xxy]+}(x,0)~ &=~ {\sf E}^a_{[xx]+}(x,0) + E^a_+(x,0; \hat{2}). 
\label{gln0}  
\end{align} 
\begin{figure}[b]
\centering
\includegraphics[scale=.8]{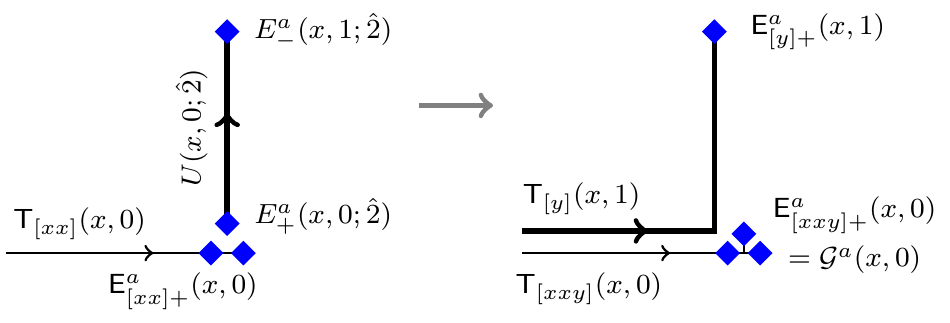}
\caption{Graphical representation of the canonical transformations (\ref{gln0}): vertical string constructions at $y=0$  in  (\ref{gln0}) and the Gauss law (\ref{glx0}) at $y=0$}. 
\label{f:can2}
\end{figure}
In (\ref{gln0}) we have defined ${\sf T}_{[xx]}(0,0) \equiv 1$ and ${\sf T}_{[xx]}({\sf N},0) \equiv {\sf T}_{[x]}({\sf N},0)$ as mentioned above. Substituting ${\sf E}^a_{[xx]+}(x,0)$ from 
(\ref{xax1}), we get:  
\begin{align} 
{\sf E}^a_{[xxy]+}(x,0) 
%=  T^a_{[x]+}(x,0) + E^a_+(x,0; \hat{2})
&= \Big({E^a_-(x,0; \hat{1}) + E^a_+(x,0; \hat{1}) + E^a_+(x,0; \hat{2})}\Big) \nonumber \\ 
&{ \equiv {\cal G}^a(x,0)} = 0.
\label{glx0}
\end{align}  
%  as their conjugate electric fields are the Gauss law generators at $(x,0)$  in (\ref{glx0}). 
Again the subscript $[xxy]$ on the string  operator ${\sf T}_{[xxy]}^a(x,0)$ denotes that its electric field at $(x,0)$ is sum of three Kogut-Susskind electric fields, two in x direction and one in y direction as in (\ref{glx0}) and represented by three squares in Figure \ref{f:can2}. %The transformations (\ref{gln0}) and the Gauss law (\ref{glx0}) are illustrated in Figure \ref{f:can2}-a.
We ignore ${\sf T}_{[xxy]}(x,0)$ from now onwards and repeat the canonical transformations (\ref{xax1}) to fuse the links in y direction along the maximal tree at fixed $x (=0,1,\cdots {\sf N})$. For this purpose, we replace ${\sf T}_{[x]}(x,0)$ 
and $U(x,0; \hat 1)$  in (\ref{xax1})
by ${\sf T}_{[y]}(x,y)$ and  $U(x,y; \hat 2)$ respectively with $y =1,2,\cdots ,({\sf N}-1)$ and  define: 
\begin{align}
{\sf T}_{[y]}(x,y+1)~ &\equiv~ {\sf T}_{[y]}(x,y) ~~U(x,y; \hat 2), \nonumber \\
{\sf T}_{[yy]}(x,y)  ~&\equiv~  {\sf T}_{[y]}(x,y),  \nonumber\\ 
{\sf E}^a_{[y]+}(x,y+1) ~&= ~E^a_-(x,y+1;\hat 2), \nonumber\\
 {\sf E}^a_{[yy]+}(x,y) ~&=~ {{\sf E}}^a_{[y]+}(x,y)+ E^a_+(x,y; \hat 2). 
\label{yay1} 
\end{align}
\begin{figure}
\centering
\includegraphics[scale=.8]{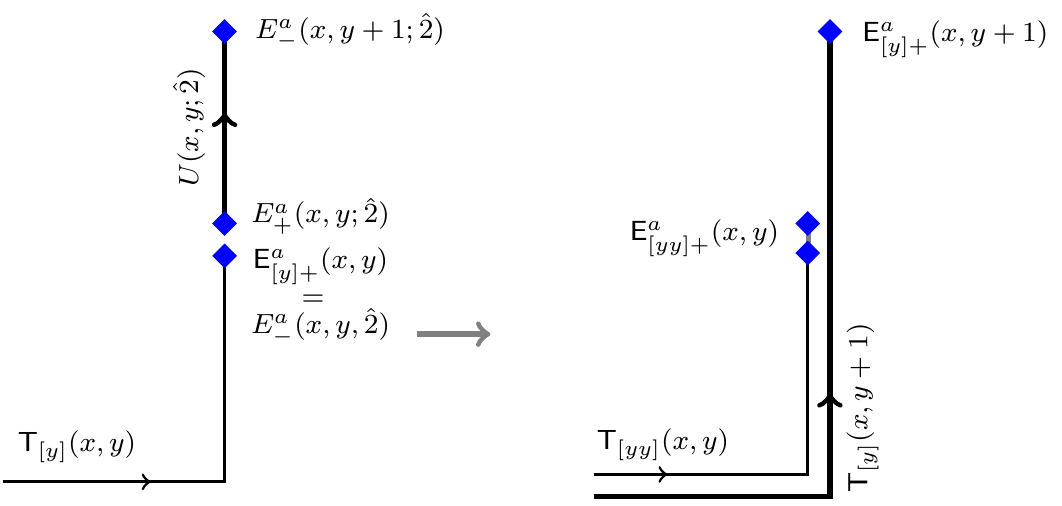}
\caption{Graphical representation of the canonical transformations (\ref{yay1}): iterative vertical string constructions in (\ref{yay1}) and the string electric field in (\ref{yyef2}). }
\label{f:can2b}
\end{figure}
In (\ref{yay1}),  the initial string operator ${\sf T}_{[y]}(x,y=1)$ is given in (\ref{gln0}).  The transformations (\ref{yay1}) are 
illustrated in  Figure \ref{f:can2b}.  Again the subscript $[yy]$ on ${\sf T}_{[yy]}(x,y)$ is to emphasize that its electric field is sum of two adjacent Kogut Susskind electric fields in the y direction: 
\begin{align}
{\sf E}^a_{[yy]+}(x,y) &= {{\sf E}}^a_{[y]+}(x,y)+ E^a_+(x,y; \hat 2)\nonumber\\
&= {E}^a_{-}(x,y; \hat 2)+ E^a_+(x,y; \hat 2).
\label{yyef2}
\end{align}
In (\ref{yyef2}) we have used (\ref{yay1}) to replace 
${\sf E}^a_{[y]+}(x,y)$ in terms of Kogut Susskind electric fields $E_-^a(x,y;\hat 2)$. 
We  again define 
${\sf T}_{[yy]}(x,{\sf N})={\sf T}_{[y]}(x,{\sf N})$ at the boundary for notational  convenience.
%$T_{[y]}(x,0) = T_{[xx]}(x,0)$.

\subsection{Plaquette loop operators}  
\begin{figure}[b]
\centering
\includegraphics[scale=.5]{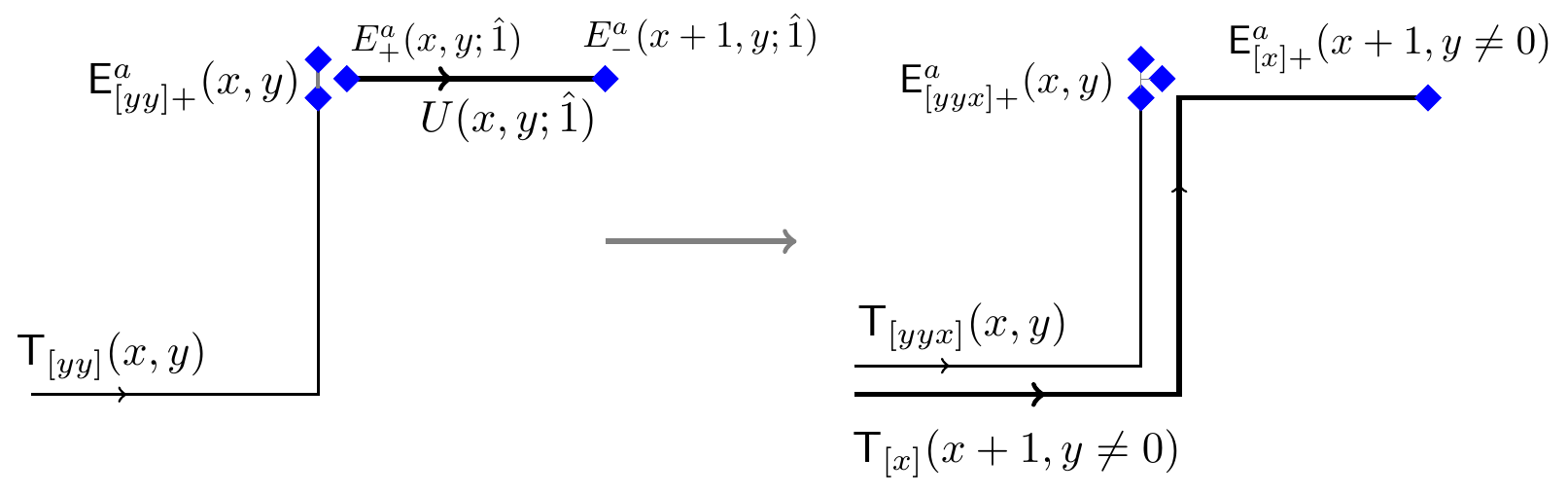}
\caption{Graphical representation of the canonical transformation in (\ref{xyxe}).}
\label{f:can3}
\end{figure}
In order to remove all local SU(N) gauge or string degrees of freedom and simultaneously obtain  SU(N) covariant loop flux operators, we now fuse the horizontal link operator $U(x,y\neq 0; \hat 1)$ with ${\sf T}_{[yy]}(x,y\neq 0)$  through the canonical transformations: 
\begin{align}
 {{\sf T}}_{[x]}(x+1,y) &~\equiv~ {\sf T}_{[yy]}(x,y) ~ U(x,y; \hat 1),\nonumber\\
  {\sf T}_{[yyx]}(x,y) &~= ~ {\sf T}_{[yy]}(x,y) \nonumber\\
{\sf E}_{[x]+}^a(x+1,y) &~=~ E^a_-(x+1,y; \hat 1),\nonumber\\ 
{\sf E}^a_{[yyx]+}(x,y)  &~=~ 
{\sf E}_{[yy]+}^a(x,y)+ E^a_+(x,y; \hat 1)  
\label{xyxe}
\end{align}
%\begin{figure}
%\centering
%\label{f:can3}
%\includegraphics[scale=.7]{5.eps}
%\caption{}
%\end{figure}
at $x=0,1,2,\cdots ,({\sf N}-1)$ and  $y=1,2,\cdots ,{\sf N}$.
The above transformations are illustrated in Figure \ref{f:can3}. Using (\ref{yay1}), the right electric field of the string flux operator ${\sf T}_{[yyx]}(x,y)$ is:
\begin{align} 
{\sf E}^a_{[yyx]+}& = {\sf E}^a_{[yy]+}(x,y) + E^a_+(x,y; \hat 1) \nonumber\\
&= E^a_-(x,y; \hat 2)+ E^a_+(x,y; \hat 2)+ E^a_+(x,y; \hat 1).
\label{pgl} 
\end{align}
 The initial loop operators $\left(W(x,y), {\mathbb E}^a(x,y)\right)$ shown in Figure-\ref{f:can4} are defined as:
 \begin{align} 
\label{loopop1}
{W}(x,y) &~\equiv~ {{\sf T}}_{[x]}(x,y \neq 0) ~{\sf T}^\dagger_{[yyx]}(x,y), \nonumber\\
{{\sf T}}_{[yyxx]}(x,y) &~\equiv~ {{\sf T}}_{[yyx]}(x,y), \nonumber\\
{\mathbb E}^a_{-}(x,y) &~=~ {\sf E}_{[x]-}^a(x,y\neq 0), \nonumber\\
{\sf E}^a_{[yyxx]+}(x,y) &~=~ 
{\sf E}^a_{[x]+}(x,y \neq 0)+ {\sf E}^a_{[yyx]+}(x,y). 
\end{align}
Above $\left(W(x,y), {\mathbb E}^a_\mp(x,y)\right)$ are canonically conjugate pairs.
We note that the conjugate electric fields of the 
string operators ${\sf T}_{[yyxx]}$ vanishes in ${\cal H}^p$ 
as: 
\begin{align} 
&{\sf E}^a_{[yyxx]+}(x,y) ~=~ 
{\sf E}^a_{[yyx]+}(x,y) +{\sf E}^a_{[x]+}(x,y \neq 0) \nonumber \\
& = \Big(E^a_-(x,y; \hat 2)+ E^a_+(x,y; \hat 2)+ E^a_+(x,y; \hat 1)
 +E^a_-(x,y; \hat 1)\Big) \nonumber\\ 
 &= {\cal G}^a(x,y) = 0. 
\label{cycle} 
\end{align}
In (\ref{cycle}), we have used (\ref{xyxe}) and (\ref{pgl})  to replace ${\sf E}_{[x]+}^a(x,y\neq 0)$ and ${\sf E}^a_{[yyx]+}(x,y)$ respectively in terms of Kogut-Susskind electric fields. The relationship (\ref{cycle}) solving the SU(N) Gauss law at $(x,y)$ is graphically illustrated  in Figure \ref{f:can4} and also earlier in Figure {\ref{flooplink}-a. 
\begin{figure}[t]
\centering
\includegraphics[scale=.54]{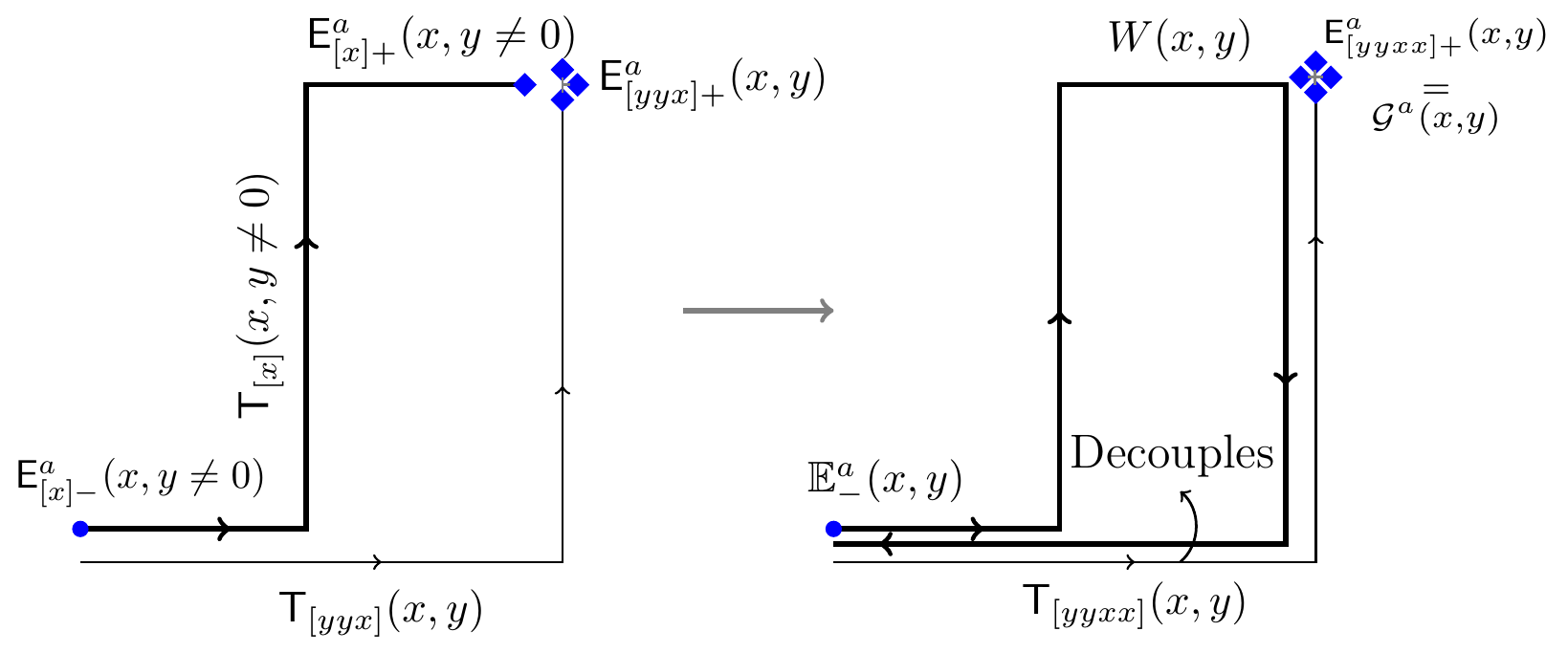}
\caption{Graphical representation of the canonical transformation in (\ref{loopop1}).}
\label{f:can4}
\end{figure}

At this stage all the local gauge degrees of freedom, 
contained in the string operators $T_{[yyxx]}(x,y)$, have been removed.  We now relabel   
${\sf T}_{[yyxx]}(x,y)$ as ${\sf T}(x,y)$ and 
${\sf E}_{[yyxx]\pm}^a(x,y)$ as ${\sf E}_\pm^a(x,y)$ 
for notational simplicity.
%We work only with the loop flux operators ${W}(x,y)$ and its conjugate electric fields ${\mathbb E}_\pm^a(x,y)$.  
To simplify the magnetic field terms in the Kogut Susskind Hamiltonian (\ref{ksham}), we further 
make  the last set 
of canonical transformations (\ref{lct}) which transform 
the loop operators $(W(x,y), {\mathbb E}^a_\pm(x,y))$ in (\ref{loopop1}) into the final  plaquette loop operators  $\left({\cal W}(x,y), {\cal E}_\mp^a(x,y)\right)$ as shown in Figure \ref{f:can5}. We define:
% ${W}(x,\sf N) = {W}(x, \sf N)$ to initiate the transformation from the top to bottom (i.e. from $y=L+1$ to $y=1$): 
\begin{align} 
{\cal W}(x,y) &~\equiv~ {W}(x,y-1)~ \bar {W}^\dagger(x,y), \nonumber\\
\bar { W}(x,y-1) &~\equiv~ {W}(x,y-1);  \nonumber\\
{\cal E}_+^a(x,y) &~=~ \bar {\mathbb E}_- ^a(x,y),\nonumber\\
 \bar {\mathbb E}_+^a(x,y-1) &~=~ {\mathbb E}_+^a(x,y-1) + \bar {\mathbb E}_+^a(x,y)
\label{lct} 
\end{align} 
Above  $\left[{\cal W}(x,y),  {\cal E}_+^a(x,y)   \right], ~  \left[\bar{W}^\dagger(x,y),  \bar {\mathbb E}_+^a(x,y)\right]$ are canonically conjugate loop operators and  $y={\sf N},~ ({\sf N}-1), \cdots, 1$.
The canonical transformation is initiated with the boundary operator $\bar W(x,y={\sf N}) 
\equiv W(x,y={\sf N})$ and at the lower boundary ${\cal W}(x,1)\equiv \bar{W}^\dagger(x,1)$.  
\begin{figure}[b]
 %\centering
 \includegraphics[scale=.57]{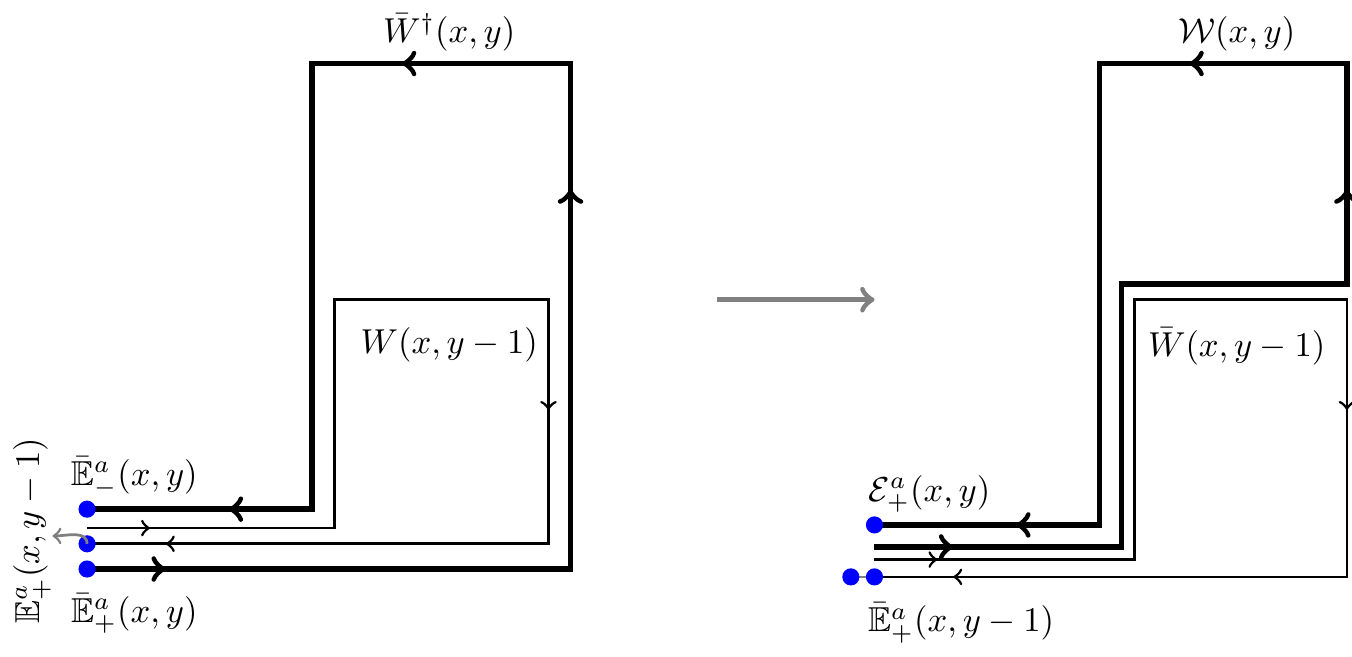}
 \caption{Graphical representation of the canonical transformation in (\ref{lct}).}
 \label{f:can5}
 \end{figure}
%THIS ALSO IMPLIES THE INITIAL BAR(W) ELECTRIC FIELDS. 
%These canonical transformations transform loop operators $\left(W(x,y)

Having constructed plaquette loop operators and conjugate electric fields using the canonical transformations (\ref{xyxe})-(\ref{lct}), we now use these relations to write the plaquette loop electric fields directly in terms of the Kogut-Susskind link electric fields. Using (\ref{lct}),
% thereby condensing the entire series of canonical transformations into a single equation. From eqn.(\ref{lct}), we have 
%{\footnotesize
\begin{align}
{\cal E}_+^a(x,y)~&=~ \bar{{\mathbb E}}_-^a(x,y)  =-R_{ab}(\bar{W}(x,y))\bar{\mathbb{E}}_+^a(x,y) \nonumber \\ 
&=-R_{ab}(\bar{W}(x,y))\left\{ \mathbb{E}_+^b(x,y)+\bar{\mathbb E}_+^b(x,y+1)\right\}
\end{align}
%}
Iterating this relation and using the relation $ \mathbb{E}_+^b(x,y') =  -R_{bc}(W^\dagger(x,y'))\mathbb{E}_-^c(x,y') $, we get 
%{\footnotesize
\begin{align}
{\cal E}_+^a(x,y)&= -R_{ab}(\bar{W}(x,y))\sum\limits_{y'=y}^{{\sf N}} \mathbb{E}_+^b(x,y') \nonumber \\ 
&=R_{ab}({W}(x,y))\sum\limits_{y'=y}^{{\sf N}}R_{bc}(W^\dagger(x,y'))\mathbb{E}_-^c(x,y')
\end{align}
%}
From eqn. (\ref{loopop1}) we have  ${\mathbb E}_-^c(x,y')={\sf E}_{[x]-}^c(x,y')=-R_{cd}({\sf T}_{[x]}(x,y'))~{\sf E}_{[x]+}^d(x,y')$ and from (\ref{xyxe}), ${\sf E}_{[x]+}^d(x,y')=E_-^d(x,y')$. Therefore,
{\footnotesize
\begin{align}
&{\cal E}_+^a(x,y) 
%&=~-R_{ab}({W}(x,y))\sum\limits_{y'=y}^{{\sf N}}-R_{bc}(W^\dagger(x,y'))R_{cd}({\sf T}_{[x]}(x,y')){\sf E}_{[x]+}^d(x,y') \nonumber \\
=-\sum\limits_{y'=y}^{{\sf N}}R_{ab}\left({W}(x,y)W^\dagger(x,y'){\sf T}_{[x]}(x,y')\right)E_-^b(x,y',\hat{1}) \nonumber \\
&=-\sum\limits_{y'=y}^{{\sf N}}R_{ab}\Big({\sf T}(x-1,y)~U(x-1,y;\hat{1})~\prod_{y''=y}^{y'}~U(x,y''; \hat{2})\Big)\nonumber \\ 
& \phantom{xxxxxxxxxxxxxxxxxxxxxxxxxxxxxxxxxxxxxxx}E_-^b(x,y',\hat{1}) \nonumber \\
&\equiv -\sum\limits_{y'=y}^{{\sf N}}~R_{ab}\left(S(x,y,y')\right) E_-^b(x,y';\hat 1).
\label{looptolinkr} 
\end{align}}
This is the relation (\ref{slef}) in the text which was 
further graphically illustrated in Figure \ref{flooplink}-b.
%}
%Also, from eqn.(\ref{cycle}), The right electric field of the string operators are given by  
%{\footnotesize
%\begin{align}
%{\sf E}^a_+(x,y) = {\sf T}_{[xxyy]+}(x,y)= \sum_{i=1}^{2} \left[ E^a_-(x,y;\hat i)+ E^a_+(x,y;\hat i)\right]   = {\cal G}^a(x,y) =0 
%\end{align}
%}

%{\bf \Large RECOVER  (\ref{slef}) HERE.} \\ \\ 

\section{From loops \& strings to links}
In this part, we  systematically  write down all Kogut-Susskind link electric fields in terms of loop flux operators and loop electric fields.
% by keeping track of the canonical transformation steps involved in their construction.
%Any link electric field can be written in terms of the plaquette loop operator and its conjugate electric fields by keeping track of the canonical transformation steps involved in their construction.
 We calculate the link electric fields in three separate cases: $a)$  $E^a(x,y=0;\hat{1})$ shown in Figure \ref{flinkloop}-a, ~$b)$ $E^a(x,y\neq 0;\hat{1})$  shown in  Figure \ref{flinkloop}-b and c) $E^a(x,y;\hat{2})$ shown in  Figure \ref{flinkloopc}.

\subsection{Case (a): ~$E_+^a(x,0;\hat{1})$} 

Consider the left electric field $E_+(x,0;\hat{1})$ of a Kogut Susskind link flux operator $U(x,0;\hat 1)$. 
%\[ E_+(0,0;\hat{1}) =-R\left(U(0,0; \hat{1})\right)E_-(1,0;\hat{1})  \
From canonical transformation (\ref{xax1}) illustrated in Figure \ref{f:can1d},  we have {\footnotesize${\sf E}_{[xx]+}^b(x,0)=E_-^b(x,0,\hat{1})+E_+^b(x,0;\hat{1})$}. Therefore,
{\footnotesize
\begin{align} E_+^a(x,0;\hat{1})&=-R_{ab}(U(x,0,\hat{1}))E_-^b(x+1,0,\hat{1})\nonumber \\
&=-R_{ab}(U(x,0,\hat{1}))\left\{{\sf E}_{[xx]+}^b(x+1,0)-E_+^b(x+1,0;\hat{1})\right\}
\end{align}
}
% ={\sf E}_{[xx]+}^a(x,0)+R_{ab}\left(U(x,0;\hat{1})\right)E_+^b(x+1,0;\hat{1})  \]
Iterating this expression, we obtain 
%I.e;
%\begin{align}
%E_+(0,0;\hat{1})&= -R\left(U(0,0; \hat{1})\right)E_-(1,0;\hat{1}) = -R\left(U(0,0; \hat{1})\right)\left\{{\sf E}_{[xx]+}(1,0)-E_+(1,0;\hat{1})\right\} \nonumber \\
%\end{align}
% Substituting for $E_+(1,0;\hat{1})$ and Iterating, we get
{\footnotesize
 \begin{align}
 E_+^a(x,0;\hat{1})
 &= R_{ab}\big({\sf T}^\dagger(x,0)\big)\sum\limits_{{\bar x}=x+1}^{{\sf N}} -R_{bc}\left({\sf T}({\bar x},0)\right) {\sf E}_{[xx]+}^c({\bar x},0) 
  \label{e+00}
\end{align}
}
Above, we have made use of the fact that {\footnotesize ${\sf T}^\dagger(x,0){\sf T}(\bar{x},0)=U(x,0;\hat{1})U(x+1,0;\hat{1})\cdots U(\bar{x}-1,0;\hat{1})$} if  $\bar{x}>x$.
%From this expression it is clear that all the ${\sf E}_{[xx]+}(x',0),~ x'>x$ are parallel transported back to the point $(x,0)$ to give $E_+(x,0; \hat{1})$. This is a general trend which will be seen at each step of canonical transformation. 
From this expression it is clear that all the $\vec{{\sf E}}_{[xx]+}({\bar x},0) ; {\bar x}>x$ are parallel transported back to the point $(x,0)$ to give $\vec{E}_+(x,0,\hat{1})$ so that the gauge transformations of link and string operators are 
consistent with (\ref{e+00}). This is a general trend which will be seen at each step of canonical transformations. In fact, the parallel transport 
is required by the SU(N) gauge transformations of the 
link and string electric fields in (\ref{e+00}). We now convert the string electric fields ${\sf E}^a_{[xx]}(x,0)$ into loop electric fields ${\cal E}^a(x',y')$ in three steps. 
%\vspace{.5 cm}\newline
%\begin{itemize}
%\item 
%\phantom{xxxxxx} 
%\textbf{$\bullet~{  \vec{\sf E}_{[xx]+}\rightarrow \vec{\sf E}_{[y]+}\rightarrow {\vec{\sf E}}_{\bf [yy]+}}$}
\subsubsection{Converting ${\vec{\sf E}_{[xx]+}\rightarrow \vec{\sf E}_{[y]+}\rightarrow {\vec{\sf E}}_{\bf [yy]+}}$}
%\newline
%\vspace{.25cm}

Writing down $\vec{\sf E}_{[xx]+}(\bar{x},0)$ in terms of $\vec{\sf E}_{[xxy]+}(\bar{x},0)$ and $\vec{\sf E}_{[y]+}(\bar{x},1)$ using canonical transformation \ref{gln0} shown in  Figure \ref{f:can2}:
{\footnotesize
 \begin{align}
{\sf E}_{[xx]+}^a(\bar{x},0)&={\sf E}_{[xxy]+}^a(\bar{x},0)-E_+^a(\bar{x},0;\hat{2}) \nonumber \\
&= R_{ab}\left(U(\bar{x},0; \hat{2})\right)\underbrace{E_-^b(\bar{x},1;\hat{2})}_{{\sf E}_{[y]+}^b(\bar{x},1)}
 \label{xx+}
\end{align}
}
%I.e,
% \begin{align}
% 
%{\sf E}_{[xx]+}(x,0)&= {\sf E}_{[xxy]+}(x,0)-R\left(U(x,0;\hat{2})\right)\underbrace{E_-(x,1;\hat{2})}_{{\sf E}_{[y]+}(x,1)}\nonumber \\
%&=R\left(U(x,0;\hat{2})\right){\sf E}_{[y]+}(x,1)
%\end{align}
We have used the fact that ${\sf E}_{[xxy]+}(\bar{x},0)=0$ by Gauss law (\ref{glx0}) at 
$(\bar{x},0)$.
But from  (\ref{yay1}) and Figure \ref{f:can2b}: 
%\[
%{\sf E}_{[yy]+}(x,1)={\sf E}_{[y]+}(x,1)+E_+(x,1;\hat{2})
%\]
{\footnotesize
\begin{align}
{\sf E}_{[y]+}^a(\bar{x},1)&={\sf E}_{[yy]+}^a(\bar{x},1)-E_+^a(\bar{x},1;\hat{2})\nonumber \\
&={\sf E}^a_{[yy]+}(\bar{x},1)+R_{ab}\left(U(\bar{x},1;\hat{2})\right)\underbrace{E_-^b(\bar{x},2;\hat{2})}_{{\sf E}_{[y]+}^b(\bar{x},2)} \nonumber \\
&= R_{ab}\left({\sf T}^\dagger(\bar{x},1)\right)\sum_{\bar{y}=1}^{\sf N} R_{bc}\left({\sf T}(\bar{x},\bar{y})\right) {\sf E}^c_{[yy]+}(\bar{x},\bar{y}).
\label{ey+}
\end{align}
}
%{\footnotesize
%\begin{align}
%{\sf E}_{[y]+}^b({\bar x},1)&={\sf E}_{[yy]+}^b({\bar x},1)-E_+^b({\bar x},1;\hat{2})={\sf E}_{[yy]+}({\bar x},1)+R_{bc}\left(U({\bar x},1;\hat{2})\right)\underbrace{E_-^c({\bar x},2;\hat{2})}_{{\sf E}_{[y]+}^c({\bar x},2)}
%\end{align}
%}
%{\footnotesize
%\begin{align}
%{\sf E}_{[xx]+}^a({\bar x},0)&=-R_{ab}\left(U({\bar x},0;\hat{2})\right){\sf E}_{[yy]+}^b({\bar x},1)-R_{ab}\left(U({\bar x},0;\hat{2})U({\bar x},1;\hat{2})\right){\sf E}_{[yy]+}^b({\bar x},2)+\cdots
%\end{align}
%}
%{\footnotesize
%\begin{align}
%E_+^a(x,0;\hat{1})
%%&=\sum\limits_{{\bar x}}-R\left(T({\bar x},0)\right)\left\{ -R\left(U({\bar x},0;\hat{2})\right){\sf E}_{[yy]+}({\bar x},1)-R\left(U({\bar x},0;\hat{2})U({\bar x},1;\hat{2})\right){\sf E}_{[yy]+}({\bar x},2)+\cdots \right\}\nonumber \\
%&=-R_{ac}\big({\sf T}^\dagger(x,0)\big)\sum\limits_{{\bar x}=x+1}^{{\sf N}}\sum\limits_{{\bar y}=1}^{{\sf N}} R_{cb}\left({\sf T}({\bar x},{\bar y})\right){\sf E}_{[yy]+}^b({\bar x},{\bar y})
%\label{e+yy+}
%\end{align}
%}
\noindent Substituting it back into eqn. (\ref{xx+}) for {\footnotesize ${\sf E}_{[xx]+}^a(\bar{x},0)$} and using {\footnotesize $U(\bar{x},0;\hat 2){\sf T^\dagger}(\bar{x},1) = {\sf T^\dagger}(\bar{x},0)$}, we get 
{\footnotesize
\begin{align}
{\sf E}_{[xx]+}^a(\bar{x},0)= R_{ab}\left({\sf T}^\dagger(\bar{x},0)\right)\sum_{\bar{y}=1}^{\sf N} R_{bc}\left({\sf T}(\bar{x},\bar{y})\right) {\sf E}^c_{[yy]+}(\bar{x},\bar{y}).
%-R_{ab}\left(U(x',0;\hat{2})\right){\sf E}_{[yy]+}^b(x',1)-R_{ab}\left(U(x',0;\hat{2})U(x',1;\hat{2})\right){\sf E}_{[yy]+}^b(x',2)+\cdots
\end{align}
}
\noindent Putting this into eqn. (\ref{e+00}) we get
{\footnotesize
\begin{align}
E_+^a(x,0;\hat{1})
%&=\sum\limits_{r}-R\left(T(r,0)\right)\left\{ -R\left(U(r,0;\hat{2})\right){\sf E}_{[yy]+}(r,1)-R\left(U(r,0;\hat{2})U(r,1;\hat{2})\right){\sf E}_{[yy]+}(r,2)+\cdots \right\}\nonumber \\
=-R_{ab}\big({\sf T}^\dagger(x,0)\big)\sum\limits_{\bar{x}=x+1}^{\sf N}\sum\limits_{\bar{y}=1}^{\sf N} R_{bc}\left({\sf T}(\bar{x},\bar{y})\right){\sf E}_{[yy]+}^c(\bar{x},\bar{y}).
\label{e+yy+}
\end{align}
}
%Using the canonical transformation eqn. (\ref{xyxe}) (Figure. \ref{f:can3}) to write down ${\sf E}_{[yy]+}^b(r,s)$ in terms of ${{\sf E}}_{[x]+}^d(r+1,s)$ , we have 
%%\[ {\sf E}_{[yyx]+}(r,s)={\sf E}_{[yy]+}(r,s)+E_+(r,s; \hat{1}) \] 
%%Therefore, 
%\begin{align}
%{\sf E}_{[yy]+}^b(r,s) ={\sf E}_{[yyx]+}^b(r,s)-E_+^b(r,s; \hat{1})= {\sf E}_{[yyx]+}^b(r,s)+R_{bd}\left(U(r,s; \hat{1})\right)\underbrace{E_-^d(r+1,s; \hat{1})}_{{{\sf E}}_{[x]+}^d(r+1,s)}
%\end{align}
%\item 
%\phantom{xxxxxx} \textbf{$\bullet ~{ \vec{\sf E}_{[yy]+}\rightarrow \vec{ \sf E}_{[x]+}\rightarrow \vec{\mathbb E}_{\pm}}$}
%\newline

%\vspace{-0.3cm}

\subsubsection{Converting ${ \vec{\sf E}_{[yy]+}\rightarrow \vec{ \sf E}_{[x]+}\rightarrow \vec{\mathbb E}_{\pm}}$}
%\newline

From canonical transformation (\ref{xyxe}) (Figure \ref{f:can3}) we have {\footnotesize ${\sf E}_{[yyx]+}^c({\bar x},{\bar y})={\sf E}_{[yy]+}^c({\bar x},{\bar y})+ E_+^c({\bar x},{\bar y},\hat{1})$} and {\footnotesize ${{\sf E}}_{[x]+}^d({\bar x}+1,{\bar y})=E_-^d({\bar x}+1,{\bar y},\hat{1})$}. Therefore,
%Using the canonical transformation eqn. (\ref{xyxe}) (Figure. \ref{f:can3}) to write down ${\sf E}_{[yy]+}^c({\bar x},{\bar y})$ in terms of ${{\sf E}}_{[x]+}^d({\bar x}+1,{\bar y})$ , we have 
%\[ {\sf E}_{[yyx]+}({\bar x},{\bar y})={\sf E}_{[yy]+}({\bar x},{\bar y})+E_+({\bar x},{\bar y},\hat{1}) \] 
%Therefore, 
{\footnotesize
\begin{align}
& {\sf E}_{[yy]+}^c({\bar x},{\bar y}) ={\sf E}_{[yyx]+}^c({\bar x},{\bar y})-E_+^c({\bar x},{\bar y},\hat{1})\nonumber\\
&= {\sf E}_{[yyx]+}^c({\bar x},{\bar y})+R_{cd}\left(U({\bar x},{\bar y},\hat{1})\right)\underbrace{E_-^d({\bar x}+1,{\bar y},\hat{1})}_{{{\sf E}}_{[x]+}^d({\bar x}+1,{\bar y})}
\label{yy+}
\end{align}
}
%Further, from the canonical transformation, eqn. \ref{loopop1} (Figure \ref{f:can4}), we have 
%$${\sf E}_{[yyx]+}^b(r,s) = {\sf E}_{[yyxx]+}^b(r,s) - {{\sf E}}_{[x]+}^b(r,s) =- {{\sf E}}_{[x]+}^b(r,s)$$
%Here, we have used the fact that ${\sf E}_{[xxyy]+}^b(r,s)=0$ by Gauss law (eqn. \ref{cycle}).
%Also, using eqn. (\ref{loopop1})
%$${{\sf E}}_{[x]+}^b(r,s)=-R_{be}({\sf T}^\dagger_{[x]} (r,s)){ {\sf E}}_{[x]-}^e(r,s)=-R_{be}({\sf T}^\dagger_{[x]}(r,s)){\mathbb E}_-^e(r,s).$$
%Therefore, 
%\begin{align}
%\label{tyy+w}
% {\sf E}_{[yy]+}^b(r,s) & = R_{be}({\sf T}^\dagger_{[x]}(r,s)){\mathbb E}_-^e(r,s) - R_{bd}\left(U(r,s; \hat{1})\right)R_{de}({\sf T}^\dagger_{[x]}(r+1,s)){\mathbb E}_-^e(r+1,s)
%\end{align}
Further, the canonical transformations (\ref{loopop1})   (Figure \ref{f:can4}) imply: 
{\footnotesize
\[ {\sf E}_{[yyx]+}^c({\bar x},{\bar y}) = {\sf E}_{[yyxx]+}^c({\bar x},{\bar y}) - {{\sf E}}_{[x]+}^c({\bar x},{\bar y}) =- {{\sf E}}_{[x]+}^c({\bar x},{\bar y})\]  
}
Here, we have used the fact that {\footnotesize ${\sf E}_{[xxyy]+}^c({\bar x},{\bar y})=0$} by Gauss law at $(\bar{x},\bar{y})$ (eqn. (\ref{cycle})).
Also, from eqn. (\ref{loopop1}), {\footnotesize ${\sf E}_{[x]-}^d(\bar{x},\bar{y})={\mathbb E}_-^d({\bar x},{\bar y})$}. Therefore,
%Also, using eqn. (\ref{loopop1})
{\footnotesize
\begin{align}
{{\sf E}}_{[x]+}^c({\bar x},{\bar y})&=-R_{cd}({{\sf T}}^\dagger_{[x]} ({\bar x},{\bar y})){ {\sf E}}_{[x]-}^d({\bar x},{\bar y})\nonumber \\
&=-R_{cd}({{\sf T}}^\dagger_{[x]}({\bar x},{\bar y})){\mathbb E}_-^d({\bar x},{\bar y})
\label{tx+}
 \end{align}
 }
\noindent Substituting for $\vec{\sf E}_{[yyx]+}, \vec{\sf E}_{[x]+}$ in eqn. (\ref{yy+}) and using the relation {\footnotesize $U({\bar x},{\bar y}; \hat{1}){\sf T}^\dagger_{[x]}({\bar x}+1,{\bar y}) = {\sf T}^\dagger({\bar x},{\bar y})$},
{\footnotesize
\begin{align}
 {\sf E}_{[yy]+}^c({\bar x},{\bar y}) & = R_{cd}\left({{\sf T}}^\dagger_{[x]}({\bar x},{\bar y})\right){\mathbb E}_-^d({\bar x},{\bar y}) \nonumber \\ &- R_{cd}\left({\sf T}^\dagger({\bar x},{\bar y})\right){\mathbb E}_-^d({\bar x}+1,{\bar y})
 \label{tyy+w}
\end{align}
}
Putting (\ref{tyy+w}) in (\ref{e+yy+}) and using the defining relations {\footnotesize\begin{align}&{\sf T}(\bar{x},\bar{y}) {\sf T}^\dagger_{[x]}(\bar{x},\bar{y}) \equiv { W}^\dagger(\bar{x},\bar{y});\nonumber \\&{\mathbb E}_+^b(\bar{x},\bar{y}) \equiv  - R_{bd}\left({W}^\dagger(\bar{x},\bar{y})\right) {\mathbb E}_-^d(\bar{x},\bar{y}). \nonumber 
%\\&{\sf T}(\bar{x},\bar{y}){\sf T}^\dagger(\bar{x},\bar{y}) = {\sf I},
\end{align}}
we get a simple relation: 
%Putting this into the expression \ref{e+yy+}  for $E_+(x,0;\hat{1})$, we get 
{\footnotesize
\begin{align} 
E^a_+(x,0;\hat{1})& = R_{ab}\big({\sf T}^\dagger_{}(x,0)\big)\nonumber \\&\sum\limits_{\bar{x}=x+1}^{\sf N}~\sum\limits_{\bar{y}=1}^{\sf N} ~\Big[{\mathbb E}_+^b(\bar{x},\bar{y}) + {\mathbb E}_-^b(\bar{x}+1,\bar{y})\Big].
\label{e+bbw+}
\end{align}
}
%\item 
%\phantom{xxxxxx} \textbf{$\bullet ~{\vec{\mathbb E}_{\pm}\rightarrow \vec{\cal E}_{\pm}}$} 
%\newline 
\vspace{-0.5cm}

\subsubsection{Converting ${\vec{\mathbb E}_{\pm}\rightarrow \vec{\cal E}_{\pm}}$} 
%\newline 
%\vspace{0.25cm}

To write ${E}_+^a(x,0;\hat 1)$ in terms of the final plaquette loop electric fields ${\cal E}_\pm^b$, we first use the canonical transformation in  equation (\ref{lct}) and shown in Figure \ref{f:can5}: 
{\footnotesize
%\begin{align}
${\mathbb E}_+^b(\bar{x},\bar{y})=\bar{{\mathbb E}}_+^b(\bar{x},\bar{y})-\bar{{\mathbb E}}_+^b(\bar{x},\bar{y}+1)$.
%\label{bbe+}
%\end{align}
}
This enables us to write down the first term in the eqn. (\ref{e+bbw+}) in terms of $\vec{\cal E}_-$ as follows: 
{\footnotesize
\begin{align} 
\sum\limits_{\bar{y}=1}^{\sf N} {\mathbb E}_+^b(\bar{x},\bar{y})&=\sum\limits_{\bar{y}=1}^{\sf N}\left[\bar{{\mathbb E}}_+^b(\bar{x},\bar{y})-\bar{{\mathbb E}}_+^b(\bar{x},\bar{y}+1)\right]= \bar{{\mathbb E}}_+^b(\bar{x},1) \nonumber\\
& = - R_{bc}\left(\bar{W}^\dagger(\bar{x},1)\right) \bar{{\mathbb E}}^c_-(\bar{x},1)\nonumber \\
&= -R_{bc}\left(\mathcal{W}(\bar{x},1)\right){\mathcal E}_+^c(\bar{x},1) = {\mathcal E}_-^b(\bar{x},1).
\label{w+}
\end{align}
}
Here, we have used the fact that 
%when summed over $\bar{y}$, all the terms in eqn.(\ref{bbe+}), except $\bar{{\mathbb E}}_+^b(\bar{x},1)$, cancel in pairs. Also, 
at the lower boundary, $\bar{W}^\dagger(\bar{x},1)=\mathcal{W}(\bar{x},1)$. 
We now write down the second term in eqn.(\ref{e+bbw+}) in terms of $\vec{\cal E}_\pm$.  Again using canonical transformation eqn.(\ref{lct}) (Figure \ref{f:can5})  as follows: 
{\footnotesize
\begin{align} 
&{\mathbb E}_-^b({\bar x}+1,{\bar y})=-R_{bc}(\bar{W}({\bar x}+1,{\bar y})){\mathbb E}_+^c({\bar x}+1,{\bar y}) \nonumber \\&= -R_{bc}(\bar{W}({\bar x}+1,{\bar y}))\left[\bar{{\mathbb E}}_+^c({\bar x}+1,{\bar y})-\bar{{\mathbb E}}^c_+({\bar x}+1,{\bar y}+1)\right] \nonumber \\
&=\bar{{\mathbb E}}_-^b({\bar x}+1,{\bar y})-R_{bc}\big(\bar{W}({\bar x}+1,{\bar y})\bar{W}^\dagger({\bar x}+1,{\bar y}+1)\big)\bar{{\mathbb E}}^c_-({\bar x}+1,{\bar y}+1)\nonumber \\
&= \bar{{\mathbb E}}_-^b({\bar x}+1,{\bar y})-R_{bc}({\mathcal W}({\bar x}+1,{\bar y}+1))\bar{{\mathbb E}}^c_-({\bar x}+1,{\bar y}+1) \nonumber \\
&= {\mathcal E}_+^b({\bar x}+1,{\bar y})+{\mathcal E}_-^b({\bar x}+1,{\bar y}+1)
\label{w-}
\end{align} 
}
Putting both the terms back into eqn. (\ref{e+bbw+}) for $E_+^a(x,0;\hat{1})$, we get
{\footnotesize
\begin{align} 
& E_+^a(x,0;\hat{1}) = R_{ab}\big({\sf T}^\dagger(x,0)\big)\nonumber \\ &\sum\limits_{{\bar x}=x+1}^{{\sf N}} \left\{{\mathcal E}_-^b({\bar x},1)+
\sum\limits_{{\bar y}=1}^{{\sf N}} \left[
{\mathcal E}_+^b({\bar x}+1,{\bar y})+{\mathcal E}_-^b({\bar x}+1,{\bar y}+1)\right]\right\} \nonumber \\
%&=R_{ab}\big({\sf T}^\dagger(x,0)\big)\bigg\{{\mathcal E}_-^b(x+1,1)+\sum\limits_{{\bar x}=x+1}^{{\sf N}}\sum\limits_{{\bar y}=1}^{{\sf N}}\bigg({\mathcal E}_-^b({\bar x}+1,{\bar y}+1)+{\mathcal E}_+^b({\bar x}+1,{\bar y})\bigg)\bigg\}\nonumber\\
&= R_{ab}\big({\sf T}^\dagger(x,0)\big)\bigg\{{\mathcal E}_-^b(x+1,1)+\sum\limits_{{\bar x}=x+2}^{{\sf N}}\sum\limits_{{\bar y}=1}^{{\sf N}}   {\sf L}^b({\bar x},{\bar y}) \bigg\}
\label{e+fin}
\end{align}
}
  Above, ${\sf L}^a(\bar{x},\bar{y}) \equiv {\cal E}_-^a(\bar{x},\bar{y}) +{\cal E}_+^a(\bar{x},\bar{y})$. 
%\end{itemize}
\subsection{Case (b):~$E_+^a(x,y \neq 0;\hat{1})$ }

%Now, consider a horizontal link electric field $E_+^a(x,y;\hat{1}); y\neq0$. 
The canonical transformation (\ref{xyxe}) and Figure \ref{f:can3}  state that {\footnotesize ${{\sf E}}_{[x]+}^b(x,y)=E_-^b(x,y;\hat{1})$}. Therefore,
%in the x direction but not along x axis. 
{\footnotesize
\begin{align}
E_+^a(x,y,\hat{1})&=-R_{ab}(U(x,y,\hat{1}))~E_-^b(x+1,y;\hat{1})\nonumber \\ &=-R_{ab}(U(x,y,\hat{1}))~{{\sf E}}_{[x]+}^b(x+1,y)
\end{align}
}
%Above, we have used the expression for ${{\sf E}}_{[x]+}(x+1,y)$ from the canonical transformation eqn. \ref{xyxe}.
 Using the relations (\ref{tx+}) and (\ref{w-})
 {\footnotesize
 \begin{align} &{{\sf E}}_{[x]+}^b(x+1,y)=-R_{bc}({{\sf T}}^\dagger_{[x]}(x+1,y)~{\mathbb E}_-^c(x+1,y);\nonumber \\ & {\mathbb E}_-^c(x+1, y)= {\mathcal E}_+^c(x+1,y)+{\mathcal E}_-^c(x+1,y+1) \end{align}}
%  we have the to write ${{\sf E}}_{[x]+}^c(x+1,y)$ in terms of ${\mathbb E}_-^c(x+1,y)$ and (\ref{w-}) to express ${\mathbb E}_-^c(x+1,y)$ in terms of ${\mathcal E}_+^c(x+1,y)$ and ${\mathcal E}_-^c(x+1,y+1)$
%  
 and relation {\footnotesize ${\sf T}^\dagger(x,y)= U(x,y,\hat{1}){{\sf T}}^\dagger_{[x]}(x+1,y)$} , we get  
{\footnotesize
\begin{align}
E_+^a(x,y,\hat{1})
%&=-R_{ai}\left(U(x,y,\hat{1})\right)\left\{-R_{ic}({{\sf T}}^\dagger_{[x]}(x+1,y)){{\sf E}}_{[x]-}^c(x+1,y)\right\} \nonumber \\
&= \left[R_{ab}(U(x,y,\hat{1}))R_{bc}({{\sf T}}^\dagger_{[x]}(x+1,y))\right]{\mathbb E}_-^c(x+1,y)\nonumber \\ &=R_{ac}\left({\sf T}^\dagger(x,y)\right){\mathbb E}_-^c(x+1,y) \nonumber \\
&= R_{ac}\left({\sf T}^\dagger(x,y)\right)\bigg\{{\mathcal E}_+^c(x+1,y)+{\mathcal E}_-^c(x+1,y+1)\bigg\}
\end{align}
}
Clubbing case (a) and case (b) together, 
%the electric field corresponding to any link in the x direction is given by 
{\footnotesize
\begin{align}
& E_+^a(x,y; \hat{1}) =  R_{ab}({\sf T}^\dagger(x,y))\nonumber \\ &\bigg({\cal E}_-^b(x+1,y+1)+{\cal E}_+^b(x+1,y) 
 +\delta_{y,0}\sum_{\bar x=x+2}^{\sf N}\sum_{\bar y=1}^{{\sf N}}{\mathbb L}^b(\bar x,\bar y)\bigg).
 \label{final1}
 \end{align}
 }
 We have defined ${\cal E}_\pm(x,0)\equiv0;~ {\cal E}_\pm(0,y)\equiv0$ for notational convenience.
 The relations (\ref{e+fin}) 
   were used in (\ref{kstowele}) and (\ref{loophamp}), (\ref{nlt}) to write down the Kogut Susskind Hamiltonian in terms of loop operators. 
 
 \subsection{Case (c):~ $E_+(x,y;\hat{2})$}
 %Now, let us calculate the electric field corresponding to a vertical link, $E_+(x,y;\hat{2})$. 
 The canonical transformations (\ref{yay1}) (Figure \ref{f:can2b}) state {\footnotesize ${\sf E}_{[y]+}^c(x,y)=E_-^c(x,y,\hat{2})$}. Therefore,
{\footnotesize
 \begin{align} 
 E_+^a(x,y;\hat{2})&=-R_{ac}\left(U(x,y;\hat{2})\right)E_-^c(x,y+1,\hat{2}) \nonumber \\ &= -R_{ac}\left(U(x,y;\hat{2})\right) {\sf E}_{[y]+}^c(x,y+1)
 \end{align}}
 Using the relation {\footnotesize ${\sf E}_{[y]+}^c(x,y)= {\sf E}_{[yy]+}^c(x,y)-E_+^c(x,y,\hat{2})$} from the canonical transformation eqn. (\ref{yay1}) (Figure \ref{f:can2b}), 
 {\footnotesize
 \begin{align}
 & E_+^a(x,y;\hat{2})= -R_{ac}(U(x,y,\hat{2}))\left\{{\sf E}_{[yy]+}^c(x,y+1)-E_+^c(x,y+1,\hat{2})\right\} \nonumber \\
 &= -R_{ac}(U(x,y,\hat{2})){\sf E}_{[yy]+}^c(x,y+1)\nonumber \\ &-R_{ac}\left(U(x,y,\hat{2})U(x,y+1,\hat{2})\right){\sf E}_{[yy]+}^c(x,y+2)-\cdots\nonumber\\
 &=-R_{ab}({\sf T}^\dagger(x,y))\sum\limits_{{\bar y}=y+1}^{\sf N} R_{bc}\left({\sf T}(x,\bar{y})\right)~{\sf E}_{[yy]+}^c(x,\bar{y})
 \end{align}
 }
 Using eqn.(\ref{tyy+w}),
{\footnotesize
%\begin{align}
% \nonumber \\
${\sf E}_{[yy]+}^c(x,\bar{y})= R_{cd}\left({\sf T}^\dagger_{[x]}(x,\bar{y})\right)~{\mathbb E}_-^d(x,\bar{y}) - R_{cd}\left({\sf T}^\dagger(x,\bar{y})\right)~{\mathbb E}_-^d(x+1,\bar{y})$
%\end{align}
}
and the expression 
{\footnotesize $W^\dagger(x,\bar{y})={\sf T}(x,\bar{y}){\sf T}^\dagger_{[x]}(x,\bar{y})$} from eqn. (\ref{loopop1}),
{\footnotesize
 \begin{align}
  E_+^a(x,y;\hat{2}) &= R_{ab}\left({\sf T}^\dagger(x,y)\right)\nonumber \\ &\sum\limits_{{\bar y}=y+1}^{\sf N}  \left[-R_{bc}\left(W^\dagger(x,\bar{y})\right){\mathbb E}_{-}^c(x,\bar{y}) + {\mathbb E}_-^b(x+1,\bar{y})\right].
 \label{e+bb}
 \end{align}
 } 
% Above, we have used the fact that $U(x,y,\hat{2})U(x,y+1,\hat{2})\cdots U(x,\bar{y}-1,\hat{2}){\sf T}^\dagger_{[x]}(x,\bar{y})={\sf T}^\dagger(x,y){ W^\dagger}(x,\bar{y})$. 
From eqn. (\ref{w-}), we have ${\mathbb E}_-^c(x,\bar{y})={\mathcal E}_+^c({x},{\bar y})+{\mathcal E}_-^c({x},{\bar y}+1)$. Therefore, {\footnotesize
${\mathbb E}_-^b(x+1,\bar{y})={\mathcal E}_+^b({x+1},{\bar y})+{\mathcal E}_-^b({x+1},{\bar y}+1)$
} and 
{\footnotesize
  \begin{align}
  &\sum\limits_{{\bar y}=y+1}^{\sf N} -R_{bc}\left(W^\dagger(x,\bar{y})\right)~{\mathbb E}_{-}^c(x,\bar{y})\nonumber \\ &=\sum\limits_{{\bar y}=y+1}^{\sf N}-R_{bc}\left(W^\dagger(x,\bar{y})\right)\Big[{\mathcal E}_+^c({x},{\bar y})+{\mathcal E}_-^c({x},{\bar y}+1)\Big]\nonumber\\
  &=\sum\limits_{{\bar y}=y+1}^{\sf N}\Big\{R_{bc}\left(W^\dagger(x,\bar{y}-1)\right){\mathcal E}_-^c({x},{\bar y})\nonumber\\&\phantom{xxxxxxxxxx}-R_{bc}\left(W^\dagger(x,\bar{y})\right){\mathcal E}_-^c({x},{\bar y}+1)\Big\}\nonumber\\
%  &=\sum\limits_{{\bar y}=y}^{\sf N}R_{bc}\left(W^\dagger(x,\bar{y})\right){\mathcal E}_-^c({x},{\bar y}+1)-\sum\limits_{{\bar y}=y+1}^{\sf N}R_{bc}\left(W^\dagger(x,\bar{y})\right){\mathcal E}_-^c({x},{\bar y}+1)\right]\nonumber\\
 &= R_{bc}\left(W^\dagger(x,y)\right){\mathcal E}_-^c({x},{ y+1})
   \end{align}}
Above, we have used the relations: {\footnotesize $W^\dagger(x,\bar{y})=W^\dagger(x,\bar{y}-1)~{\cal W}(x,\bar{y})$} and {\footnotesize$R_{cd}\left({\cal W}(x,\bar{y})\right){\cal E}_+^d(x,\bar{y})={\cal E}_-^c(x,\bar{y})$}.
Putting these two terms back into eqn. (\ref{e+bb}),
 {\footnotesize
  \begin{align}
   E_+^a(x,y;\hat{2})=& R_{ab}\left({\sf T}^\dagger(x,y)\right)\Bigg\{R_{bc}\left({ W}^\dagger(x,y)\right){\mathcal E}_-^c(x,y+1) \nonumber\\& ~+ \sum\limits_{{\bar y}=y+1}^{{\sf N}} \left[{\mathcal E}_+^b(x+1,{\bar y})+{\mathcal E}_-^b(x+1,{\bar y}+1)\right]\Bigg\}
  \end{align}
  }
  Therefore,
%   a general link left electric field in the y direction is 
{\footnotesize
\begin{align}
   E_+^a(x,y; \hat{2}) &= 
R_{ab}({\sf T}^\dagger(x,y)) \bigg({\cal E}_+^b(x+1,y+1)
\nonumber\\&+ R_{bc}({{\cal W}_{xy}(x,y)}){\cal E}_-^c(x,y+1)
+\sum\limits_{\bar y=y+2}^{{\sf N}}{\sf L}^b(x+1,\bar y) 
  \bigg)
  \label{final2}
  \end{align}
  }
  Above, ${\cal W}_{xy}(x,y))\equiv{\cal W}(x,1){\cal W}(x,2)\cdots{\cal W}(x,y)$ as defined in (\ref{partran}).  The relation (\ref{final2}) 
  was stated in (\ref{kstowele}) and used later 
  in (\ref{loophamp}) to get the SU(N) loop Hamiltonian.
  % = R_{bc}\big(
  %\prod_{\bar y=y}^{1}
  %{\cal W}(x,1){\cal W}(x,2)\cdots {\cal W}(x,y)
  %\big)$ }. We define {\footnotesize ${\cal W}(0,y)={\cal W}(x,0)={\sf T}(0,0)\equiv{\sf 1}$} and {\footnotesize ${\cal E}_\pm^a(0,y)={\cal E}_\pm^a(x,0)={\sf E}_\pm^a(0,0)\equiv{\sf 0}$} for convenience of notation.
%  and ${\mathbb L}^a(x,y) \equiv \left({\cal E}_-(x,y) +{\cal E}_+(x,y)\right)$
  
%\section{The global SU(N) Gauss law}

  Once the string operators decouple from the theory (as shown in the previous section), the only remaining or residual  Gauss law is at the origin. 
%  As shown in the previous section, the right electric field of string operators ${\sf T}(x,y)$ equals the generators of gauge transformation ${\cal G}^a(x,y)$ at $(x,y)$. Therefore, the string operators decouple from the physical gauge invariant sector, leaving behind a Gauss law only at the origin where there are no string operators.
%  After removing the unphysical string operators from the theory, the only Gauss law that remains is at the origin. The Gauss law at the origin 
This Gauss law at the origin states: {\footnotesize $$E_+^a(0,0;\hat{1})+ E_+^a(0,0;\hat{2}) =0.$$ }
When rewritten in terms of the plaquette electric fields using the above relations (\ref{final1}) and (\ref{final2}) it takes the form: 
  {\footnotesize
%  \sum\limits_{{\bar x},{\bar y}=1}^{{\sf N}}  {\mathcal E}_-^a({\bar x},{\bar y})+ {\mathcal E}_+^a({\bar x},{\bar y}) =0 
$$\sum\limits_{{x},{ y}=1}^{{\sf N}} {\sf L}^a({ x},{ y}) ~=~0.$$}
  This is the residual  SU(N)  Gauss law at the origin. 

\section{Mathieu equation}
\label{matheqn}    
We now exploit the simple action of the magnetic field term  on the hydrogen atom basis (\ref{mft}) to construct the dual magnetic basis where this magnetic term is diagonal. We define $\ket{ j} \equiv \ket {\tt n}_{\tt n \equiv 2j+1}$ and
{\footnotesize
\begin{eqnarray}
\ket{\omega} = \sum_{j}~ \chi_j(\omega) ~\ket{j}.
\label{dmf} 
\end{eqnarray}}
In (\ref{dmf}),  $\chi_j(\omega) = \frac{sin\left(2j+{1}\right)\frac{\omega}{2}}{sin\left(\frac{\omega}{2}\right)}$ are the SU(2) characters.
Using the recurrence relations \cite{varshalovich}:
{\footnotesize $$\chi_{j+\frac{1}{2}}(\omega)  + \chi_{j-\frac{1}{2}}(\omega) = 2 cos\left(\frac{\omega}{2}\right) \chi\left(\omega\right),$$ }
we get 
{\footnotesize
\begin{eqnarray} 
H_B \ket{\omega} = 
%\frac{1}{g^2} 
~ \frac{1}{g^2}~\left(Tr {\cal W}\right)~ \ket{\omega} 
= 
%\frac{1}{g^2} 
~\frac{2}{g^2}  ~cos\left(\frac{\omega}{2}\right)~\ket{\omega}.
\label{mev}
\end{eqnarray}
}
Note that $\omega$ is a gauge invariant angle. 
We now  use the differential equation of the SU(2) character \cite{varshalovich}: 
{\footnotesize
$$\frac{d^2\chi_j}{d\omega^2}+ cot\left(\frac{\omega}{2}\right)\frac{d\chi_j(\omega)}{d\omega}+j(j+1)\chi_j(\omega)=0$$} to convert $H_E$ in (\ref{lb}) into differential operator in $\omega$. Finally 
the Schr\"odinger equation $H\ket{\psi}_\epsilon  = \epsilon\ket{\psi}_\epsilon$ in this gauge invariant loop basis is  the Mathieu equation:
{\footnotesize
\begin{eqnarray} 
\Big[\frac{d^2}{d\omega^2} +\frac{1}{4}\Big]\phi_{\epsilon}(\omega)+ \frac{\kappa}{4} 
\Big[\epsilon- 2\kappa\left(1-cos\left(\frac{\omega}{2}\right)\right)\Big]\phi_{\epsilon}\left(\omega\right) =0.
\label{schreqn}
\end{eqnarray}
}  
In (\ref{schreqn}) we have defined $\kappa \equiv \frac{1}{g^2}$ and $\phi_\epsilon(\omega) \equiv sin\frac{\omega}{2} ~\psi_\epsilon(\omega)$ where 
$~\psi_\epsilon(\omega) \equiv \langle\omega|\psi\rangle_\epsilon$. The Mathieu equation (\ref{schreqn}) and its discrete solutions has been  extensively discussed in the past in the context of single plaquette lattice gauge theory \cite{bishop,robson,kolawa,mathieu}.

\end{document}